\documentclass[twocolumn,times]{aastex63}

\usepackage{CJKutf8}

\usepackage{amsmath}
\usepackage{amssymb}
\usepackage{amsfonts}
\usepackage{graphicx}
\usepackage{wasysym}
\usepackage{multirow}
\usepackage{mdframed}
\usepackage[T1]{fontenc}

\hypersetup{linkcolor=violet,citecolor=blue,filecolor=cyan,urlcolor=violet}

\newcommand{\amm}{\mbox{NH$_3$}}

\newcommand{\ammthree}{\mbox{\amm{}\,(3,\,3)}}

\newcommand{\kms}{\mbox{km\,s$^{-1}$}}

\newcommand{\sqc}{\mbox{cm$^{-2}$}}
\newcommand{\cc}{\mbox{cm$^{-3}$}}

\newcommand{\msol}{\mbox{$M_\odot$}}
\newcommand{\msolpyr}{\mbox{$M_\odot$\,yr$^{-1}$}}

\newcommand{\mjypbm}{\mbox{mJy\,beam$^{-1}$}}
\newcommand{\jypbm}{\mbox{Jy\,beam$^{-1}$}}

\newcommand{\hii}{\mbox{H\,{\sc ii}}}
\newcommand{\vlsr}{\mbox{$V_\text{lsr}$}}

\newcommand{\ctw}{the 20~\kms{} cloud}
\newcommand{\cfi}{the 50~\kms{} cloud}

\newcommand{\gzp}{G0.253+0.016}

\newcommand{\coo}{CO\,0.02$-$0.02}

\received{- -, --}
\revised{- -, --}
\accepted{- -, --}
\submitjournal{ApJ}

\shorttitle{Magnetic fields in the Central Molecular Zone}
\shortauthors{Lu et al.}

\begin{document}
\begin{CJK}{UTF8}{gbsn}

\title{Magnetic Fields in the Central Molecular Zone Influenced by Feedback and Weakly Correlated with Star Formation}
\correspondingauthor{Xing Lu, Junhao Liu}
\email{xinglu@shao.ac.cn, junhao.liu@nao.ac.jp}

\author[0000-0003-2619-9305]{Xing Lu (吕行)}
\affiliation{Shanghai Astronomical Observatory, Chinese Academy of Sciences, 80 Nandan Road, Shanghai 200030, P.\ R.\ China}

\author[0000-0002-4774-2998]{Junhao Liu (刘峻豪)}
\affiliation{National Astronomical Observatory of Japan, 2-21-1 Osawa, Mitaka, Tokyo, 181-8588, Japan}
\affiliation{East Asian Observatory, 660 N.\ A`oh\={o}k\={u} Place, University Park, Hilo, HI 96720, USA}

\author{Thushara Pillai}
\affiliation{Haystack Observatory, Massachusetts Institute of Technology, 99 Millstone Road, Westford, MA 01886, USA}

\author{Qizhou Zhang}
\affiliation{Center for Astrophysics | Harvard \& Smithsonian, 60 Garden Street, Cambridge, MA 02138, USA}

\author{Tie Liu (刘铁)}
\affiliation{Shanghai Astronomical Observatory, Chinese Academy of Sciences, 80 Nandan Road, Shanghai 200030, P.\ R.\ China}
\affiliation{Key Laboratory of Radio Astronomy and Technology, Chinese Academy of Sciences, A20 Datun Road, Chaoyang District, Beijing, 100101, P.\ R.\ China}

\author{Qilao Gu (顾琦烙)}
\affiliation{Shanghai Astronomical Observatory, Chinese Academy of Sciences, 80 Nandan Road, Shanghai 200030, P.\ R.\ China}

\author{Tetsuo Hasegawa}
\affiliation{National Astronomical Observatory of Japan, 2-21-1 Osawa, Mitaka, Tokyo, 181-8588, Japan}

\author{Pak Shing Li}
\affiliation{Shanghai Astronomical Observatory, Chinese Academy of Sciences, 80 Nandan Road, Shanghai 200030, P.\ R.\ China}

\author{Xindi Tang}
\affiliation{Xinjiang Astronomical Observatory, Chinese Academy of Sciences, 830011 Urumqi, P.\ R.\ China}

\author[0000-0003-0946-4365]{H Perry Hatchfield}
\affiliation{Jet Propulsion Laboratory, California Institute of Technology, 4800 Oak Grove Drive, Pasadena, CA, 91109, USA}
\affiliation{University of Connecticut, Department of Physics, 196A Auditorium Road, Unit 3046, Storrs, CT 06269 USA}

\author{Namitha Issac}
\affiliation{Shanghai Astronomical Observatory, Chinese Academy of Sciences, 80 Nandan Road, Shanghai 200030, P.\ R.\ China}

\author{Xunchuan Liu}
\affiliation{Shanghai Astronomical Observatory, Chinese Academy of Sciences, 80 Nandan Road, Shanghai 200030, P.\ R.\ China}

\author{Qiuyi Luo}
\affiliation{Shanghai Astronomical Observatory, Chinese Academy of Sciences, 80 Nandan Road, Shanghai 200030, P.\ R.\ China}
\affiliation{School of Astronomy and Space Sciences, University of Chinese Academy of Sciences, No.\ 19A Yuquan Road, Beijing 100049, P.\ R.\ China}

\author{Xiaofeng Mai}
\affiliation{Shanghai Astronomical Observatory, Chinese Academy of Sciences, 80 Nandan Road, Shanghai 200030, P.\ R.\ China}
\affiliation{School of Astronomy and Space Sciences, University of Chinese Academy of Sciences, No.\ 19A Yuquan Road, Beijing 100049, P.\ R.\ China}

\author{Zhiqiang Shen}
\affiliation{Shanghai Astronomical Observatory, Chinese Academy of Sciences, 80 Nandan Road, Shanghai 200030, P.\ R.\ China}
\affiliation{Key Laboratory of Radio Astronomy and Technology, Chinese Academy of Sciences, A20 Datun Road, Chaoyang District, Beijing, 100101, P.\ R.\ China}

\begin{abstract} % =<250 words
Magnetic fields of molecular clouds in the Central Molecular Zone (CMZ) have been relatively under-observed at sub-parsec resolution. Here we report JCMT/POL2 observations of polarized dust emission in the CMZ, which reveal magnetic field structures in dense gas at $\sim$0.5~pc resolution. The eleven molecular clouds in our sample including two in the western part of the CMZ (Sgr C and a far-side cloud candidate), four around the Galactic longitude 0 (\cfi{}, \coo{}, the `Stone' and the `Sticks \& Straw' among the Three Little Pigs), and five along the Dust Ridge (\gzp{}, clouds b, c, d, and e/f), for each of which we estimate the magnetic field strength using the angular dispersion function method. The morphologies of magnetic fields in the clouds suggest potential imprints of feedback from expanding \hii{} regions and young massive star clusters. A moderate correlation between the total viral parameter versus the star formation rate and the dense gas fraction of the clouds is found. A weak correlation between the mass-to-flux ratio and the star formation rate, and a weak anti-correlation between the magnetic field and the dense gas fraction are also found. Comparisons between magnetic fields and other dynamic components in clouds suggest a more dominant role of self-gravity and turbulence in determining the dynamical states of the clouds and affecting star formation at the studied scales.
\end{abstract}


%%%%%%%%%%%%%%%%%%%%%
\section{INTRODUCTION}\label{sec:intro}

\begin{figure*}[p]
\sbox0{
  \begin{tabular}{c}
   \includegraphics[width=1.3\textwidth]{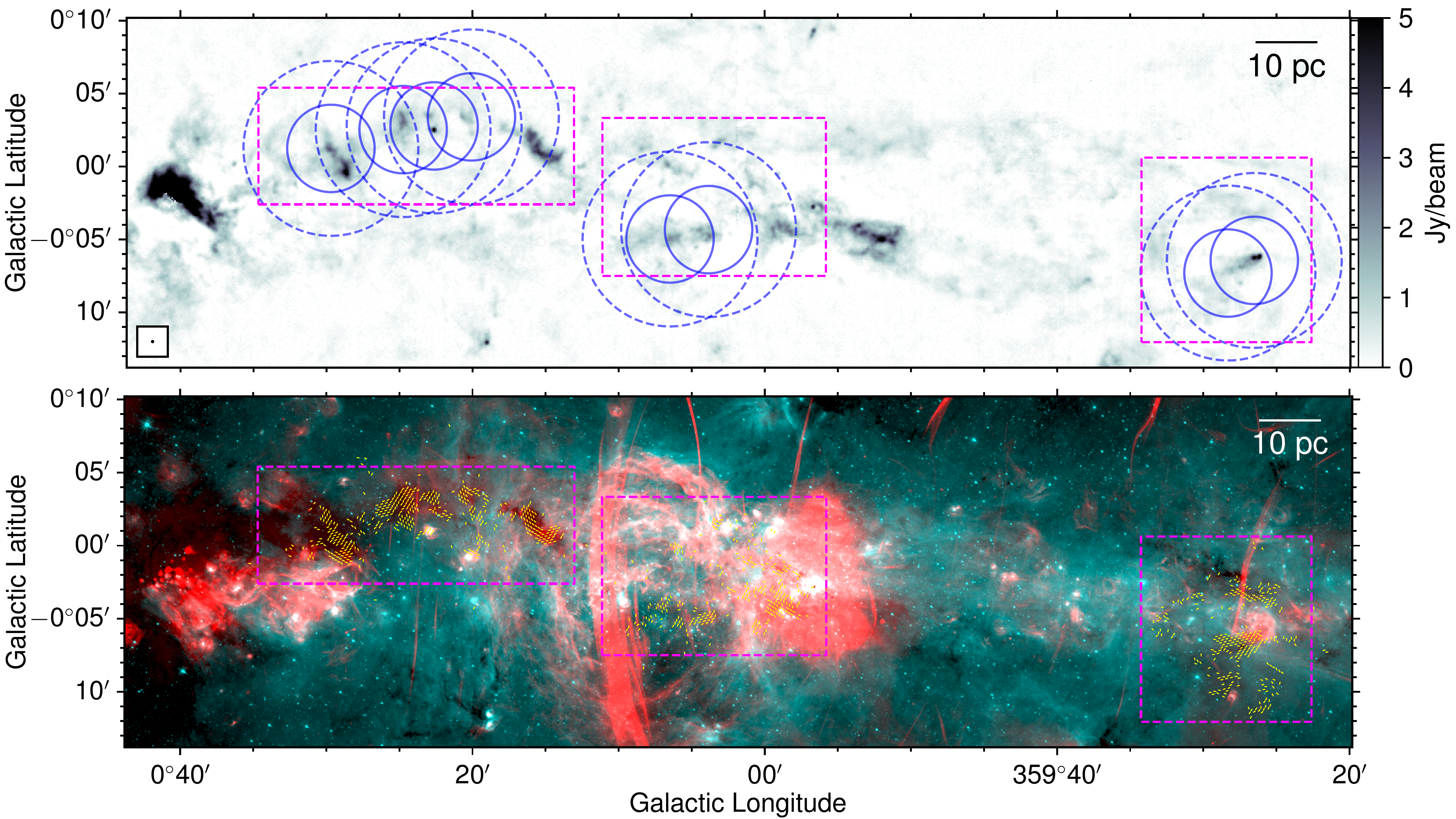}
  \end{tabular}
}
\rotatebox{90}
{
\begin{minipage}[c][\textwidth][c]{\wd0}
  \usebox0
  \caption{\textbf{Top}: Overview of the CMZ and the observed fields. The background image shows 850~\micron{} continuum emission after removing potential CO contamination \citep{parsons2018}. The concentric circles mark the fields observed by JCMT, with the inner solid circles showing radii of $3'$ and the outer dashed ones showing radii of $6'$. The magenta boxes correspond to the regions plotted in Figures~\ref{fig:clouds}--\ref{fig:diffbpa_map}. \textbf{Bottom}: False-color maps created with the MeerKAT 1.28~GHz continuum (red; \citealt{heywood2022}) and \textit{Spitzer} 8~\micron{} (cyan; \citealt{stolovy2006}) data. The yellow segments show orientations of magnetic fields derived from JCMT/POL2 observations with signal-to-noise ratios greater than 3 in their polarization percentages ($p$/d$p$$>$3).}
  \label{fig:allcmz}
 \end{minipage}
}
\end{figure*}

The central $\sim$500~pc of our Galaxy contains several $10^7$~\msol{} of molecular gas and thus is named the Central Molecular Zone \citep[CMZ;][see \autoref{fig:allcmz}]{mills2017,henshaw2023}. CMZ is a prominent high-mass ($>$8~\msol{}) star-forming environment. Several massive molecular clouds of $\gtrsim$$10^5$~\msol{} have been identified in the CMZ \citep[e.g., G0.253+0.016 the ‘Brick’, \ctw{}, Sgr C;][]{longmore2012,kauffmann2017a,kauffmann2017b,lu2015b,lu2017,lu2019a,lu2019b,lu2020,lu2021,walker2021}.

Despite the large amount of molecular gas, the star formation rates (SFRs) both for the whole CMZ and for individual clouds in the CMZ are found to be lower by an order of magnitude than the dense gas star formation relation \citep{longmore2013a,kauffmann2017a,barnes2017, lu2019a,lu2019b}. Here, the dense gas star formation relation refers to the linear correlation found between dense gas masses and SFRs of objects ranging from nearby molecular clouds to external galaxies \citep{gao2004,lada2010,lada2012}. The extreme physical conditions in the CMZ, e.g., the strong turbulence (FWHM linewidth $\sim$$10$--$10^2$~\kms{}), have been suggested to suppress star formation in the CMZ and be responsible for the low SFRs \citep{kruijssen2014}. \citet{kauffmann2017b} found that several massive CMZ clouds are unbound or only marginally bound through a virial analysis that takes turbulent linewidths into account, but at smaller scales ($<$1~pc) the gas may be bound and suitable for star formation \citep{lu2019a}.

However, among these unbound or marginally bound clouds, some turn out to have higher SFRs that are consistent with the dense gas star formation relation \citep[e.g., Sgr C, Dust Ridge cloud~c;][]{ginsburg2015,kauffmann2017a,walker2018,lu2019a}. These clouds also show a higher fraction of gas mass confined in dense cores of 0.1 pc scales than other relatively quiescent clouds such as \ctw{} and \gzp{} \citep[3--10\% vs.\ $<$1\%;][]{lu2019a,battersby2020}. It is unclear how clouds of similar virial status show such different dense gas fractions and SFRs. The magnetic field could be the culprit, which may support the clouds in synergy with turbulence against gravitational collapse. Weaker magnetic fields in some clouds may lead to a preferable environment for collapse, fragmentation, and subsequent star formation.

To this end, it is critical to investigate the correlation between the magnetic field, dense gas fractions, and SFRs of the massive clouds in the CMZ. Only a few CMZ clouds have been mapped in polarized dust emission with sufficiently high angular resolutions, including the Three Little Pigs \citep{chuss2003}, \gzp{} \citep{pillai2015}, and the Sgr~A complex \citep[the circumnuclear disk and the 20/50~\kms{} clouds;][]{chuss2003,hsieh2018}. A more comprehensive database of the magnetic field in a large number of CMZ clouds with uniform sensitivity and resolution is in need for a robust statistical analysis. A recent polarization survey using the 214~\micron{} band of the Stratospheric Observatory for Infrared Astronomy (SOFIA) mapped the whole CMZ at a resolution of 20\arcsec{} \citep{butterfield2023}, which is a promising dataset for statistical analyses.

Here we report observations of polarized dust emission at 850~\micron{} in a sample of 11 massive clouds in the CMZ at a resolution of 14\arcsec{}. Assuming that the short axes of dust grains are aligned with the magnetic field \citep{lazarian2007a,lazarian2007b}, we are able to use polarized dust emission to trace the plane-of-sky component of magnetic field orientations. We further infer magnetic field strengths through a statistical approach and compare them to the SFRs of the clouds. These are by far the highest angular resolution observations of magnetic fields in molecular clouds in the CMZ. Throughout the paper, we adopt the parallax distance of 8.1~kpc to the CMZ \citep{reid2019}.

%%%%%%%%%%%%%%%%%%%%%
\section{OBSERVATIONS AND DATA REDUCTION}\label{sec:obs}

The polarized dust emission at 850~\micron{} was observed with SCUBA2 \citep{holland2013} and POL2 \citep{friberg2016} mounted on the James Clark Maxwell Telescope (JCMT) between June 14 and August 1 2020, under the project M20AP023. The targets include eight fields in the CMZ, which are plotted in \autoref{fig:allcmz}, each with an on-source time of 1 hour. The observations were carried out with the POL2 Daisy mode\footnote{\url{https://www.eaobservatory.org/jcmt/instrumentation/continuum/scuba-2/pol-2/\#Observing_mode}}, leading to maps with high signal-to-noise ratios (SNRs) in the inner radius of $3'$, although we found that SNRs at $6'$ away from the field centers can still be sufficiently high thanks to bright continuum emission in the CMZ. The effective beam size is 14\farcs{1} ($\sim$0.55~pc at the distance of the CMZ).

Data reduction was done with the SMURF \citep{jenness2013} package in Starlink \citep{currie2014}. We used a pixel size of 7\arcsec{}, and combined adjacent fields to create three mosaicked maps marked by the magenta boxes in \autoref{fig:allcmz}.

The measured rms of the total intensity (Stokes I) maps varies from 20 to 50~\mjypbm{}, which is much higher than the thermal noise  of 4--8~\mjypbm{} (the square root of the variance of Stokes I) in the central $3'$ of the maps, likely due to significant foreground and background emission in the maps. We took a median value of 40~\mjypbm{} as the canonical rms, and chose a 5$\sigma$ level (i.e., 0.2~\jypbm{}) as the threshold for estimating physical parameters in the following sections.

The CMZ has been mapped with SCUBA2 at 850~\micron{} using the Pong mapping mode at the same angular resolution as ours \citep[][see \autoref{fig:allcmz} top panel]{parsons2018}. To validate our data reduction, we compare the 850~\micron{} total intensity maps from our data with those without the CO-contamination correction in \citet{parsons2018} after regridding them to a common pixel frame. The intensities are consistent within 40\% between the two datasets. The difference is likely due to the different mapping modes, where the Pong mapping mode provides more uniform coverage and the data can be reduced with a larger spatial filter of $\gtrsim$300\arcsec{}, while the default spatial filter for POL2 data using the Daisy mode is 150\arcsec{}. We note that results in this paper should be rarely affected by the spatial filtering issue of the Daisy mode, as the clouds in our sample are generally $\lesssim$150\arcsec{}, and we have limited our statistical analyses of magnetic fields to $<$90\arcsec{} (see \autoref{subsec:results_Bfield}).

The polarized intensities were binned to a sampling interval of 14\arcsec{} and were debiased, and the polarization position angles were derived following the procedures in \citet{liujh2019}. In the following, only polarization detections where SNRs of the polarization percentages are higher than 3 ($p$/d$p>3$), corresponding to an error in polarization position angles $<$10\arcdeg{}, are considered for further analyses. When considering polarization signals in clouds, we further enforce the threshold of 0.2~\jypbm{} (SNR $>$ 5 for total intensities) to select polarization detections. Catalogs of the derived polarization segments are publicly available at\dataset[10.5281/zenodo.8409806]{https://doi.org/10.5281/zenodo.8409806}.

%%%%%%%%%%%%%%%%%%%%%
\section{RESULTS}\label{sec:results}

\begin{figure*}[!thpb]
\centering
\includegraphics[width=0.5\textwidth]{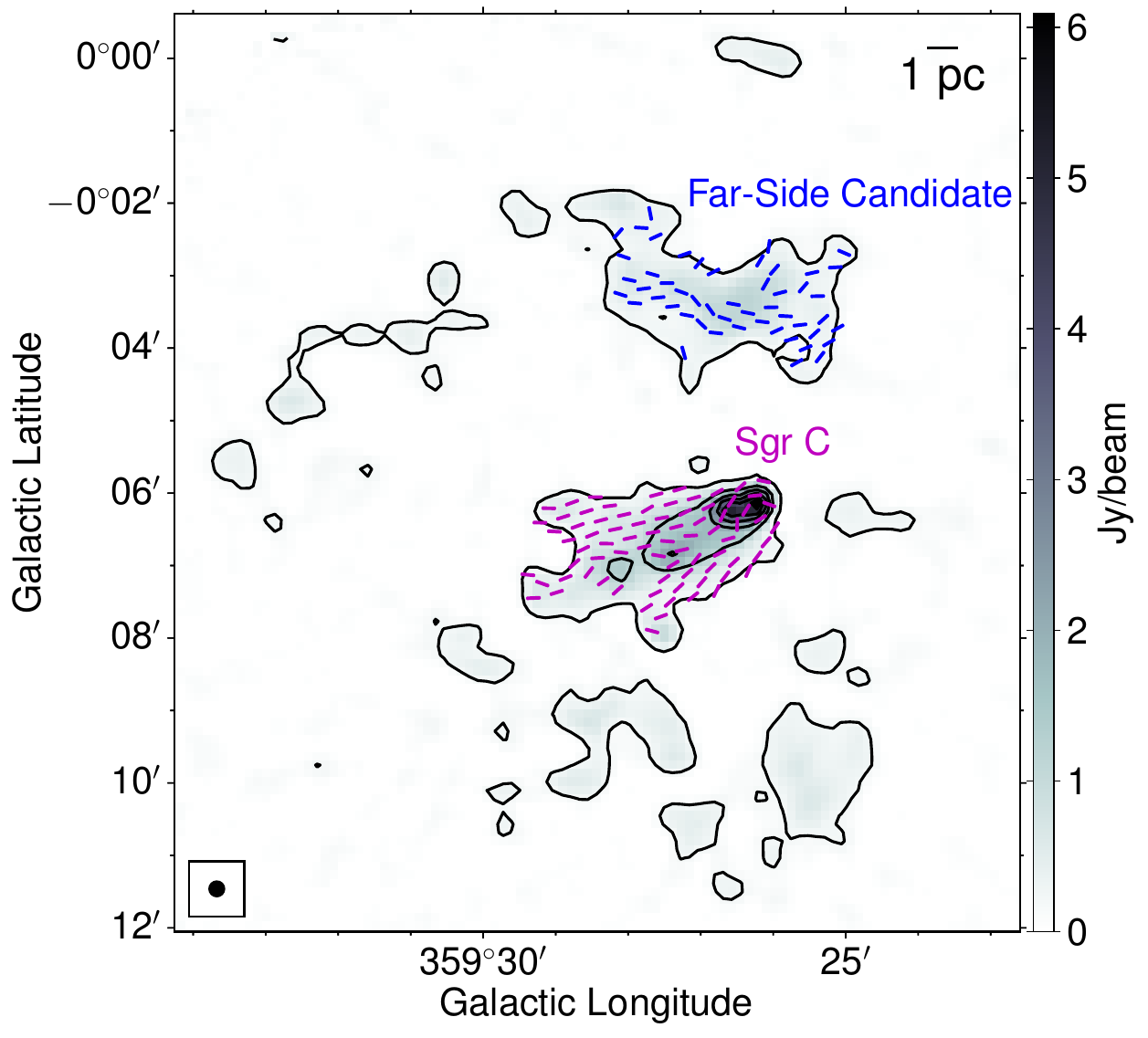} \\
\includegraphics[width=0.6\textwidth]{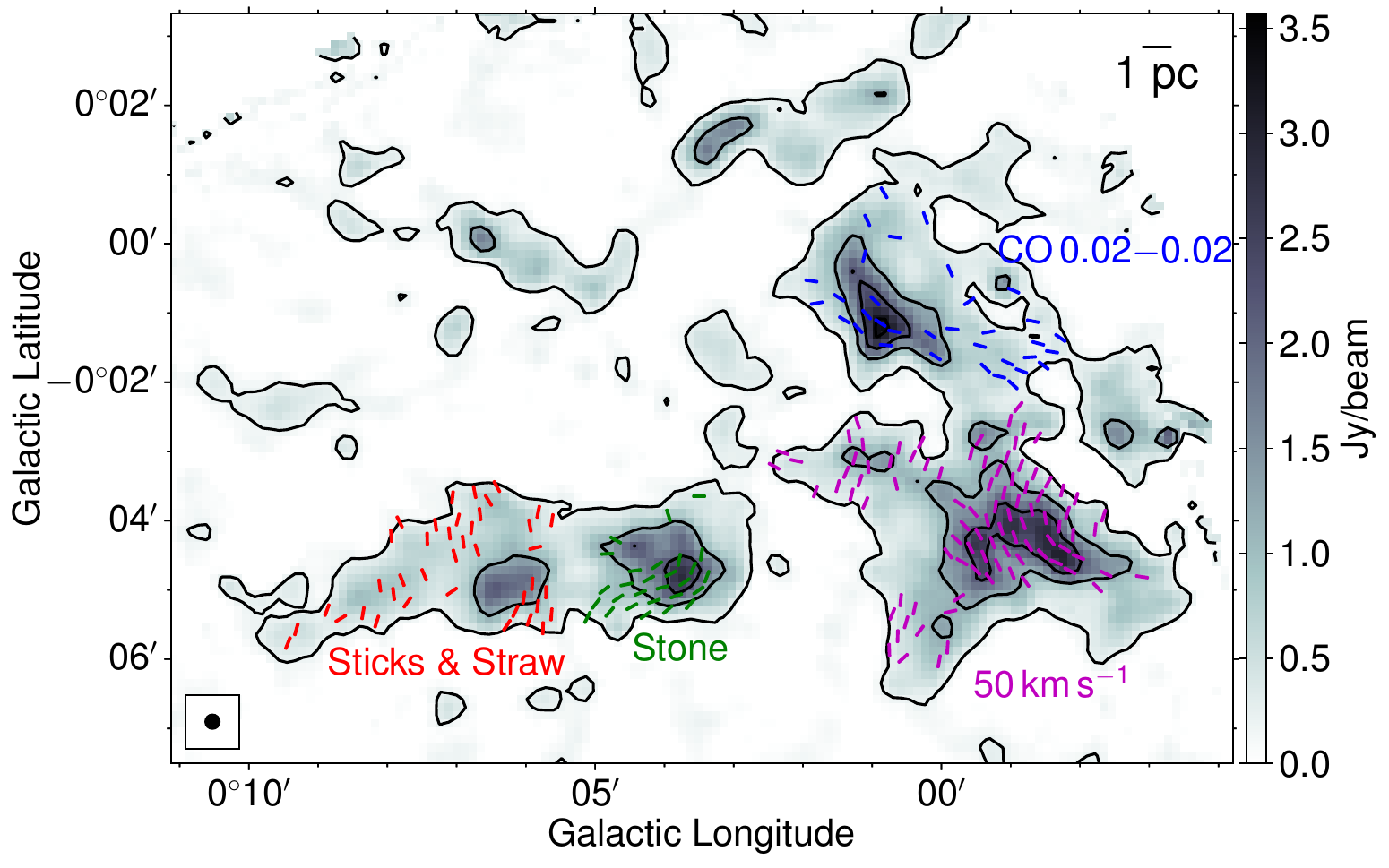} \\
\includegraphics[width=0.72\textwidth]{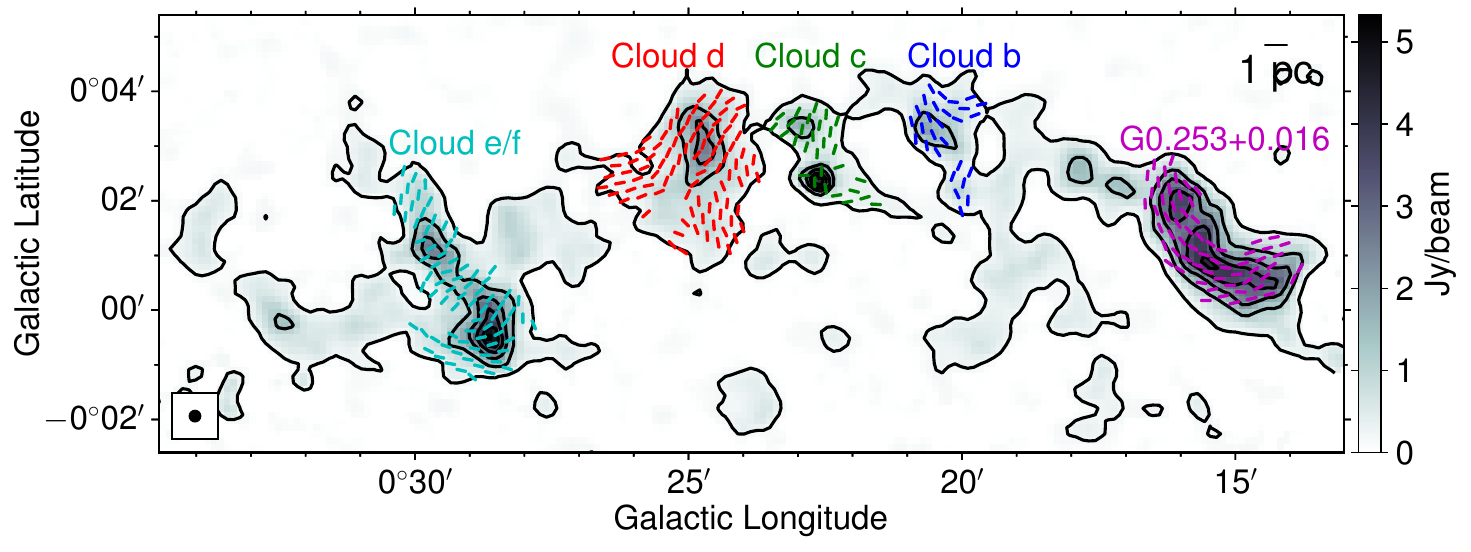}
\caption{Magnetic fields in individual clouds. The segments are color-coded to represent the 11 clouds in our sample. Comparing with \autoref{fig:allcmz}, all segments lying outside of the 5$\sigma$ total intensity contours as well as not belonging to any of the clouds are dropped.}
\label{fig:clouds}
\end{figure*}

We defined a sample of 11 massive clouds that have been studied by e.g., \citet{lu2019b} and \citet{battersby2020}, and labeled them in \autoref{fig:clouds}. The sample includes the most massive clouds in the CMZ, except Sgr~B2 and \ctw{} whose magnetic fields have been studied \citep{novak2000,chuss2003,butterfield2023}. The clouds have sufficient numbers ($\geq$28) of independent polarization measurements allowing for statistical analyses \citep{liu2021}, i.e., for a round region, there should be at least three independent polarization segments along its radius. Two adjacent clouds defined in the literature, the `Sticks' and the `Straw', have 19 independent polarization measurements each. Their mean magnetic field position angles do not differ much (87.0$\pm$22.1\arcdeg{} vs.\ 86.9$\pm$20.8\arcdeg{}). Therefore, we considered them as as a single cloud to boost the number of polarization segments.

\subsection{Masses and Densities of the Clouds}\label{subsec:results_mass}

Fundamental properties of the clouds, including their masses, column densities, and number densities, can be estimated using the 850~\micron{} continuum emission, assuming that it is entirely from thermal dust emission. These properties will be used for estimating magnetic field strengths later.

Assuming optically thin dust emission, H$_2$ column densities of the clouds are calculated as
\begin{equation}\label{equ:column}
N(\text{H}_2)=\frac{I_\nu}{B_\nu(T_\text{dust})\kappa_\nu}\frac{1}{\mu_\text{H$_2$} m_\text{H}},
\end{equation}
where $I_\nu$ is the dust emission intensity at frequency $\nu$, $B_\nu(T_\text{dust})$ is the Planck function at the dust temperature $T_\text{dust}$, $\kappa_\nu=0.1(\nu/1 \text{THz})^\beta$~cm$^2$\,g$^{-1}$ is the dust opacity assuming a gas-to-dust ratio of 100 and with an opacity index $\beta$ \citep{hildebrand1983}, and $\mu_\text{H$_2$}=2.8$ is the molecular weight per H$_2$.

We used the continuum intensity from the SCUBA2 data with the CO-contamination correction \citep{parsons2018} instead of the Stokes~I data in our POL2 observations, as the latter has filtered out more continuum emission and included contamination from the CO\,3--2 line emission.

We took $T_\text{dust}$ and $\beta$ maps of 14\arcsec{} resolutions from \citet{tang2021} and calculated the column densities pixel by pixel. Note that  \citet{tang2021} derived $T_\text{dust}$ and $\beta$ from a combination of \textit{Planck}, \textit{Herschel}, and the Large Millimeter Telescope (LMT) observations, which did not remove any foreground or background emission as the POL2 data did. The derived column density maps are presented in \autoref{appd_sec:density}.

Cloud masses were then obtained by summing up the column densities over the pixels in the clouds above the threshold of 0.2~\jypbm{} in our total intensity maps: 
\begin{equation}\label{equ:mass}
M=\mu_\text{H$_2$} m_\text{H}\sum N(\text{H}_2)A_\text{pix}, 
\end{equation}
where $A_\text{pix}$ is the area of one pixel determined by the pixel angular size (7\arcsec) and the distance to the CMZ.

Number densities of the clouds are more uncertain, as the 3D geometries of the clouds are unknown. To get an order-of-magnitude estimate of the number densities, we approximated all the clouds as spheres and adopted an effective radius $r_\text{eff}$ of $\sqrt{A/\pi}$, where $A$ is the area of the clouds above the total intensity threshold, from which the volumes of the spheres were derived. Then the number densities were obtained by dividing the cloud masses by the volumes.

The cloud masses $M$ and mean number densities $n(\text{H}_2)$ are tabulated in \autoref{tab:stats}. The uncertainties in the masses and densities are typically 50\% \citep[e.g.,][]{sanhueza2017}, mainly contributed by the uncertainties in the gas-to-dust ratio \citep[could be lower than 100 in the CMZ;][]{giannetti2017}, the opacity index, and the dust temperature \citep{tang2021}.

\begin{deluxetable*}{cccccccccccccccc}[!tbh]
\tabletypesize{\scriptsize}
\tablecaption{Measured physical parameters and estimated magnetic fields of the clouds.\label{tab:stats}}
\tablewidth{0pt}
\tablehead{
\colhead{Cloud\tablenotemark{a}} & $M$ & $r_\text{eff}$ & $n$(H$_2$) & $\sigma_v$ & $\left(\langle B_\text{t}^{\phantom{\text{t}}2}\rangle/\langle B^2\rangle\right)^{0.5}$ & $\sigma_\theta$ & $B$ & $B_\text{classical}$ & $v_\text{A}$ & $\mathcal{M}_\text{A}$ & $\lambda$ & $\alpha_\text{vir}$ & $\alpha_{\text{vir},B}$ & SFR\tablenotemark{b} & CDGF\tablenotemark{b} \\
 & (10$^4$~\msol{}) & (\arcsec{}, pc) & (10$^3$~\cc{}) & (\kms{}) & & (\arcdeg{}) & (mG) & (mG) & (\kms{}) & & & & & (10$^{-3}$~\msolpyr{})
}
\startdata
Sgr~C                        &  4.0 & 76, 3.0   & 5.1 & 5.9   & 0.32 & 22.2 & 0.21 & 0.40 & 4.8    & 2.1 & 2.3 & 1.8   & 1.9   & 65.4$\pm$39.2 & 0.11\\
Far-Side Candidate  & 1.5  & 76, 3.0   & 1.9  & 5.4   & 0.36 & 32.8 & 0.10 & 0.14 & 3.9    & 2.4 & 1.7 & 4.2   & 4.3   & \nodata  & \nodata \\
% updated
50~\kms{}                  & 6.8 & 121, 4.8  & 2.2 & 16.0 & 0.25 & 32.9 & 0.48 & 0.44 & 16.7 & 1.7 & 0.7 & 12.5  & 12.7 & 50.6$\pm$30.6 & 0.017 \\
CO\,0.02$-$0.02       & 0.9  & 90, 3.5    & 0.7 & 8.8   & 0.37 & 26.0 & 0.10 & 0.18 & 6.2    & 2.4 & 0.8 & 22.3 & 22.4 & \nodata & \nodata \\
% updated
Stone                        &  2.8 & 70, 2.8    & 4.6 &11.2  & 0.23 & 27.3 & 0.52 & 0.56 & 12.6 & 1.5 & 0.8  & 8.7   & 8.9   & 6.4$\pm$3.8     & 0.022 \\
Sticks \& Straw         &  3.7 & 82, 3.2    & 3.8 & 14.5 & 0.33 & 22.1 & 0.44 & 0.84 & 11.5  & 2.2 & 0.9 & 12.9 & 13.0 & 0.4$\pm$0.3     & 0.010 \\
% updated
\gzp{}                        &  9.3 & 86, 3.4    & 8.4 & 18.5 & 0.16 & 51.7 & 1.73 & 0.51 & 30.7 & 1.0 & 0.5 & 8.7   & 9.2   & 2.3$\pm$1.4     & 0.019 \\
% updated
Cloud b                     &  1.4 & 60, 2.4    & 3.7 &12.5  & 0.35 & 37.6 & 0.35 & 0.38 & 9.5   & 2.3 & 0.8 & 17.9 & 18.0 & $<$0.3\tablenotemark{c} & 0.019 \\
Cloud c                     & 1.9  & 55, 2.1    & 6.7 & 4.2   & 0.36 & 37.4 & 0.15 & 0.17 & 3.0   & 2.4 & 3.0 & 1.4   & 1.5  & 38.7$\pm$23.2  & 0.077 \\
Cloud d                     & 6.4  & 83, 3.3    & 6.3 & 11.4 & 0.28 & 35.4 & 0.53 & 0.49 &10.7 & 1.8 & 1.2 & 4.6   & 4.8  & 2.7$\pm$1.6      & 0.014 \\
% updated
Cloud e/f                   & 11.4  & 83, 3.3  &11.4 & 9.6  & 0.21 & 39.7 & 0.77 & 0.47 & 11.7  & 1.4 & 1.5 & 1.8   & 2.1  & 16.9$\pm$10.1  & 0.016 \\
\enddata
\tablenotetext{a}{References of the cloud naming: \citet{lu2019b,battersby2020}.}
\tablenotetext{b}{Reference of the SFR: Hatchfield et al.\ (2023, in prep.). Reference of the CDGF: \citet{battersby2020}.}
\tablenotetext{c}{The SFR of this cloud cannot be constrained using the method of Hatchfield et al.\ (2023, in prep.), therefore an upper limit is given \citep{lu2019a}.}
\tablecomments{Boundaries of the clouds can be found in \autoref{fig:clouds}. Uncertainties of the parameters are discussed in \autoref{appd_sec:uncertainty}.}
\end{deluxetable*}

\subsection{Magnetic field morphologies}\label{subsec:results_morph}
The orientation of the magnetic field is inferred by rotating the detected polarization orientations by 90\arcdeg{}. The bottom panel of \autoref{fig:allcmz} shows an overview of the magnetic field morphologies in the observed regions. \autoref{fig:clouds} shows magnetic field orientations in the individual clouds. Several clouds, e.g., \gzp{} and \cfi{}, lie beyond the central $3'$ of the DAISY fields. However, their polarized emission is sufficiently strong ($p$/d$p$$>$3). Therefore, we included them in the analysis as well.

\begin{figure*}[!thpb]
\centering
\includegraphics[width=0.5\textwidth]{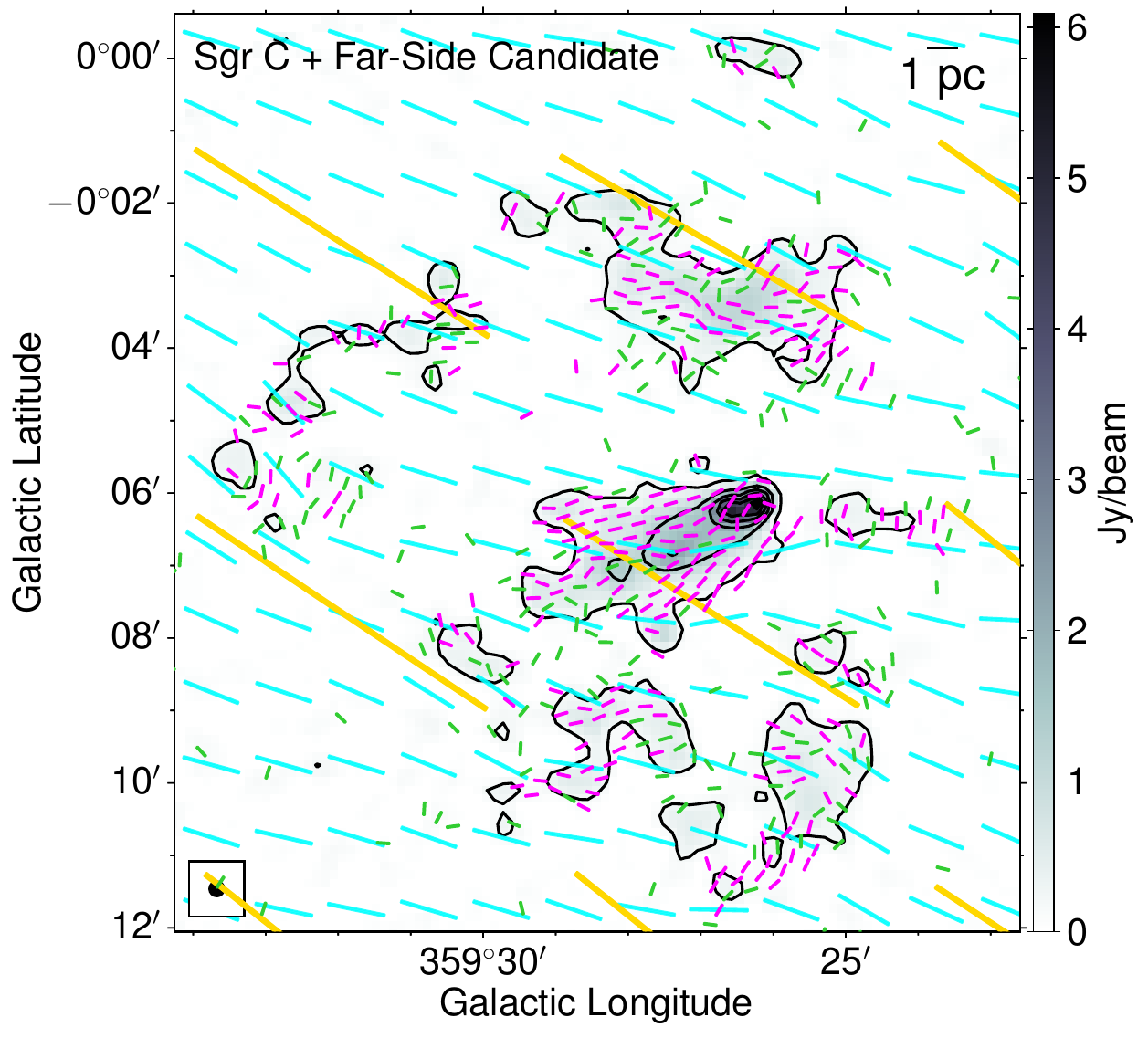} \\
\includegraphics[width=0.6\textwidth]{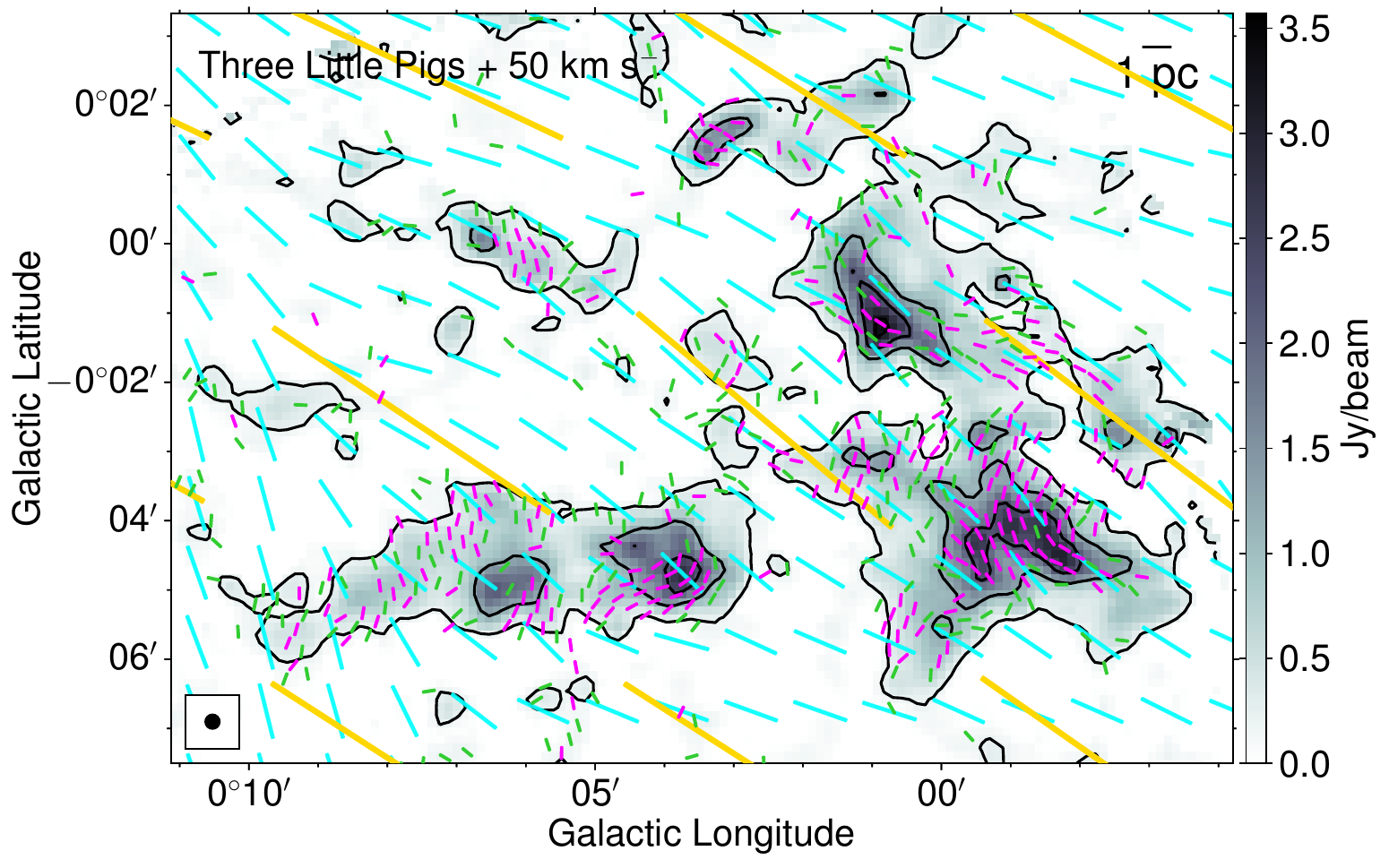} \\
\includegraphics[width=0.72\textwidth]{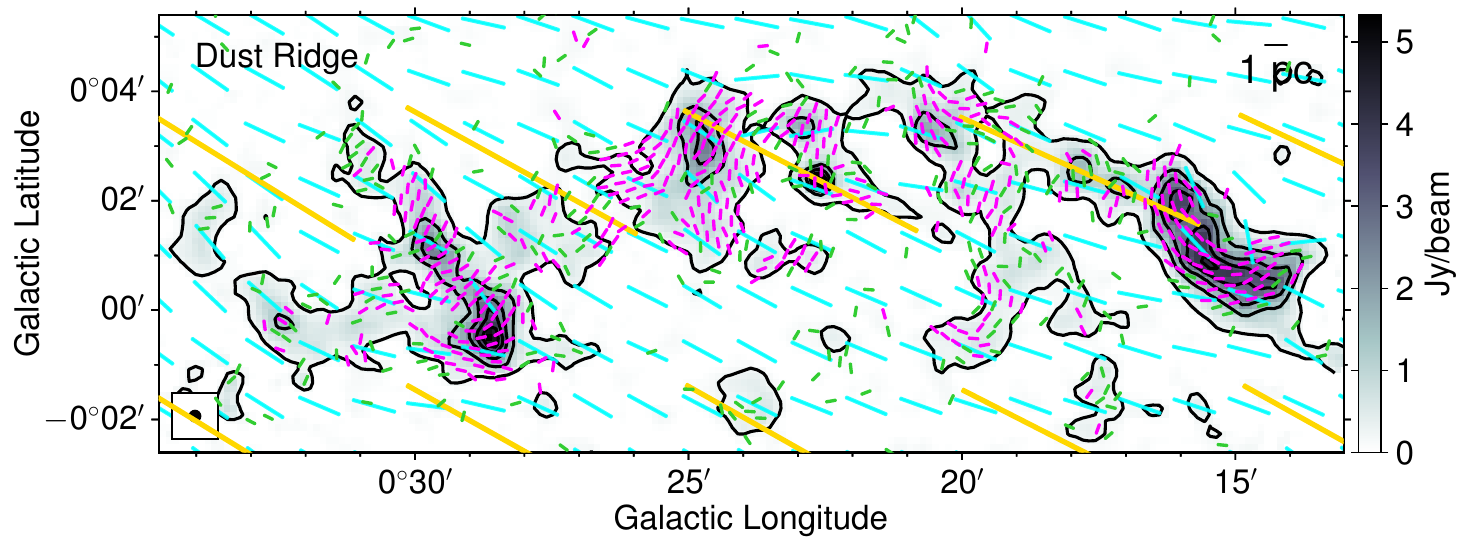}
\caption{Magnetic fields of different spatial scales in the three CMZ fields. The magenta segments show orientations of magnetic fields derived from JCMT/POL2 observations with $p$/d$p$$>$3, while the green ones are those with 2$<$$p$/d$p$$<$3 that are only for visual presentation but not for quantitative analyses. The cyan segments show orientations of magnetic fields from ACT observations at a resolution of $1'$, while the yellow ones from \textit{Planck} 353~GHz at a resolution of $5'$.}
\label{fig:act}
\end{figure*}

To investigate magnetic field morphologies at difference spatial scales, we compared our JCMT POL2 observations with data from the Atacama Cosmology Telescope (ACT) at the 1.36~mm (220~GHz) band that have an angular resolution of $1'$ \citep{guan2021} and from \textit{Planck} at 353~GHz that have an angular resolution of $5'$ \citep{planck2015}, which are able to probe magnetic fields at larger scales than the JCMT POL2 data. In \autoref{fig:act}, we overlaid magnetic field orientations estimated from the ACT 1.36~mm band and from the \textit{Planck} 353~GHz data. The orientations based on ACT and \textit{Planck} are generally consistent with each other, suggesting that both data likely trace the same large-scale magnetic field.

\begin{figure*}[!thpb]
\centering
\includegraphics[width=0.5\textwidth]{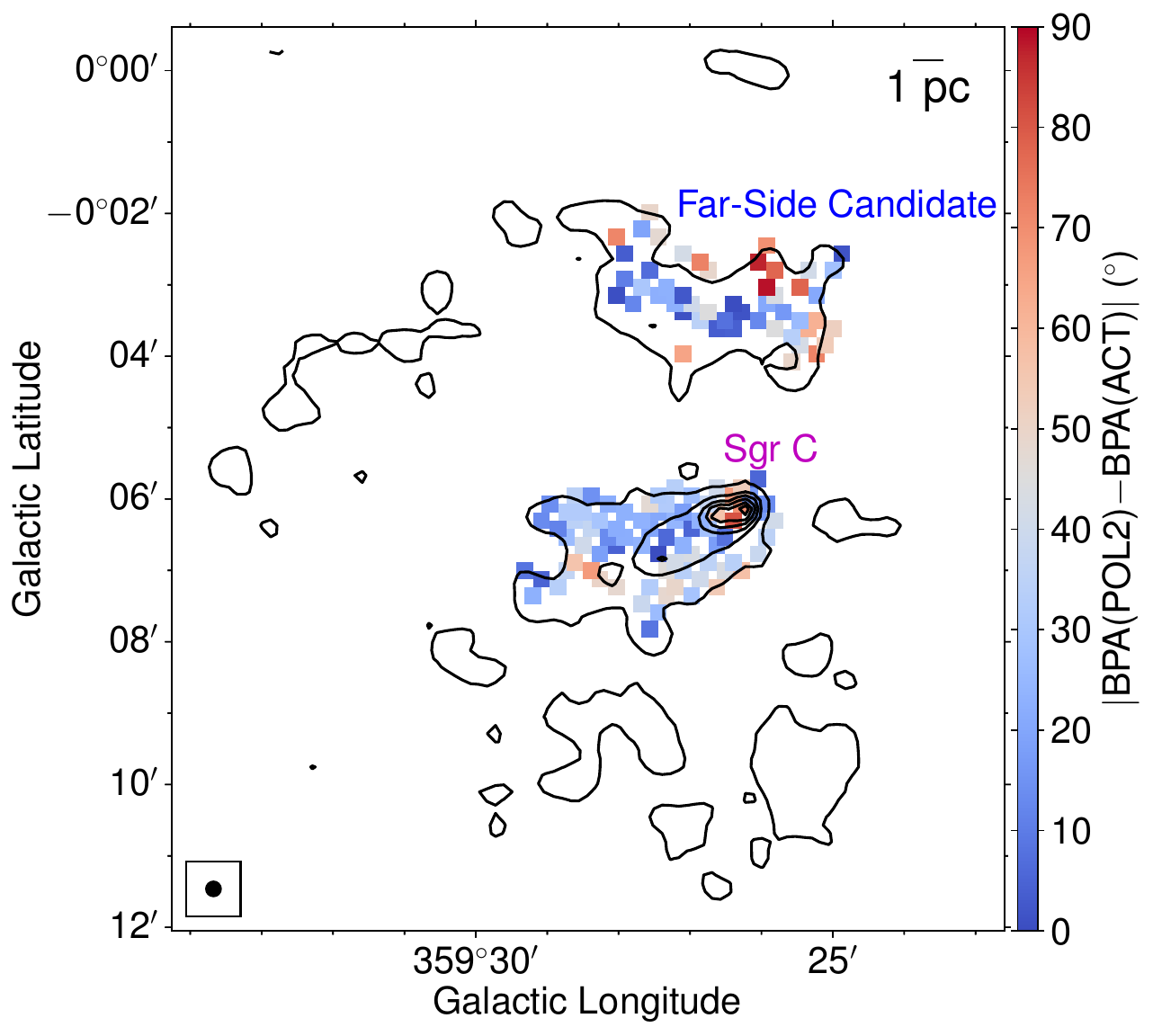} \\
\includegraphics[width=0.6\textwidth]{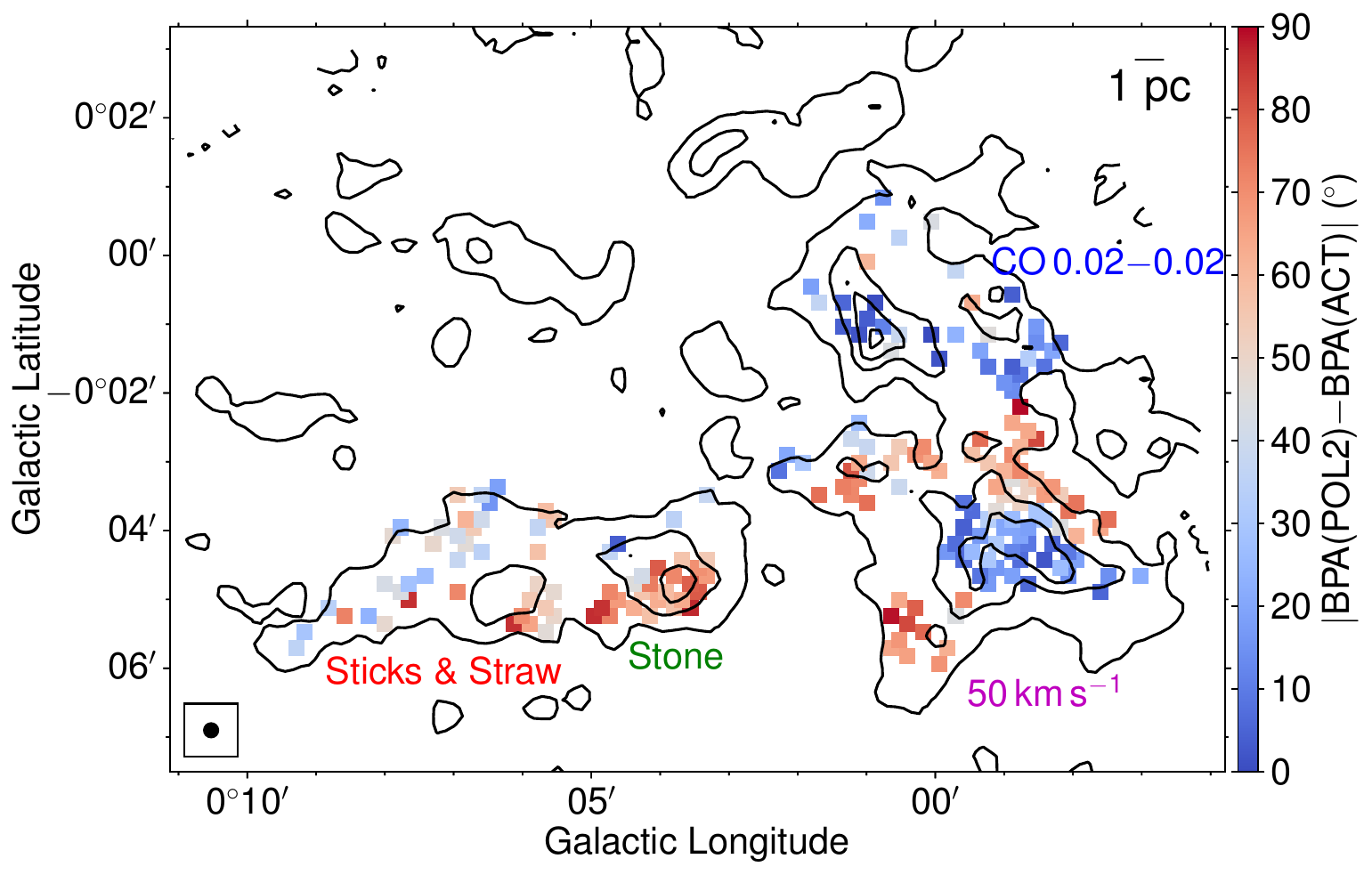} \\
\includegraphics[width=0.72\textwidth]{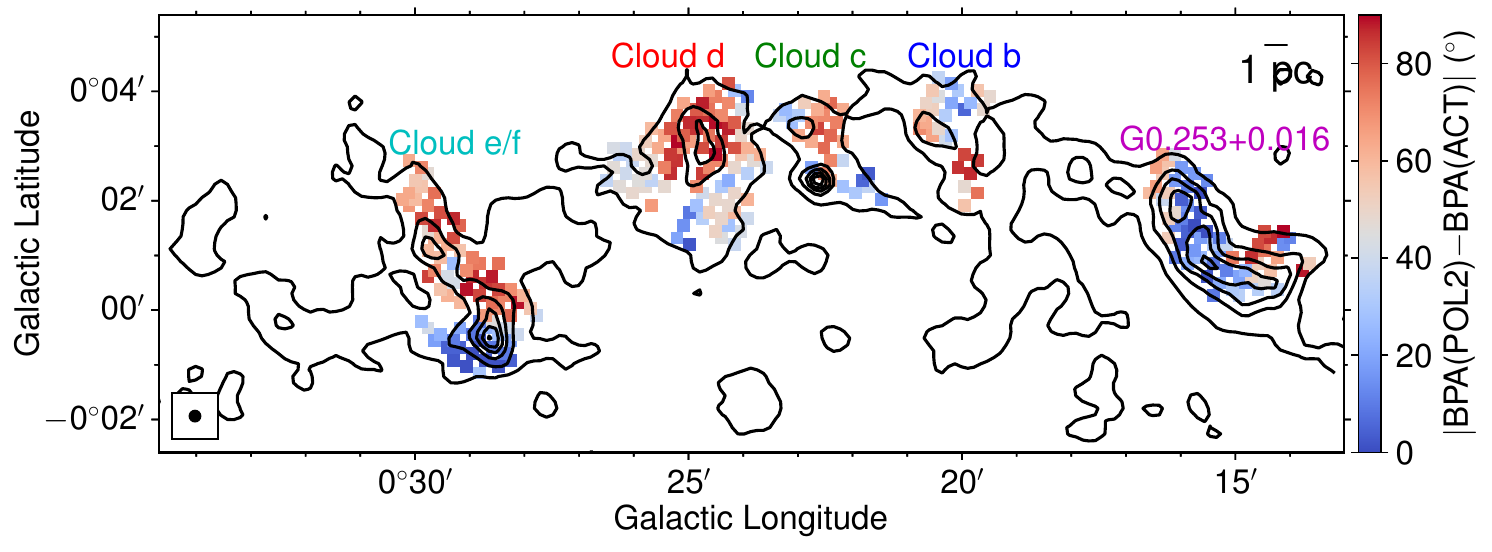}
\caption{Relative orientation between small-scale ($\sim$0.5~pc) magnetic fields probed by JCMT/POL2 and large-scale ($\gtrsim$2.5~pc) magnetic fields probed by ACT. The image is color-coded to quantify whether the small-scale magnet fields tend to be parallel (blue) or perpendicular (red) to the large-scale ones. Only polarization detections within the clouds in our sample are plotted (cf.\ \autoref{fig:clouds}). Contours are the same as in \autoref{fig:allcmz}, representing the JCMT 850~\micron{} total intensity emission.}
\label{fig:diffbpa_map}
\end{figure*}

\autoref{fig:diffbpa_map} further displays maps of the difference between magnetic field position angles probed by our JCMT/POL2 observations and those probed by ACT. To produce this map, we did not smooth or convolve the two datasets. Instead, for each independent measurement of polarized emission in the JCMT/POL2 data that derives a magnetic field orientation (with a resolution of 14\arcsec{}), we found the nearest magnetic field segment of the ACT data (with a resolution of $1'$), and calculated the difference of position angles. Since we only know the orientations, not directions, of magnetic fields, the angle difference was limited in a range of 0\arcdeg{}--90\arcdeg{}, with 0\arcdeg{} representing the two field lines being parallel, and 90\arcdeg{} representing the two field lines being perpendicular. We also produced similar maps using POL2 data smoothed and regridded to the same frame as the ACT data, which are presented in \autoref{appd_sec:bpa}. The two datasets clearly show different magnetic field position angles in most clouds. This could be because the POL2 observations have filtered out signals above the scale of 150\arcsec{} and therefore they are not able to probe large-scale magnetic fields in the CMZ as are the ACT data \citep{juvela2018}. However, we cannot exclude the possibility that the ACT data have included foreground emission along the line of sight, and thus represent a superposition of different components of magnetic fields between the CMZ and us. In such a case, the difference between the field position angles proved by POL2 and ACT would be a natural result as they trace different ISM components along the line of sight.

If we assume that the ACT data are tracing magnetic fields inside the CMZ, then it can be seen from Figures~\ref{fig:act} \& \ref{fig:diffbpa_map} that the large-scale magnetic fields traced by the ACT polarization data are twisted inside the massive clouds as revealed by the POL2 data. The deviation of the orientations of small versus large-scale magnetic fields will be discussed in \autoref{subsec:disc_PAs}.

\subsection{Magnetic field strengths in the clouds}\label{subsec:results_Bfield}
We inferred the magnetic field strengths using the Angular Dispersion Function (ADF) method \citep{hildebrand2009,houde2009,houde2016}. For summaries of this method, see e.g., \citet{liu2021,liu2022a,liu2022b}.

The analysis routine is outlined as follows:

\textit{a).} Derive the ADF, $1-\langle\cos[\Delta\Phi(l)]\rangle$, following the definition in \citet{houde2009}. Here, $\Delta\Phi(l)$ is the difference in position angles of two magnetic field segments separated by a distance $l$.

\textit{b).} Fit the ADF with the following form \citep{houde2009}:
\begin{equation}\label{equ:ADF}
\begin{split}
1-\langle\cos[\Delta\Phi(l)]\rangle&\simeq b(l)+a'_2l^2 \\
&=\frac{\langle B_\text{t}^{\phantom{\text{t}}2}\rangle}{\langle B^2\rangle}\times(1-e^{-l^2/2(l_\delta^2+2W^2)})+a'_2l^2,
\end{split}
\end{equation}
where $\left(\langle B_\text{t}^{\phantom{\text{t}}2}\rangle/\langle B^2\rangle\right)^{0.5}$ is the turbulent-to-total magnetic field strength ratio in the plane of sky, $l_\delta$ is the correlation length scale for the local turbulent magnetic field, and $W$ is the standard deviation of the Gaussian beam. 

The numerical study by \citet{liu2021} suggested that the ADF method may not work well in accounting for the effect of line-of-sight (LOS) signal integration. Thus, $\left(\langle B_\text{t}^{\phantom{\text{t}}2}\rangle/\langle B^2\rangle\right)^{0.5}$ adopted here does not consider the LOS signal integration effect. 

The fitting results of the 11 clouds are plotted as blue curves in \autoref{appd_sec:adf}. The best-fit $\left(\langle B_\text{t}^{\phantom{\text{t}}2}\rangle/\langle B^2\rangle\right)^{0.5}$ is tabulated in \autoref{tab:stats}.

\textit{c).} Estimate the mean densities ($\rho$) and turbulent velocity dispersions ($\sigma_v$) of the clouds. The estimate of the mean densities $\rho$ has been elaborated in \autoref{subsec:results_mass}.

As for the velocity dispersions, the \ammthree{} line data from the SWAG survey \citep{krieger2017} were used \citetext{J.\ Ott, priv.\ comm.}. \amm{} has a critical density of $\sim$$10^3$~\cc{} \citep{shirley2015}, which is close to the mean densities of the clouds. The morphologies of the \amm{} emission match well with those of the clouds seen in the JCMT 850~\micron{} continuum \citep{krieger2017}. Therefore, \amm{} is appropriate for tracing turbulent motions of dense gas in the clouds. For each cloud, the mean \ammthree{} spectrum was fitted with a Gaussian to obtain the velocity dispersion $\sigma_v$.

We do not consider the satellite lines that are usually much weaker than the main hyperfine line at this transition \citep{ho1983}. We do not subtract the thermal line width from the measured velocity dispersion because the former is much narrower than the latter in the CMZ environment and therefore makes no difference to the measurement. The fitting results are presented in \autoref{appd_sec:ammonia}. We note that velocity gradients and/or multiple velocity components along the line of sight exist inside the clouds (\autoref{appd_sec:ammonia}), and therefore the measured velocity dispersions should be treated as upper limits. An extreme case is \gzp{}, where the mean velocity dispersion was measured to be 4.4~\kms{} based on ALMA observation after four independent velocity components were decomposed \citep{henshaw2019}, and therefore the estimated magnetic field strengths would be four times lower. A detailed kinematic analysis using high angular resolution observations is necessary to assess the impact of multiple velocity components in the clouds and more accurately determine the velocity dispersion, which is beyond the scope of the current work. For consistency within our sample, we still adopt the velocity dispersions measured from the SWAG \ammthree{} line, but note the caveat that the measurements may have large uncertainties.

\textit{d).} Estimate the total magnetic field strength ($B$) in the plane of the sky using the Davies-Chandrasekhar-Fermi (DCF) method \citep{liu2021, liu2022a}:
\begin{equation}\label{equ:DCF}
B \sim 0.21\sqrt{\mu_0\rho}\,\sigma_v\left(\frac{\langle B_\text{t}^{\phantom{\text{t}}2}\rangle}{\langle B^2\rangle}\right)^{-0.5},
\end{equation}
where $\mu_0$ is the permeability of vacuum that is $4\pi$ under the cgs metric system. Here we assume isotropic turbulence and equipartition between turbulent kinetic and magnetic energies. \citet{liu2022b} have demonstrated that the uncertainty brought by the anisotropic turbulence is a minor issue for the DCF method in self-gravitating regions. We adopt the numerically derived correction factor 0.21 (with 45\% uncertainty at $>$0.1 pc scales) from \citet{liu2021} to account for the LOS signal integration effect. The uncertainty in the estimated magnetic field strengths is not straightforward to quantify, as the DCF method itself has inherent uncertainties (see \autoref{appd_sec:uncertainty}). We adopt the typical uncertainty of a factor of 2 \citep{liu2021}, which is derived by applying the DCF method to numerical simulations and comparing estimated values to input models. This uncertainty should be treated as a lower limit, because unlike numerical simulations, in realistic situations there should be even more sources of errors. The estimated $B$ values of the clouds are listed in \autoref{tab:stats}.

We also estimated the magnetic field strength following the classical DCF method \citep[e.g.,][]{falceta2008,lips2022}:
\begin{equation}\label{equ:cDCF}
B_\text{classical} \sim 0.5\sqrt{\mu_0\rho}\frac{\sigma_v}{\tan\sigma_\theta},
\end{equation}
where $\sigma_\theta$ is the dispersion of the magnetic field orientations. The correction factor of 0.5 is adopted from the numerical simulations of \citet{ostriker2001}. The estimated magnetic field strengths are consistent with those based on the ADF method within a factor of 2 (\autoref{tab:stats}), except for \gzp{} whose $\sigma_\theta$ is up to $>$50\arcdeg{}. \citet{chen2022} have pointed out that $\tan\sigma_\theta$ does not correlate well with the turbulent-to-total magnetic field strength ratio in the plane of sky, especially when $\sigma_\theta$ is high. Therefore, \autoref{equ:cDCF} may not work well for \gzp{}. There also exist other variations of the classical DCF method \citep[e.g.,][]{skalidis2021} that better suit certain conditions, which we do not explore further in this work. In the following, we adopt the estimates of magnetic field strengths from the ADF method for further analyses.

One cloud in our sample, \gzp{}, has been mapped in polarized dust emission that enables a DCF analysis. \citet{pillai2015} estimated a plane-of-the-sky magnetic field strength (the `total' magnetic field strength in their definition divided by 1.3) of 4.2~mG, which was corrected to 1.7~mG by \citet{federrath2016b} after adopting the correct density. The latter value is consistent with our estimate of 1.73 mG. However, we caution that the consistency is coincidental, as we adopted different velocity dispersions, densities, and correction factors for \autoref{equ:DCF}. This highlights the large uncertainties in the estimate of magnetic field strengths through the DCF/ADF method.

\subsection{Comparing turbulence, magnetic field, and gravity}\label{subsec:results_energies}
To quantify the relative importance between the turbulence, magnetic field, and self-gravity of individual clouds, we follow the framework in Section 2.4 of \citet{liu2022b} to estimate the Alfv\'{e}nic Mach numbers, the mass-to-flux ratios, and the virial parameters of the clouds. The results are tabulated in \autoref{tab:stats}.

\subsubsection{Alfv\'{e}nic Mach numbers}\label{subsubsec:results_mach}
The relative importance between the turbulence and magnetic field can be parameterized by the Alfv\'{e}nic Mach number
\begin{equation}
\mathcal{M}_\text{A}=\sigma_{v,\text{3D}}/v_\text{A},
\end{equation}
where the 3D velocity dispersion $\sigma_{v,\text{3D}}$ is $\sqrt{3}\sigma_v$ for isotropic turbulence. The Alfv\'{e}nic speed is $v_\text{A}=B_\text{3D}/\sqrt{\mu_0\rho}$, where the 3D magnetic field strength $B_\text{3D}$ is estimated by multiplying the magnetic field strength in the plane of the sky by 1.25, a scaling factor derived for a randomly distributed 3D mean field orientation \citep{liu2022b} that estimates $B_\text{3D}$ with an uncertainty of a factor of 2 \citep{liu2021}.

\subsubsection{Mass-to-flux ratios}\label{subsubsec:results_masstoflux}
The relative importance between the magnetic field and self-gravity can be parameterized by the magnetic critical parameter $\lambda$, which is the mass-to-flux ratio normalized against its critical value:
\begin{equation}
\lambda=\frac{M/\Phi}{(M/\Phi)_\text{cr}}.
\end{equation}
Here the mass-to-flux ratio is estimated using:
\begin{equation}
M/\Phi=\frac{\mu_{\text{H$_2$}}m_\text{H}N(\text{H}_2)}{B},
\end{equation}
and the critical value is
\begin{equation}
(M/\Phi)_\text{cr}=\frac{1}{2\pi\sqrt{G}},
\end{equation}
following \citet{nakano1978}. Note that when calculating the column densities, we have selected the same area where polarized emission is detected and thus where magnetic field segments can be plotted \citep{crutcher2004}.

\subsubsection{Virial parameters}\label{subsubsec:results_virial}
Finally, the dynamical equilibrium of the clouds, assuming a radial density profile of $\rho(r)\propto r^{-2}$ \citep[self-gravitating gas in hydrostatic equilibrium;][]{shu1987}, can be characterized by the ratio between the virial mass $M_\text{vir}=3\sigma_v^2r/G$ and the cloud mass $M$, i.e., the virial parameter:
\begin{equation}
\alpha_\text{vir}=\frac{M_\text{vir}}{M}=\frac{3\sigma_v^2r}{GM},
\end{equation}
where $r=\sqrt{A/\pi}$ is the effective radius, and $G$ is the gravitational constant. If we additionally take the magnetic field support into account, then the total virial mass can be expanded to
\begin{equation}
M_{\text{vir},B}=\sqrt{M^2_B+\left(\frac{M_\text{vir}}{2}\right)^2}+\frac{M_\text{vir}}{2},
\end{equation}
where the magnetic virial mass is given by
\begin{equation}
M_B=\frac{Br^2}{6\sqrt{G/10}}.
\end{equation}
The total virial paramter $\alpha_{\text{vir},B}$ is then the ratio of $M_{\text{vir},B}$ and $M$.

%%%%%%%%%%%%%%%%%%%%%%%%%
\section{DISCUSSION}\label{sec:disc}

\begin{figure*}[!thpb]
\centering
\includegraphics[width=0.5\textwidth]{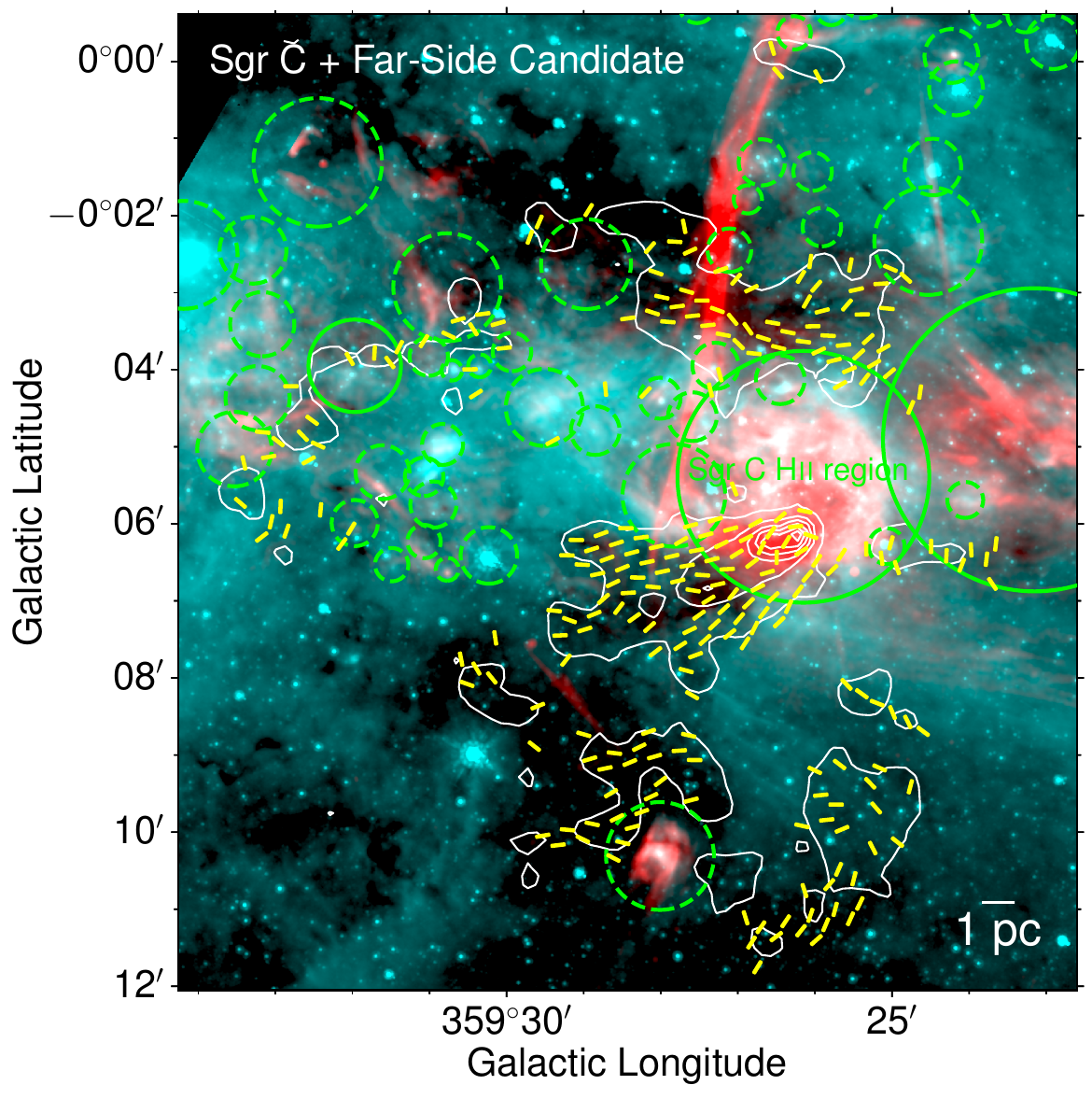} \\
\includegraphics[width=0.6\textwidth]{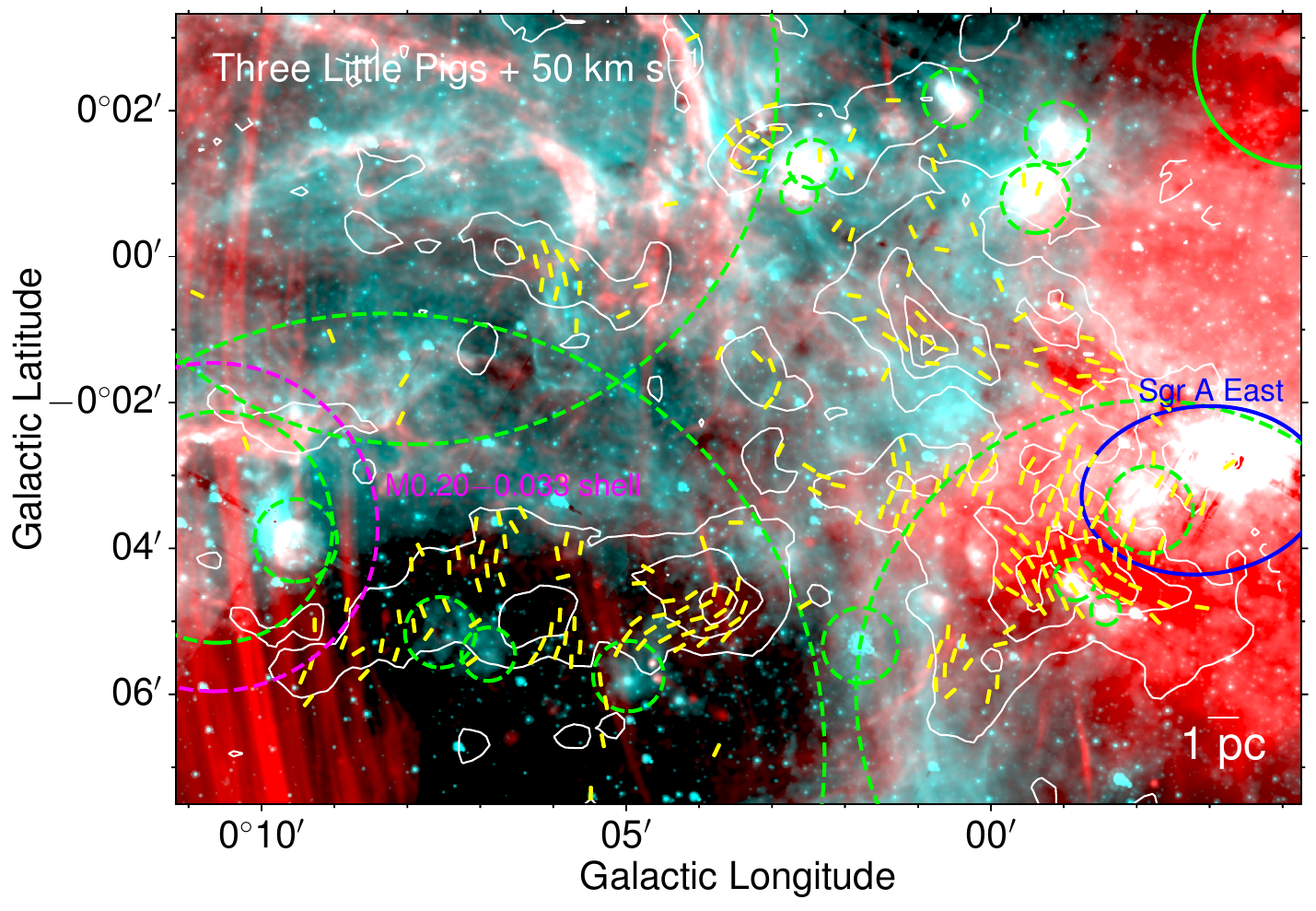} \\
\includegraphics[width=0.72\textwidth]{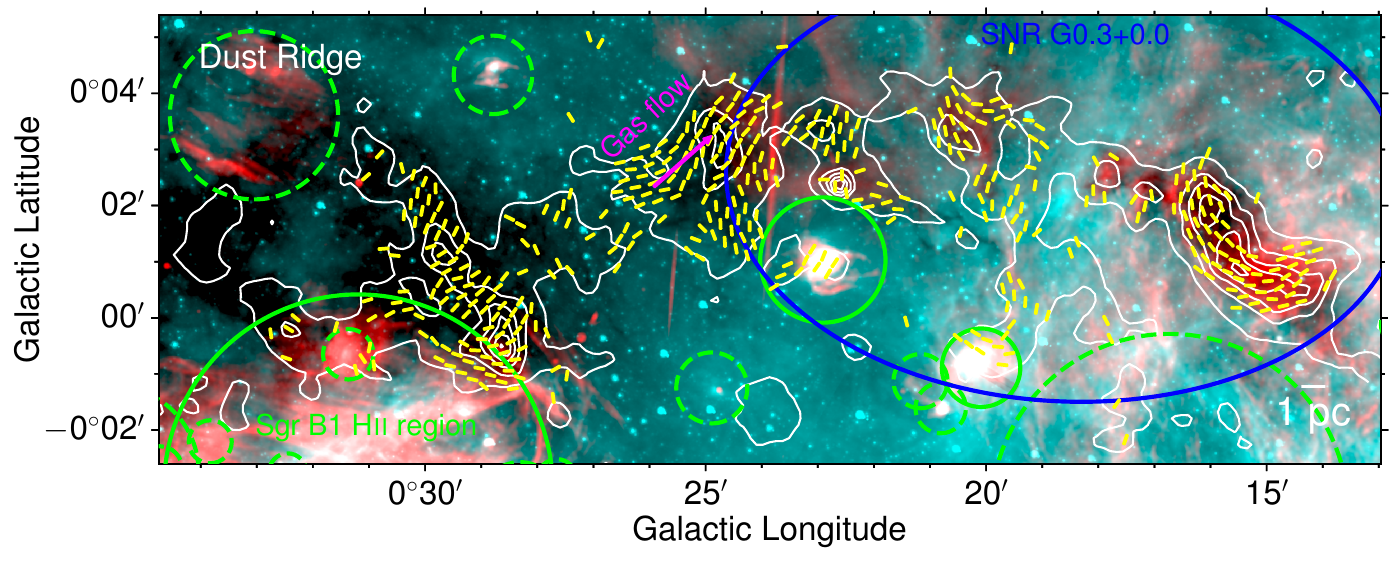}
\caption{False-color maps created with the MeerKAT 1.28~GHz continuum (red; \citealt{heywood2022}) and \textit{Spitzer} 8~\micron{} (cyan; \citealt{stolovy2006}) data. Contours and segments are the same as in \autoref{fig:allcmz}, representing the JCMT/POL2 850~\micron{} total intensities and the magnetic field orientations, respectively. The solid and dashed green circles denote the known and candidate \hii{} regions from the WISE catalog of Galactic \hii{} regions V2.4, respectively. The blue ellipses denote the two known SNRs \citep{green2022}. Other objects of interest, including the M0.20$-$0.033 molecular shell and the potential gas flow into cloud d, are labeled with magenta symbols and are discussed in \autoref{subsec:disc_environ}.}
\label{fig:env}
\end{figure*}

\subsection{Impacts of environments on magnetic fields}\label{subsec:disc_environ}

Given the complicated physical environments in the CMZ, e.g., young massive star clusters \citep{lang2005,simpson2018}, expanding supernova remnants (SNRs) and \hii{} regions \citep{hankins2020,heywood2022}, and pc-scale gas dynamics including cloud-cloud interactions \citep{hasegawa1994,dale2019}, it is expected that the clouds, as well as magnetic fields in them, would be affected, although exactly how is unclear owing to a lack of observations. Here we do not elaborate on every cloud, but focus on four cases where we find signatures of impacts of environments on magnetic fields in clouds. The following discussions are accompanied by the illustrations in \autoref{fig:env}, where we overlay \hii{} regions from the WISE catalog of Galactic \hii{} regions V2.4\footnote{\url{http://astro.phys.wvu.edu/wise/}} \citep{anderson2014} and SNRs from the catalog of \citet{green2022}.

\subsubsection{Cometary magnetic fields in Sgr C: signature of a cloud-\hii{} interaction event?}\label{subsubsec:disc_sgrc}

The Sgr~C molecular cloud is known to lie adjacent to an \hii{} region of $\sim$5~pc in size that has been seen in radio continuum \citep{lang2010} and mid-infrared emission \citep{hankins2020}, which reveals itself as a red patch in \autoref{fig:env}. In the mid-infrared map, the cloud appears as a silhouette against the bright emission from the \hii{} region \citep{hankins2020}. Therefore, the cloud should be in the foreground of the \hii{} region. The interaction between the cloud and the \hii{} region has been suggested in e.g., \citet{kendrew2013} and \citet{lu2019a}.

Magnetic fields in the Sgr~C cloud show a cometary morphology, which could be a consequence of the interaction with the expanding \hii{} region. Radiative feedback from massive stars in the \hii{} region as well as high-pressure ionized gas may have compressed molecular gas in the cloud, leading to aligned magnetic fields on the surface of the cloud. This scenario is similar to the results of magnetohydrodynamic simulations where an expanding \hii{} region erodes surrounding molecular gas, creating molecular pillars as well as cometary magnetic fields (\citealt{krumholz2007}; \citealt{arthur2011}, see their Figs.~22 \& 23; \citealt{mackey2011}, see their Figs.~2 \& 3). The simulations have shown that in such a scenario the magnetic field cannot be dynamically important compared to turbulence or thermal pressure, which indeed is the case for Sgr~C ($\mathcal{M}_\text{A}\sim2.1$, see \autoref{subsec:results_energies} \& \autoref{tab:stats}).

The putative interaction event is consistent with the star formation activities in the Sgr~C cloud. In the interaction scenario, the `head' of the cometary cloud should be firstly impacted by the \hii{} region, therefore should have the most evolved phase of star formation. The `tail' of the cloud then should contain subsequently less evolved phases of star formation. High-resolution interferometer observations do reveal such an evolutionary trend, where two massive protostellar objects associated with ultra-compact \hii{} regions and powerful outflows are found in the `head', and progressively less evolved protostellar activities including masers and weak outflows are observed further away from the  \hii{} region in the cloud \citep{kendrew2013,lu2019a,lu2021,lu2022}.

\subsubsection{Curved magnetic fields in the Three Little Pigs: perturbed by an expanding shell and interaction between clouds?}

The Three Little Pigs clouds have similar gas masses and column densities but show different fragmentation levels in high-resolution observations: the Stone cloud is highly substructured, while the Sticks and the Straw clouds are only scantly to moderately substructured \citep{battersby2020}. The origin of such a difference is unclear. The clouds are adjacent to the Quintuplet cluster as well as an expanding shell seen in molecular line emission \citep{butterfield2018,butterfield2022}. \citet{butterfield2018} have suggested a scenario where these clouds are interacting with the shell likely originated from the Quintuplet cluster. Here we investigate possible impacts on the magnetic fields in the three clouds by their environments.

The magenta dashed circle in \autoref{fig:env} illustrates the spatial extent of the expanding shell proposed by \citet{butterfield2018}, which is concentric with an \hii{} region candidate likely excited by the Quintuplet cluster. The magnetic field in the Straw cloud (G0.145$-$0.086) seems to show a curved morphology along the southwestern edge of the shell, which resembles that produced in simulations of expanding shells \citep{arthur2011,klassen2017}. The curved magnetic field could be a consequence of the interaction between the shell and the cloud. The same interaction may have enhanced turbulence in the Straw cloud, which is observed as a broader linewidth and a higher Alfv\'{e}nic Mach number in the Sticks \& Straw cloud than in the Stone cloud (\autoref{tab:stats}), thus making it less prone to fragmentation as compared to the Stone cloud on the western side \citep[e.g.,][]{federrath2015}.

The Sticks cloud (G0.106$-$0.082) is another candidate that may interact with adjacent gas components \citep[the filamentary cloud M0.11$-$0.11 or the Radio Arc bubble;][]{butterfield2022} and thus have its magnetic field morphology affected. Part of the Radio Arc bubble, a candidate \hii{} region, can be seen as the large dashed green circle surrounding the Three Little Pigs in \autoref{fig:env} \citep{rodriguezfernandez2001}. However, we have limited detections of magnetic field segments toward the Sticks cloud. It is difficult to compare its magnetic field structure to different interaction scenarios. More sensitive polarization observations are needed to study the relation between the environment of the Three Little Pigs clouds and their magnetic fields.

\subsubsection{Curved magnetic fields in Dust Ridge cloud d: tracing converging gas flows?}

\citet{williams2022} have proposed a scenario of large-scale ($\sim$pc) converging gas flows for cloud d in the Dust Ridge. One of the gas flows is suggested to come from the southeastern sides of the cloud, which is marked by an arrow in \autoref{fig:env}. Interestingly, magnetic fields as revealed by our observations seem to be curved in the same direction on the southeastern side of the cloud.

It is unclear whether the alignment between the gas flow and the magnetic field is caused by strong magnetic field regulating flowing gas, or the other way around, i.e., by strong turbulence dragging magnetic fields with in. The Mach number in cloud d, $\mathcal{M}_\text{A}\sim1.8$, is moderate (\autoref{subsec:results_energies}), and thus we are not able to tell whether or not the magnetic field here is dynamically important with respective to turbulence. If it is the gas flow that affects the magnetic field, curving the latter to the observed morphology, then it is a scenario similar to the alignment of gas flows and magnetic fields observed toward massive filaments in Galactic disk clouds \citep[e.g.,][]{pillai2020} and reproduced in numerical simulations \citep[e.g.,][]{gomez2018}.

\subsubsection{Magnetic fields across the filament and in the PDR in Dust Ridge cloud e/f}\label{subsubsec:dref}

The Dust Ridge cloud e/f presents a filamentary morphology that extends $\sim$10~pc with a total gas mass of 1.1$\times$$10^5$~\msol{}. Previous high-resolution interferometer observations have revealed signatures of star formation including masers and protostellar outflows in this cloud \citep[][]{lu2019a,lu2019b,lu2021}. However, its star formation efficiency is still $\sim$10 times lower than expected from the dense gas-star formation relation \citep{lu2019a}.

The magnetic field is overall perpendicular to the major axis of the filament. This is expected when the gas density is sufficiently high and thus the gas is channelled by the magnetic field to accumulate to form a filament \citep{lihb2013,lips2019}. Similar magnetic field geometry has been found in dense gas filaments in nearby star forming clouds \citep[e.g.,][]{pattle2017,pillai2020}, suggesting the important role of magnetic fields in regulating gas dynamics in dense filaments. An alternative explanation is that the filament is formed by the large-scale interaction of magnetic fields and turbulence \citep[e.g.,][]{liut2018}. In such a scenario, the small-scale perpendicular magnetic fields in the filament could be attributed to converging flows \citep{inoue2018}.

In addition, as shown in \autoref{fig:env}, the cloud is spatially adjacent to the Sgr~B1 region to the south, where a cluster of \hii{} regions and photodissociation regions (PDRs) have been detected \citep{simpson2018,simpson2021}. The southern edge of the cloud is facing the Sgr~B1 \hii{} regions, and therefore is likely a PDR. The magnetic field on this edge is well aligned. This alignment could be interpreted similarly to the case of Sgr~C (\autoref{subsubsec:disc_sgrc}), where PDRs are compressed by the ionized gas from \hii{} regions and thus have ordered magnetic fields along the interface.

\subsection{Orientations of local and global magnetic fields}\label{subsec:disc_PAs}

The angular resolutions of the JCMT and ACT observations are 14\arcsec{} and 1$'$, respectively, corresponding to small scales of $\sim$0.5~pc and large scales of $\gtrsim$2.5~pc. Therefore, the difference between the position angles of the two measurements could be used to trace the change of magnetic field orientations from large to small spatial scales, although we caution that the ACT data may include foreground signals and a better removal of foreground polarization emission is necessary (\autoref{subsec:results_morph}). The change in the orientations of magnetic fields can be attributed to the motion of partially ionized molecular gas on which the magnetic field is frozen. Figures~\ref{fig:act} \& \ref{fig:diffbpa_map} already illustrate the orientations of global and local magnetic fields. Here we present further discussions.

First of all, the observed change in the orientations of magnetic fields at the two spatial scales is unlikely attributed to difference grain sizes or dust temperatures. Although grain growth in molecular clouds is possible \citep[e.g.,][]{andersson2015}, there has not been observational evidence of grain growth between scales of 0.5--2.5~pc as well as its effect on magnetic fields. For spatial scales greater than disks, grain sizes are unlikely to vary significantly, given a lack of mechanisms to modulate them. As for dust temperatures, typical values in the clouds are 20--30~K \citep{tang2021}, with the spectral energy distribution (SED) peaking at $\sim$100--150~\micron{}. The wavelengths of the JCMT and ACT data (850~\micron{} vs.~1.3~mm) are both faraway from the SED peak, and therefore the temperature effect should not be significant.

\begin{figure*}[!thpb]
\centering
\includegraphics[width=0.3\textwidth]{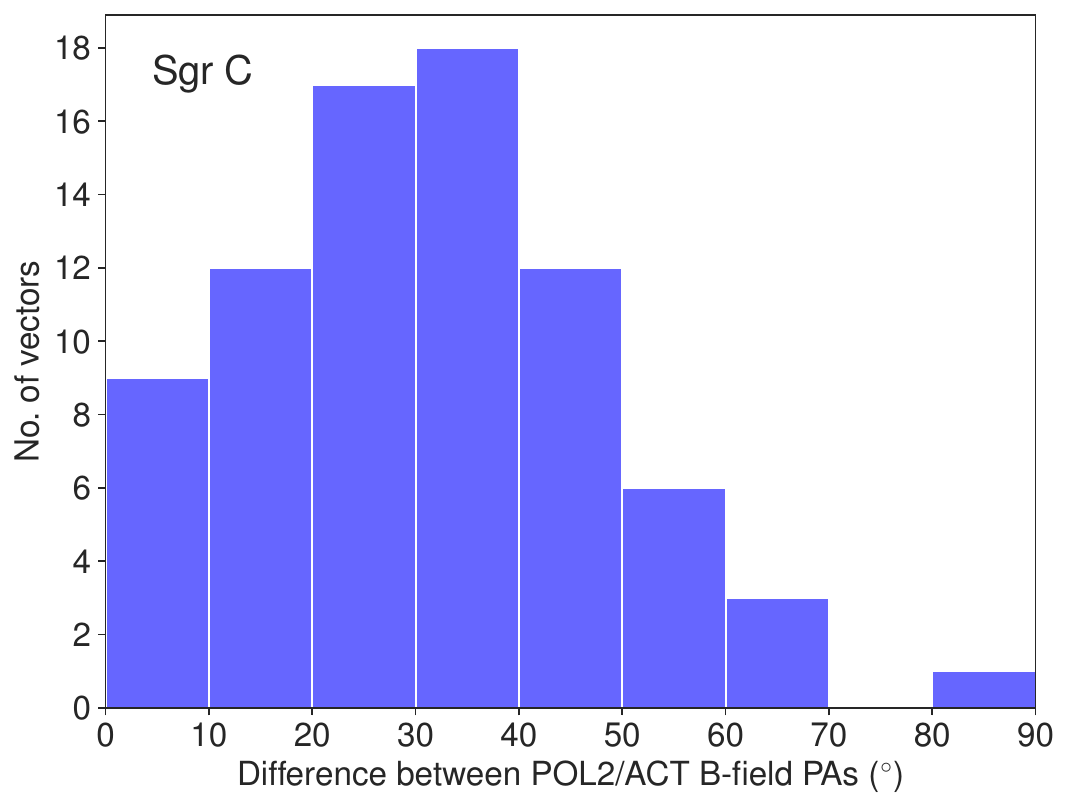}
\includegraphics[width=0.3\textwidth]{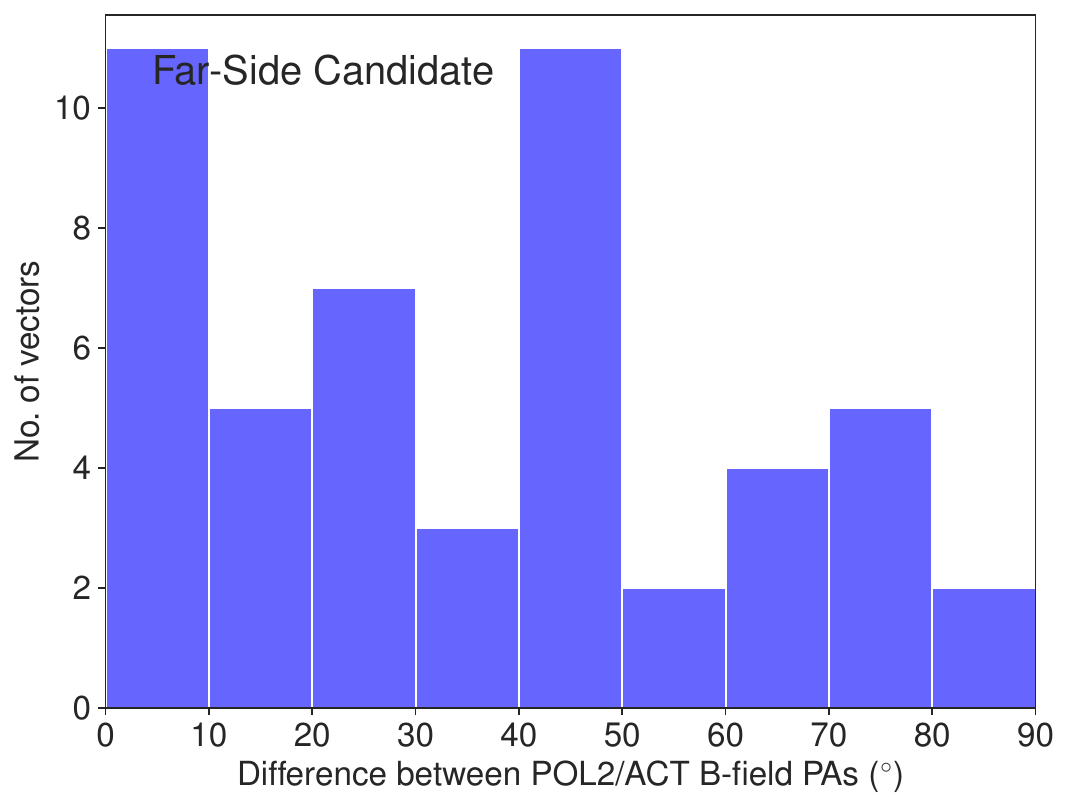} 
\includegraphics[width=0.3\textwidth]{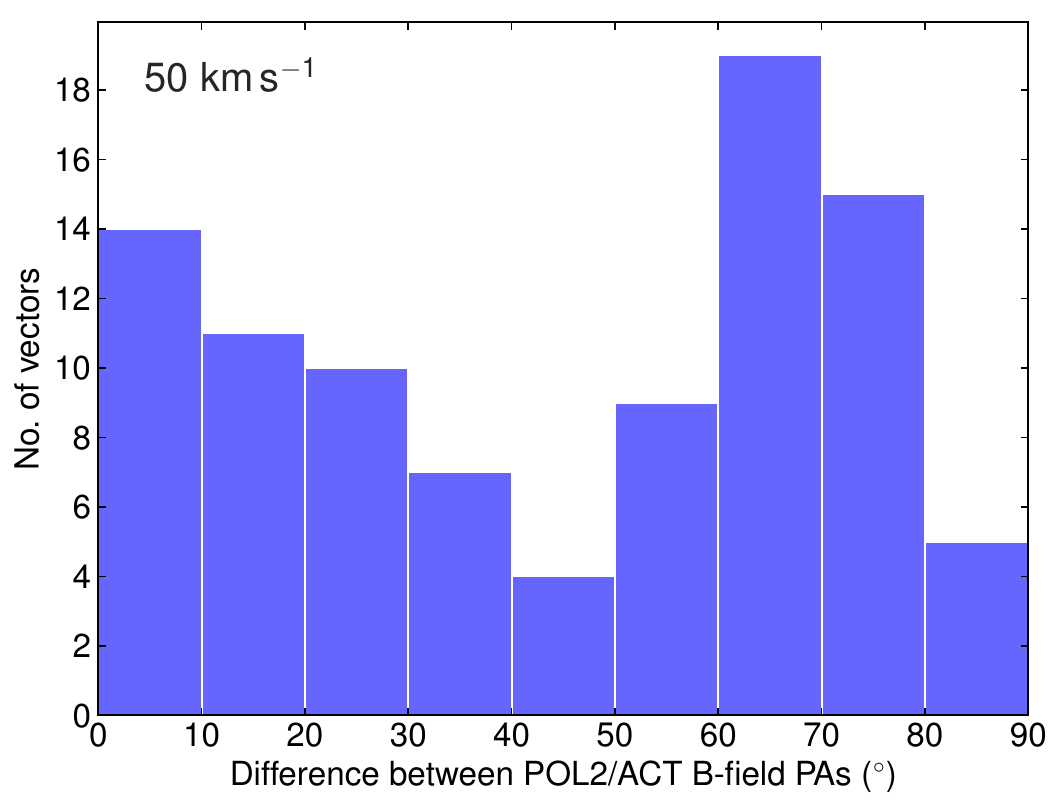} \\
\includegraphics[width=0.3\textwidth]{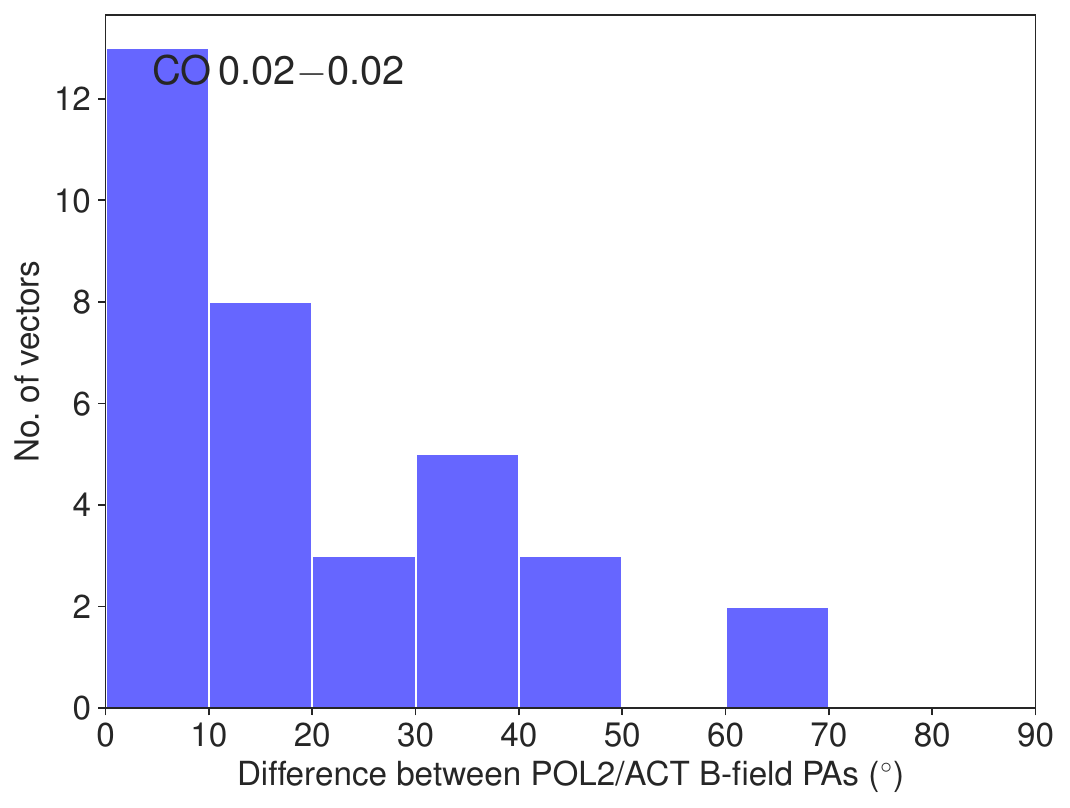}
\includegraphics[width=0.3\textwidth]{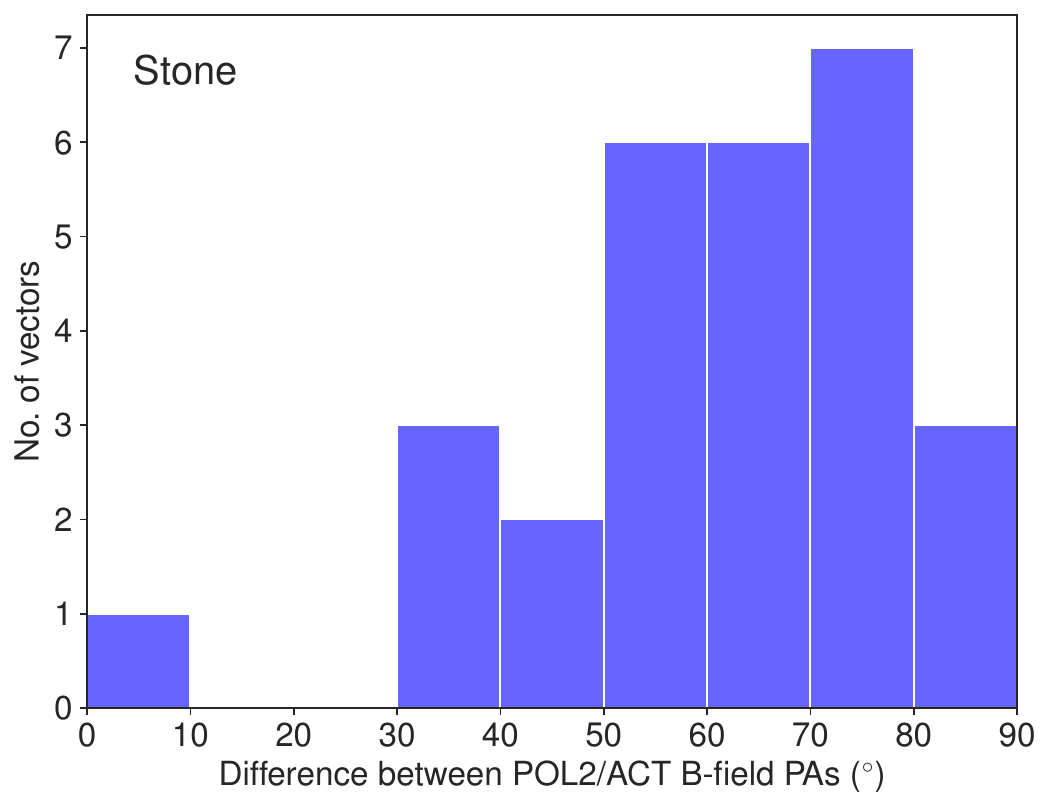} 
\includegraphics[width=0.3\textwidth]{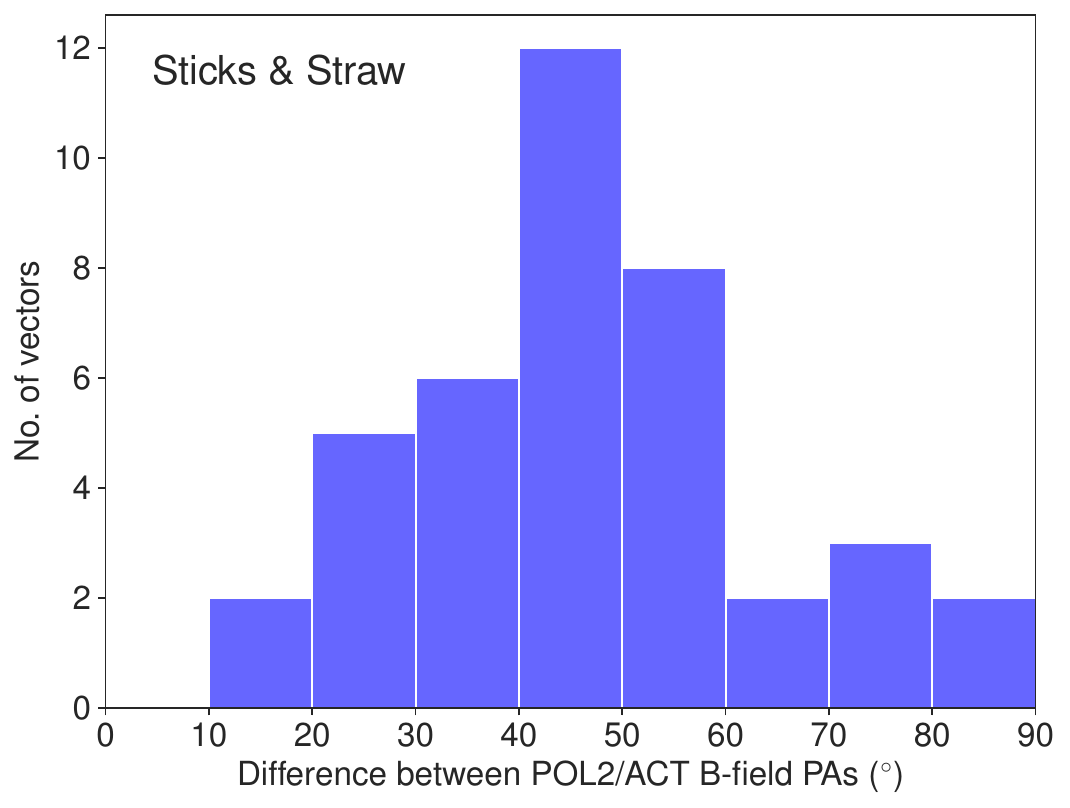} \\
\includegraphics[width=0.3\textwidth]{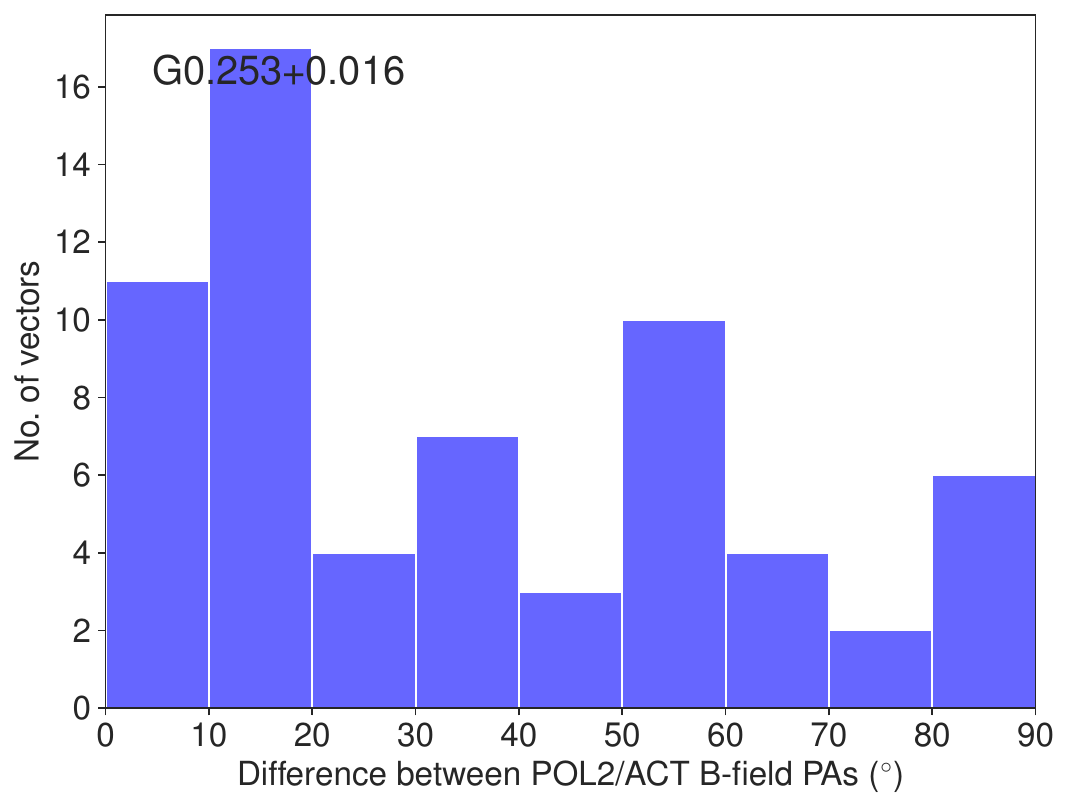}
\includegraphics[width=0.3\textwidth]{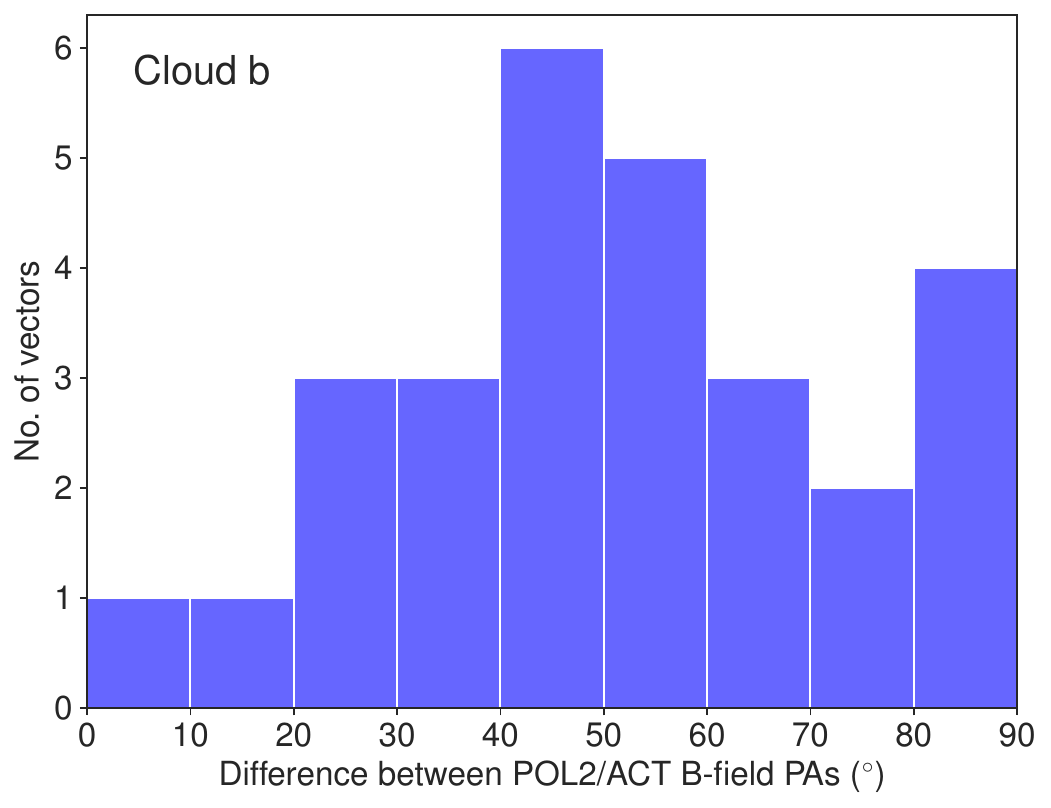} 
\includegraphics[width=0.3\textwidth]{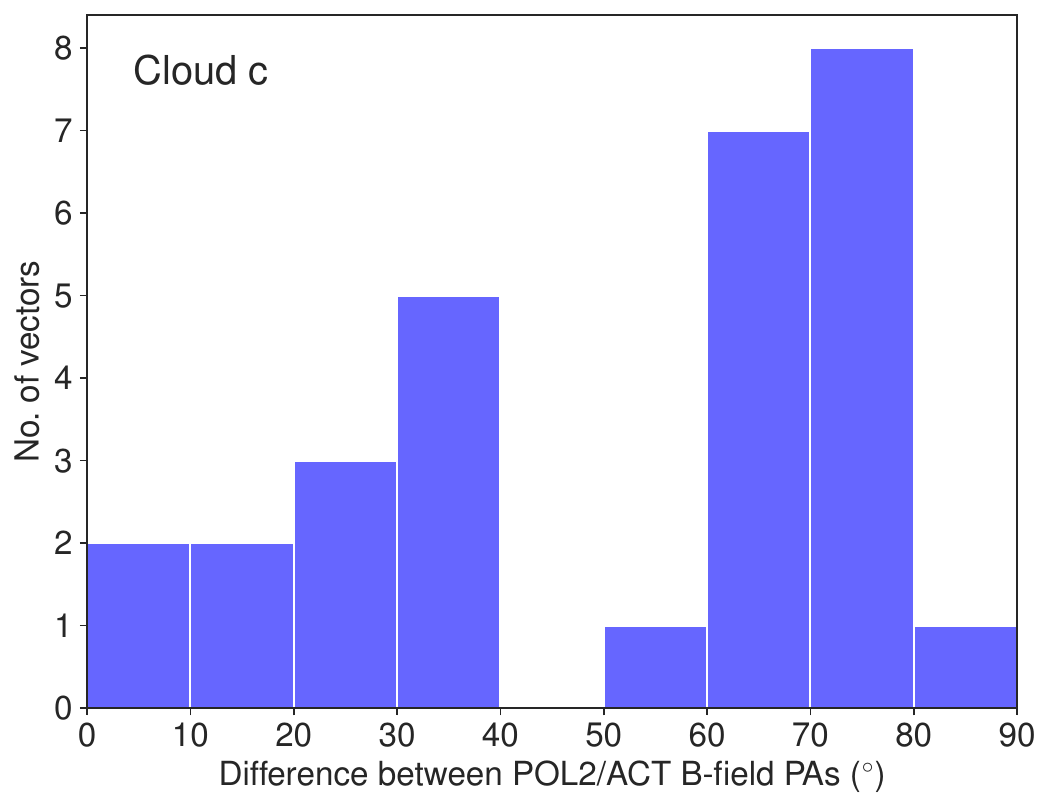} \\
\includegraphics[width=0.3\textwidth]{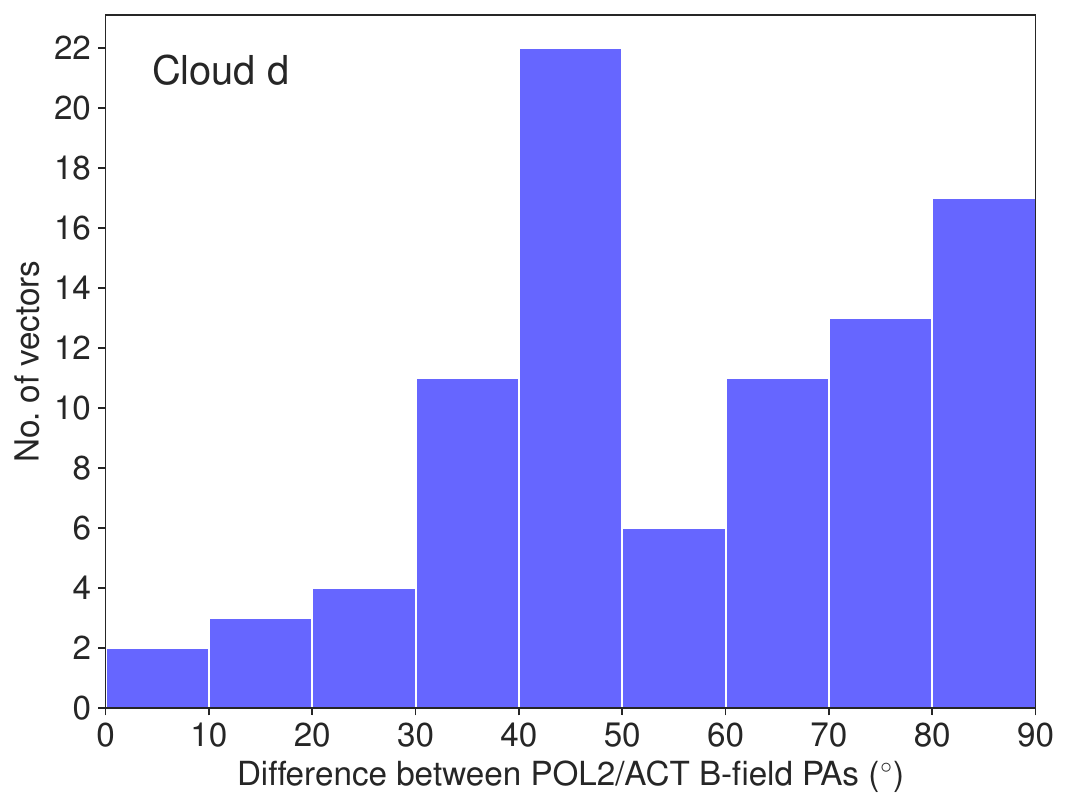}
\includegraphics[width=0.3\textwidth]{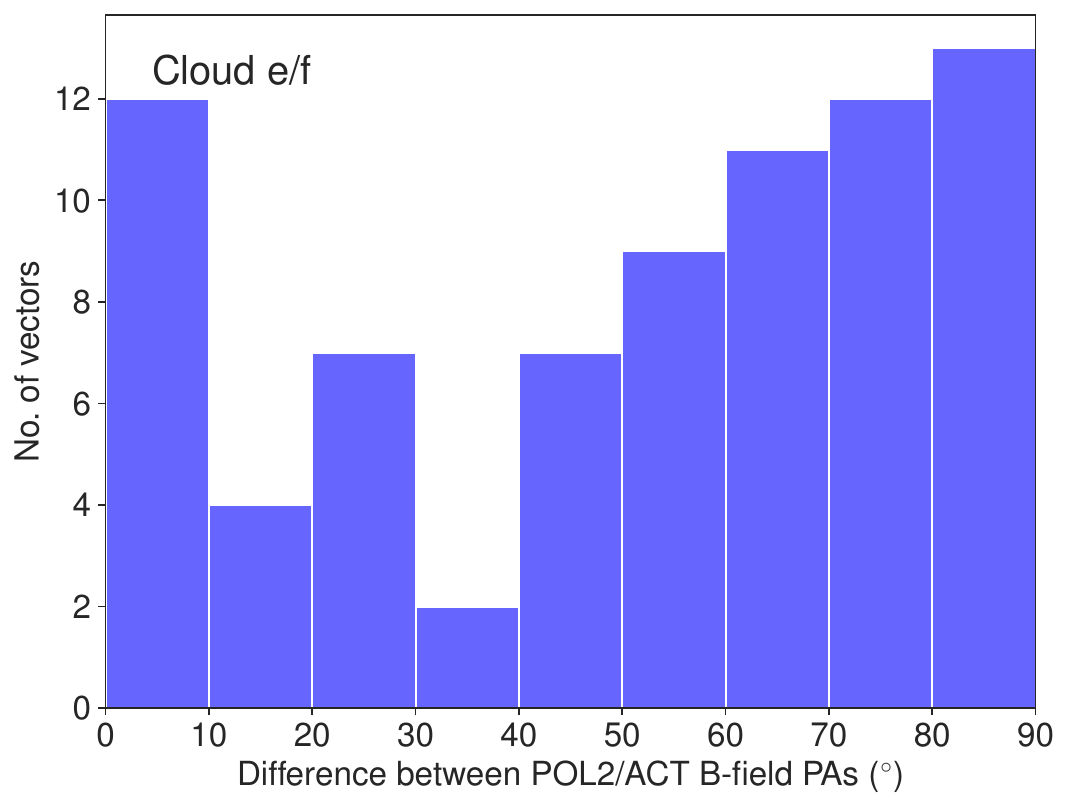}
\includegraphics[width=0.3\textwidth]{ 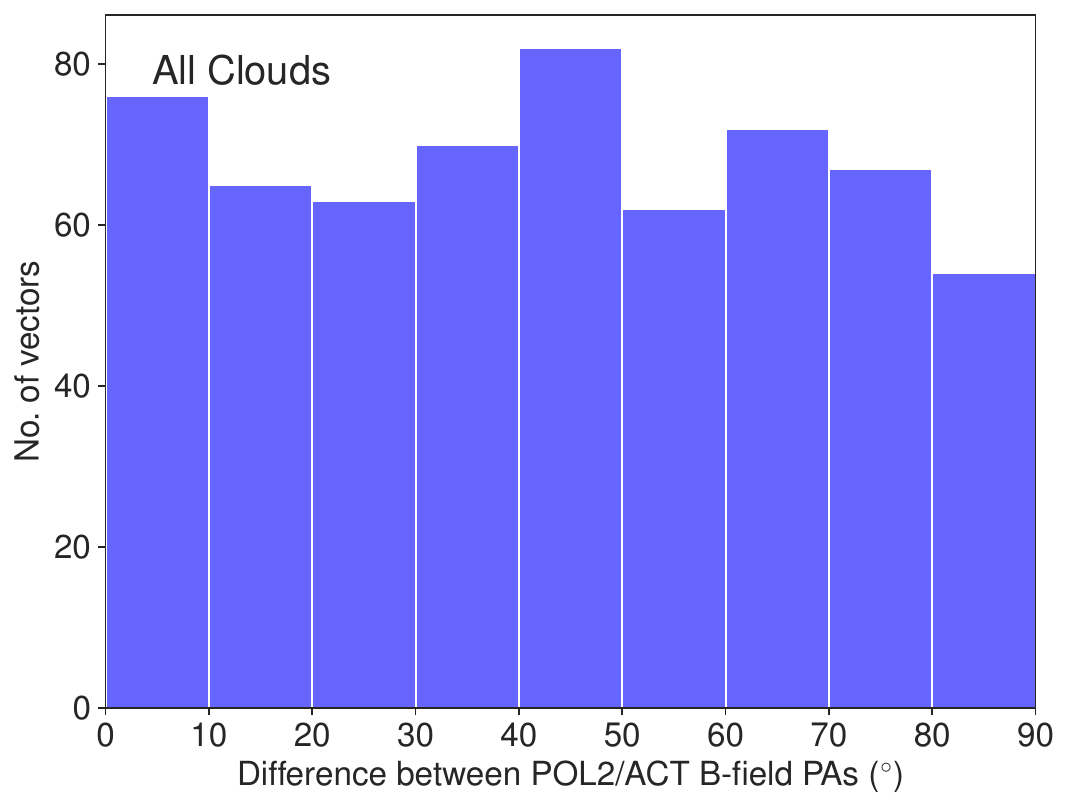}
\caption{The histograms show distributions of relative orientations between magnetic field position angles as probed by JCMT/Pol2 and ACT in each cloud (the first 11 panels) as well as for all the clouds (the last panel). The range of angles is limited between 0\arcdeg{} and 90\arcdeg{} as the difference between two positions angles is always smaller than 90\arcdeg{}.}
\label{fig:diffbpa_histo}
\end{figure*}

The distributions of the difference between magnetic field position angles probed by our JCMT/POL2 observations and those probed by ACT are illustrated as histograms in \autoref{fig:diffbpa_histo}. There are regions where the local magnetic fields tend to be perpendicular to the global ones, including the peak of Sgr~C, the Stone cloud, cloud c, cloud d, and cloud e/f. Note that some of these trends are not evident in \autoref{fig:diffbpa_histo} because the histograms include all measurements of position angles across a cloud, but such trends can be better illustrated in \autoref{fig:diffbpa_map}. These regions all have high column densities ($\gtrsim$10$^{23}$~\sqc{}, see \autoref{app_fig:column}). However, there also exist high column density regions where the local magnetic fields are aligned to the global ones, e.g., most of Sgr~C, the peak of \cfi{}, and \gzp{}. Lastly, orientations of local magnetic fields in several regions are likely affected by environments such as feedback. For example, the bimodal distribution of local magnetic field orientations in cloud e/f is due to the two different magnetic field geometries in the dense filament and the southern PDR in this cloud (see discussions in \autoref{subsubsec:dref}).

Therefore, qualitatively, we do not find a clear correlation between the orientations of local and global magnetic fields. It is possible that the local magnetic fields are affected by local environments such as expanding shells (Sgr~C, Stone) or photoionization from massive stars (cloud e/f), whose orientations are thus detached from the global ones and are forced to be aligned to the local environments such as PDR interfaces. It is also possible that the orientations of local magnetic fields probed by JCMT POL2 are affected by spatial filtering \citep{juvela2018}.

Lastly, in the last panel of \autoref{fig:diffbpa_histo}, we plot the integrated distribution of position angle differences between JCMT/POL2 and ACT observations for all the clouds in the sample. The distribution appears to be uniform between 0\arcdeg{} and 90\arcdeg{}, which is expected if the two position angles are randomly distributed with respect to each other. We run both the Kolmogorov-Smirnov (K-S) test and the Watson test from circular statistics with the null hypothesis that the observed position angle differences are distributed according to the uniform distribution between 0\arcdeg{} and 90\arcdeg{}. The resulting $p$-values are relatively high ($>$0.05). Therefore, we cannot reject the hull hypothesis, and the observed JCMT/POL2 and ACT position angles are likely randomly oriented with respect to each other. Previous studies of magnetic fields in Galactic high-mass star forming regions have found a bi-modal rather than random distribution for the position angle differences of different spatial scales \citep[cores at 0.03~pc scales vs.\ clumps at 0.1--0.6~pc scales;][see their Figure~2]{zhang2014}. The bi-modal distribution indicates that the magnetic field plays a dynamically important role in the formation of the dense cores. In the CMZ, however, strong turbulence could dominate gas dynamics and randomize the orientations of magnetic fields in individual clouds. Whether the random distribution of position angle differences continues to even smaller spatial scales in these CMZ clouds can be tested with high angular resolution observations from interferometers such as ALMA.

\begin{figure*}[!thpb]
\centering
\includegraphics[width=1\textwidth]{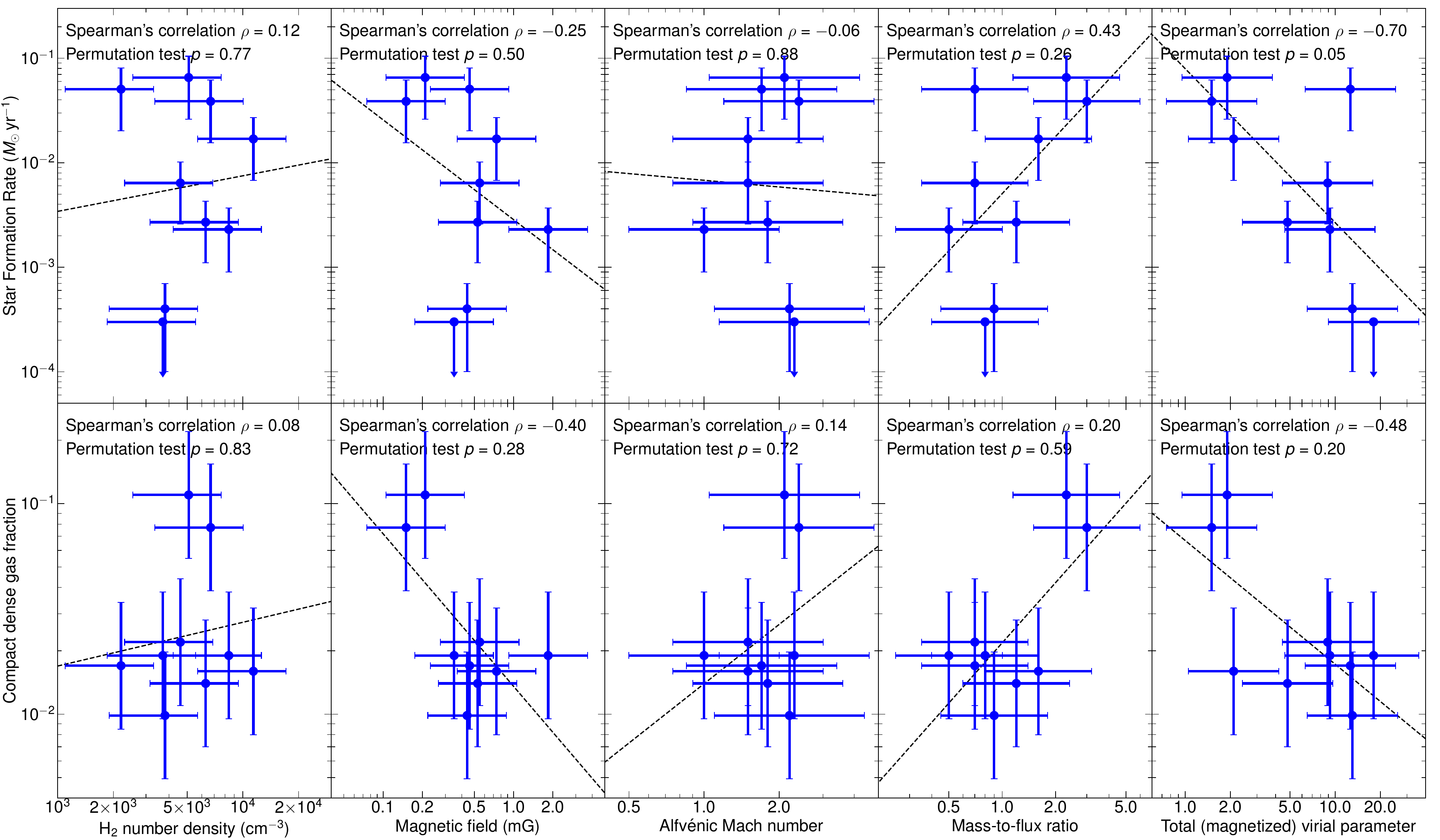}
\caption{Correlations between physical properties of nine clouds in our sample. Two clouds, the Far-Side Candidate and \coo{}, are not covered by CMZoom, and thus are not included here. The x-axis is one of the physical properties derived from our observations: H$_2$ number density, magnetic field strength, Alfv\'{e}nic Mach number, mass-to-flux ratio, and total virial parameter. The y-axis is either the CDGF \citep{battersby2020} or the SFR of the clouds (Hatchfield et al.\ 2023, in prep.). The Spearman's rank correlation coefficient $\rho$ and the $p$-value derived from a permutation test are labeled in each panel. The dashed line is a linear fit to the data points.}
\label{fig:correlation}
\end{figure*}

\subsection{What physical properties are correlated with the dense gas fractions and SFRs?}\label{subsec:disc_Bfield}

The impact of strong magnetic fields on star formation in the CMZ has been discussed in \citet{pillai2015}. Here, we extend the discussion to a wider sample of the 11 clouds and search for correlations between physical properties (including magnetic fields) and star formation activities.

To characterize star formation activities of the clouds, we choose two parameters: the compact dense gas fraction (CDGF), and the SFR. The clouds in our sample are included in the CMZoom survey (\citealt{battersby2020,hatchfield2020}; Hatchfield et al.\ 2023, in prep.), in which the CDGFs and SFRs of the clouds are quantified. The CDGF is defined as the ratio of gas masses in compact substructures on 0.1--2~pc scales based on the SMA observations and cloud masses based on the \textit{Herschel} observations \citep[see Table 4 of][`method2']{battersby2020}. It can be used as a measure of `dense' gas fraction of the clouds that is directly related to star formation \citep{lu2019a}. Typical uncertainties in the CDGFs are a factor of 2 and potentially higher \citep[Section 5.5 of][]{battersby2020}. The SFR of a cloud is estimated from its total molecular gas mass divided by its free-fall time, corrected by factors accounting for a $\sim$25\% star formation efficiency over one free-fall time and an initial mass function (Hatchfield et al.\ 2023, in prep.). The uncertainties of the SFRs derived by Hatchfield et al.~(2023, in prep.) are listed in \autoref{tab:stats}, with a typical value of 50\% and potentially higher.

The physical properties of the clouds that we have measured include the number density $n$(H$_2$), the magnetic field strength $B$, the Alfv\'{e}nic Mach number $\mathcal{M}_\text{A}$, the mass-to-flux ratio $\lambda$, and the virial parameter $\alpha_\text{vir,B}$ (see \autoref{tab:stats}). 

In \autoref{fig:correlation}, we plot these physical properties against the two characteristics of star formation activities, the SFR and the CDGF. We derive Spearman's rank correlation coefficients $\rho$ for the pairs of parameters. $\rho=1$ suggests a strong correlation, $\rho=-1$ a strong anti-correlation, and $\rho=0$ no correlation. Among all the relations, we only find moderate anti-correlations between the total virial parameter $\alpha_\text{vir,B}$ and the SFR/CDGF, with $\rho=-0.70$ and $-0.48$, respectively, and relatively lower $p$-values. There are also a weak correlation between the mass-to-flux ratio $\lambda$ and the SFR, and a weak anti-correlation between the magnetic field $B$ and the CDGF, both with $\textbar\rho\textbar\sim0.4$. For all the other pairs of parameters, we do not find clear correlations ($\textbar\rho\textbar\lesssim0.2$).

The anti-correlation between $\alpha_\text{vir,B}$ and the SFR/CDGF is expected when star formation in these clouds is regulated by the dynamical states of the clouds co-determined by the self-gravity, turbulence, and magnetic field. In our sample, the magnetic field is a minor contribution in the virial parameter, which is dominantly determined by the self-gravity and turbulence (see \autoref{tab:stats}: virial parameters without or with taking the magnetic field into account do not vary much). Combining with the fact that only weak to no correlations are detected between the magnetic field and the SFR/CDGF, we suggest that the magnetic field plays a minor role in regulating star formation in these clouds as compared to self-gravity and turbulence.

\citet{palau2021} studied magnetic fields at $\lesssim$0.1~pc scales in a sample of high-mass star forming cores in the Galactic disk, and found a tentative correlation between their mass-to-flux ratios and fragmentation levels. Whether the same trend holds for the spatial scale of dense cores ($\lesssim$0.1~pc) in the CMZ clouds need further investigations using interferometers such as ALMA. If such a trend is corroborated, it would indicate that the magnetic field in the CMZ plays a non-negligible albeit probably minor role in influencing gas dynamics and star formation at the spatial scales of clouds to cores.

We also note that only two clouds in the sample, Sgr~C and Dust Ridge cloud c, have low virial parameters ($\alpha_\text{vir,B}<2$) that suggest marginally gravitationally bound gas. These two clouds have high CDGFs ($\gtrsim$0.1) and are known to be forming high-mass stars \citep{kendrew2013,ginsburg2015,walker2018,lu2019a,lu2019b,lu2020,lu2021}. All the other clouds have virial parameters higher than 2 and therefore are likely unbound in the absence of external pressure. However, these clouds are unlikely to be dispersing, given the compact dense substructures detected inside them \citep[e.g,][]{battersby2020,lu2020,walker2021}. Previous observations have found evidence of strong external pressure confining dense clouds in the CMZ \citep{myers2022,callanan2023}. Our results suggest that for these clouds the external pressure could be critical for keeping them in dynamic equilibrium, although the conclusion is subject to large uncertainties in the virial parameters (at least a factor of 2; see \autoref{appd_sec:uncertainty}).

The robustness of the conclusions suffers from the limited sample size. Future full-CMZ surveys of the magnetic field \citep[e.g.,][]{butterfield2023} would enlarge the sample size and lead to more robust results regarding correlations between the magnetic field and star formation. For example, correlations between magnetic field strengths and SFRs of giant molecular clouds have been suggested in external galaxies \citep[e.g.,][]{tabatabaei2018}, although the magnetic field therein was estimated from a different approach (basing on the non-thermal radio continuum and assuming equipartition between the energy densities of the magnetic field and cosmic rays). With a larger sample toward the CMZ, we will be able to examine whether the same correlations hold.

Lastly, we note that the correlation in \autoref{fig:correlation} could be alternatively interpreted as an effect of cloud evolution. Quiescent clouds at the earliest stages of star formation could be gravitationally unbound and magnetically subcritical. As more gas is accumulated from the environment \citep[e.g., through large-scale gas inflow;][]{williams2022}, the clouds could become gravitationally bound and supercritical, and therefore start to collapse and form stars.

%%%%%%%%%%%%%%%%%%%%%%%%%
\section{CONCLUSIONS}\label{sec:conclusions}

We present JCMT/POL2 observations of polarized dust emission in the CMZ at a resolution of 14\arcsec{} (0.55~pc linear resolution). The observations cover three large areas and sample 11 clouds. This is by far the largest sample of high-resolution ($<$1~pc, thus spatially resolving the clouds) studies of magnetic fields in molecular gas in the CMZ. The results from this study are:
\begin{itemize}
\item The morphologies of magnetic fields in the clouds projected on the plane of the sky are inferred from the polarized dust emission. By comparing to larger scale magnetic field morphologies traced by the $1'$ resolution ACT observations, we find that the large and small scale magnetic fields in these clouds may not be aligned. The misalignment may be partly due to the impact of local environments, since we find evidence of local magnetic fields aligned to expanding shells around \hii{} regions and PDR interfaces.
\item The magnetic field strengths in the clouds are estimated using the ADF method to be between 0.1 and 1.7~mG, with an uncertainty of at least a factor of 2. The Alfv\'{e}nic Mach number, mass-to-flux ratio, and total (magnetized) virial parameter are then derived. All the clouds but Sgr~C and Dust Ridge cloud c have virial parameters higher than 2, suggesting that they would be gravitationally unbound if the external pressure is not considered.
\item Correlations between the five physical properties of the clouds derived from our observations (H$_2$ number density, magnetic field strength, Alfv\'{e}nic Mach number, mass-to-flux ratio, and total virial parameter) and the two characteristics of star formation (SFR and CDGF) are explored. A moderate correlation is found between the total virial parameter and the SFR/CDGF. Weak (anti-)correlations are found between the mass-to-flux ratio and the SFR, and between the magnetic field and the CDGF. This may suggest a minor role of the magnetic field in regulating star formation in these clouds.
\item To summarize, the magnetic field alone is unable to explain the different star formation states of the 11 clouds in our sample. Self-gravity and turbulence seem to be dominant in determining the star formation states of the clouds. In certain clouds (e.g., Sgr~C, cloud e/f), star formation as well as the magnetic field could be subject to feedback from \hii{} regions and massive star clusters.
\end{itemize}

\acknowledgments
We thank the anonymous referee for constructive comments that have improved the quality of the paper. This work has been supported by the National Key R\&D Program of China (No.\ 2022YFA1603101). This work has been sponsored by Natural Science Foundation of Shanghai (No.\ 23ZR1482100). X.L.\ acknowledges support from the National Natural Science Foundation of China (NSFC) through grant No.\ 12273090, and the Chinese Academy of Sciences (CAS) `Light of West China' Program (No.\ xbzg-zdsys-202212). Q.Z.\ gratefully acknowledges support from the National Science Foundation under Award No.\ 2206512. T.L.\ acknowledges the supports by National Natural Science Foundation of China (NSFC) through grants No.\ 12073061 and No.\ 12122307, the international partnership program of Chinese Academy of Sciences through grant No.\ 114231KYSB20200009, and Shanghai Pujiang Program 20PJ1415500. We thank Prof.\ Yu Gao for his efforts in supporting the operation of JCMT. We thank J\"{u}rgen Ott for sharing the ATCA \ammthree{} data, and Yuping Tang for helpful discussions on the LMT data. We also thank Bethan Williams, Natalie Butterfield, and Yuhei Iwata for helpful discussions. This research has made use of NASA’s Astrophysics Data System. The James Clerk Maxwell Telescope is operated by the East Asian Observatory on behalf of The National Astronomical Observatory of Japan, Academia Sinica Institute of Astronomy and Astrophysics, the Korea Astronomy and Space Science Institute, the National Astronomical Observatories of China and the Chinese Academy of Sciences (Grant No.\ XDB09000000), with additional funding support from the Science and Technology Facilities Council of the United Kingdom and participating universities in the United Kingdom and Canada. 

\vspace{5mm}
\facilities{JCMT (SCUBA2, POL2), ATCA, LMT}

\software{Starlink \citep{berry2005,chapin2013}, APLpy \citep{aplpy2012}, Astropy \citep{astropy2018,astropy2022}}

\clearpage

\appendix
\section{Column Density Maps and Estimates of Number Densities}\label{appd_sec:density}

\begin{figure*}[!thpb]
\centering
\includegraphics[width=0.5\textwidth]{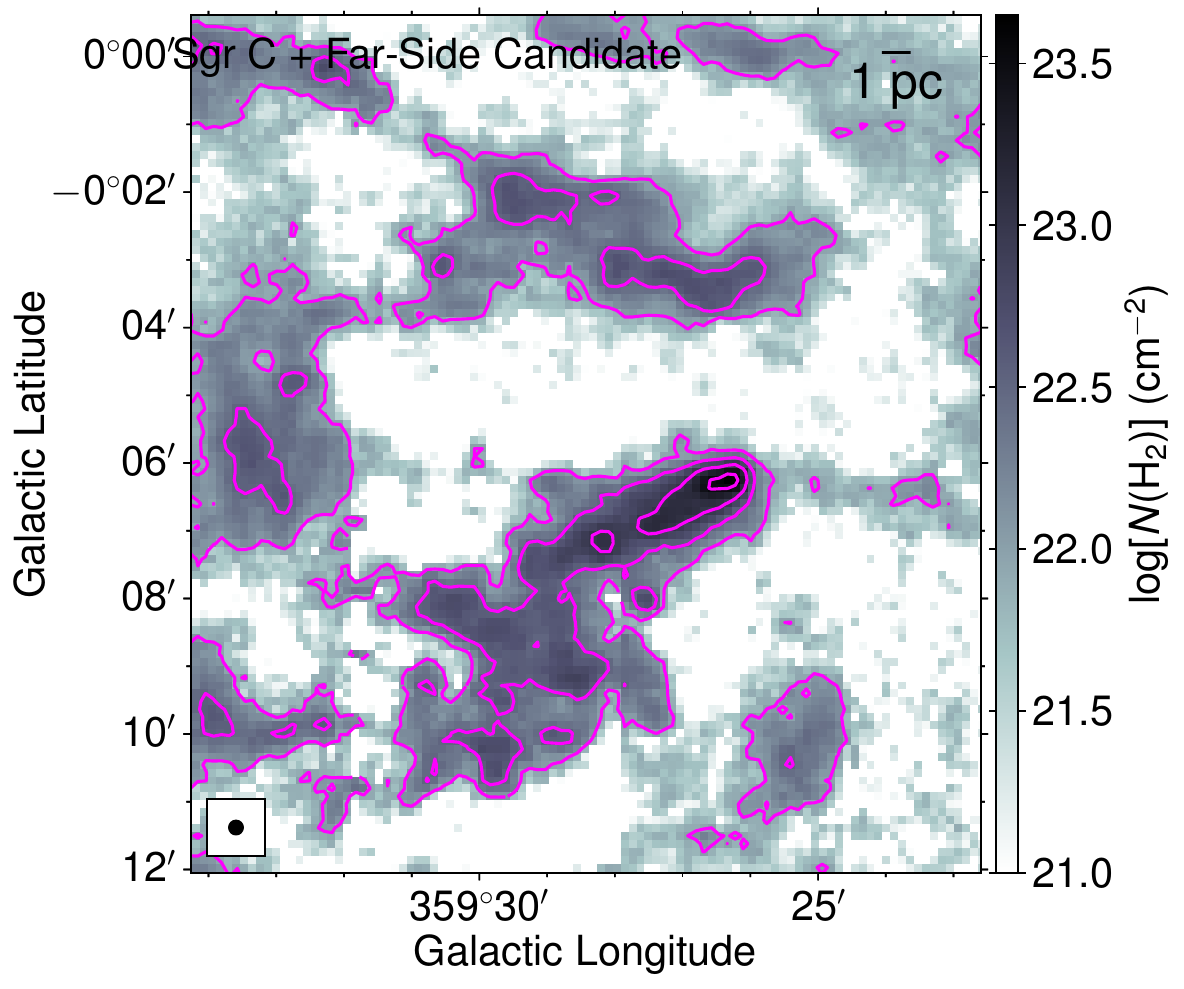} \\
\includegraphics[width=0.6\textwidth]{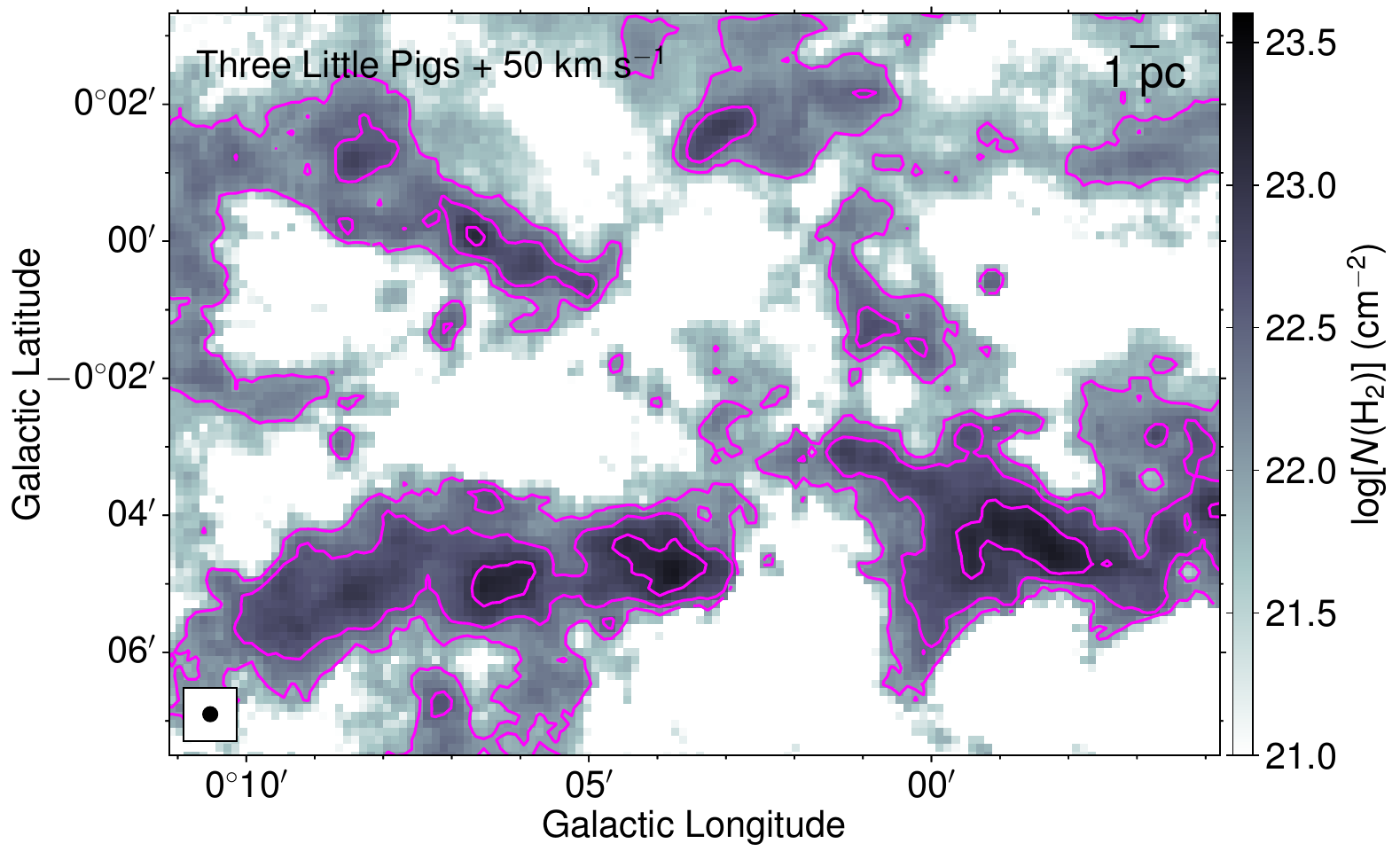} \\
\includegraphics[width=0.72\textwidth]{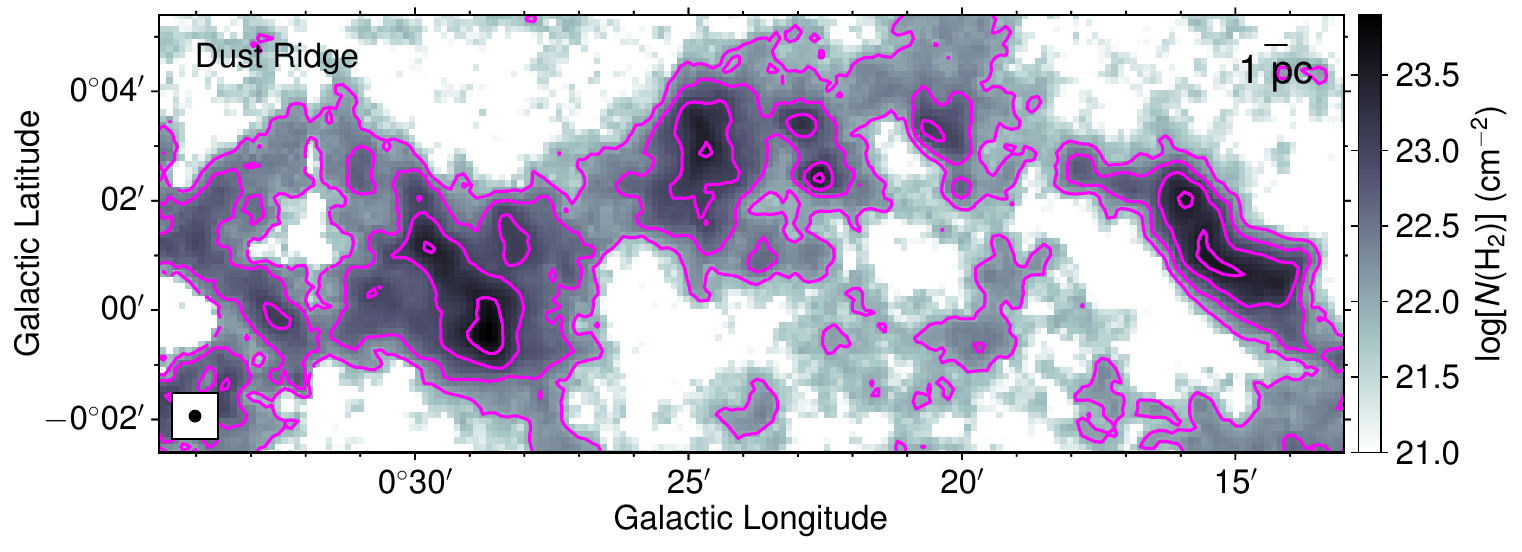}
\caption{Molecular hydrogen (H$_2$) column densities of the three CMZ fields derived from the JCMT 850~\micron{} continuum of \citet{parsons2018}. The contours start at $10^{22}$~\sqc{}, with an increment of 0.5 in the logarithmic scale.}
\label{app_fig:column}
\end{figure*}

Column density maps of the three fields, following the methods in \autoref{subsec:results_mass}, are presented in \autoref{app_fig:column}. Note that the 850~\micron{} continuum emission total intensities are from the SCUBA2 observations of \citet{parsons2018} that provide a better sampling of diffuse dust emission and removal of CO contamination.

\section{Angular dispersion functions and fitting results}\label{appd_sec:adf}

\begin{figure*}[!thpb]
\centering
\includegraphics[width=0.3\textwidth]{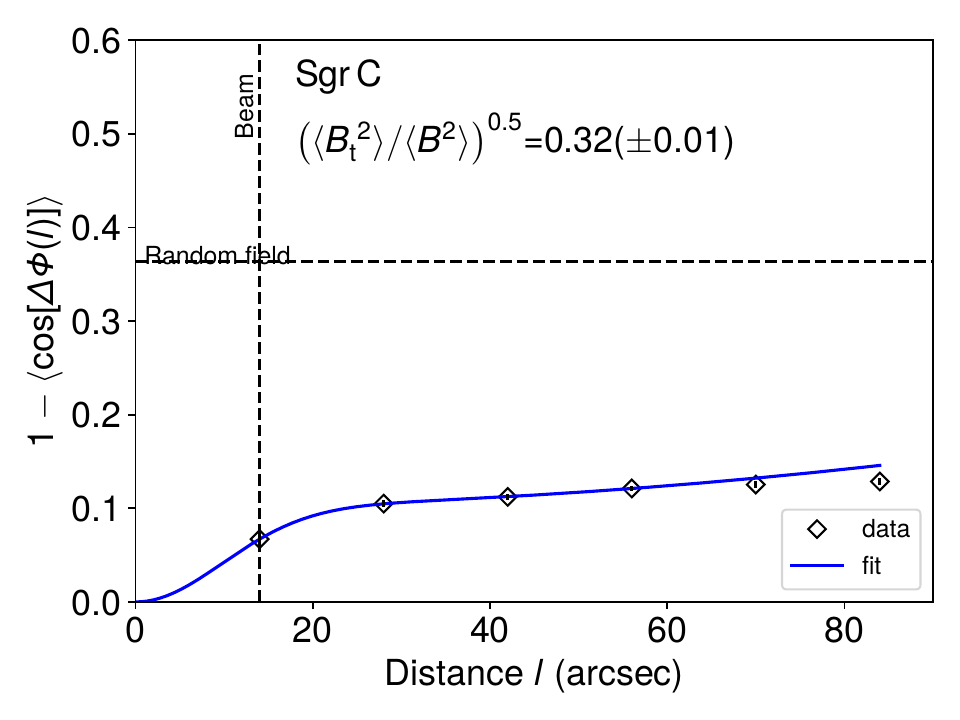}
\includegraphics[width=0.3\textwidth]{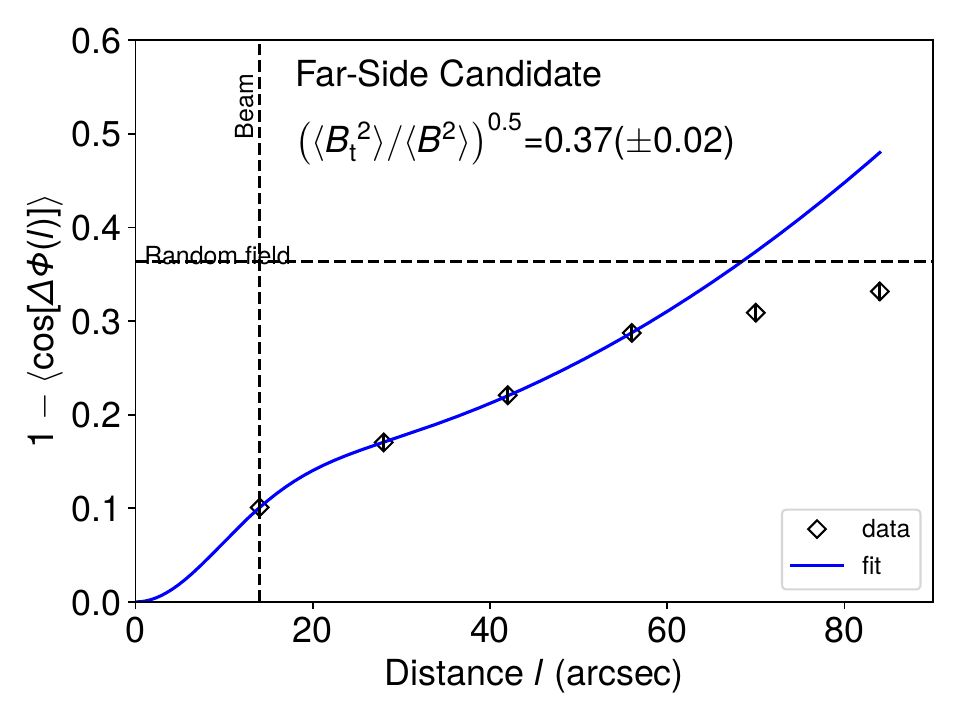}
\includegraphics[width=0.3\textwidth]{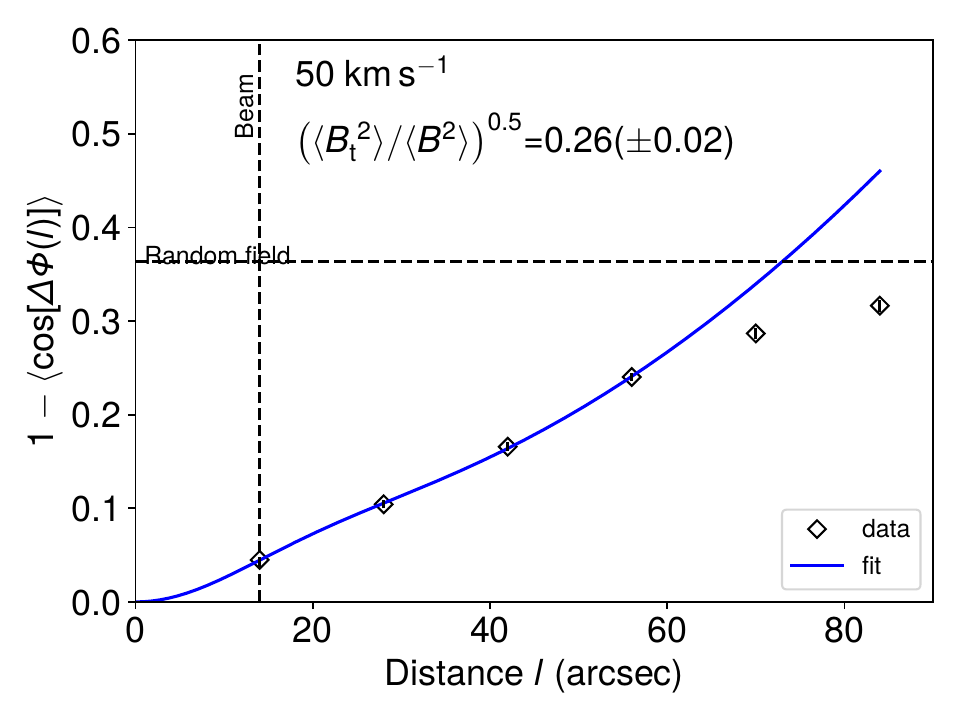} \\
\includegraphics[width=0.3\textwidth]{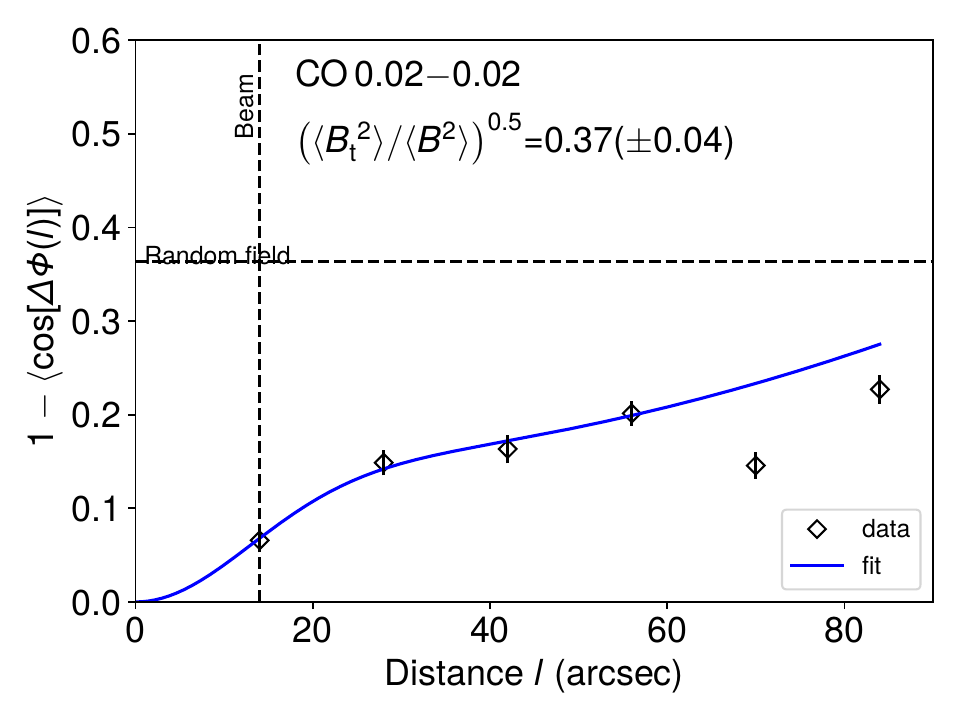}
\includegraphics[width=0.3\textwidth]{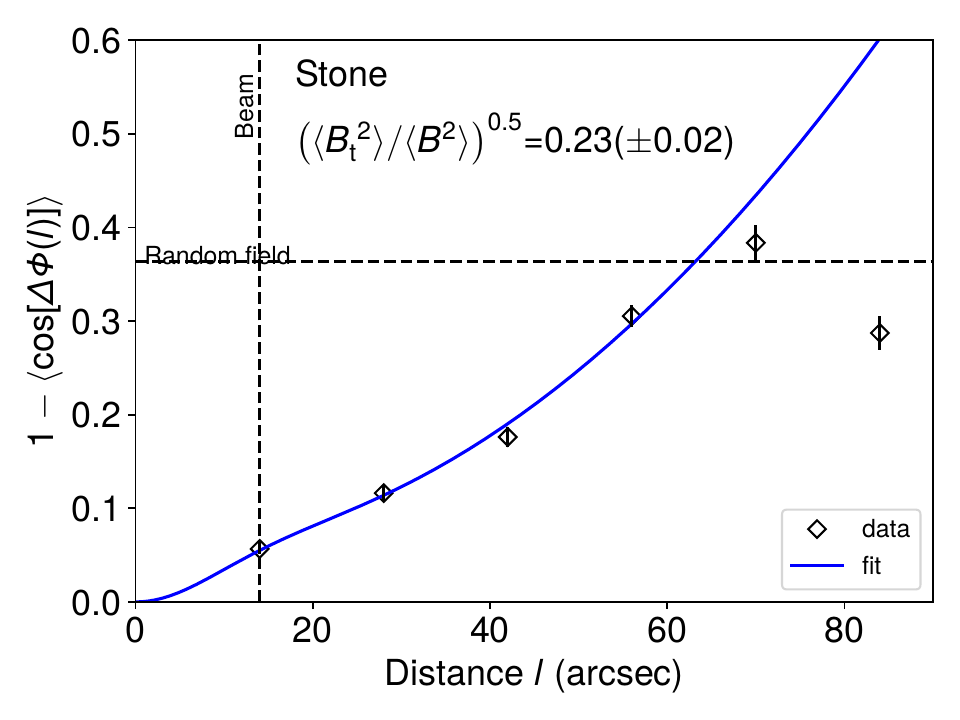}
\includegraphics[width=0.3\textwidth]{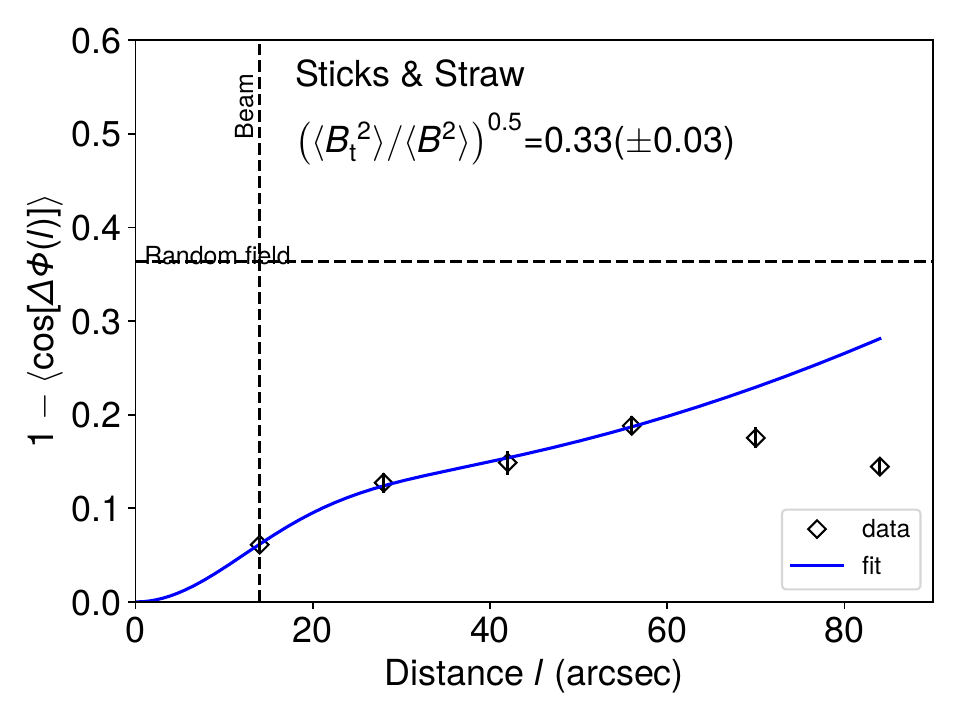} \\
\includegraphics[width=0.3\textwidth]{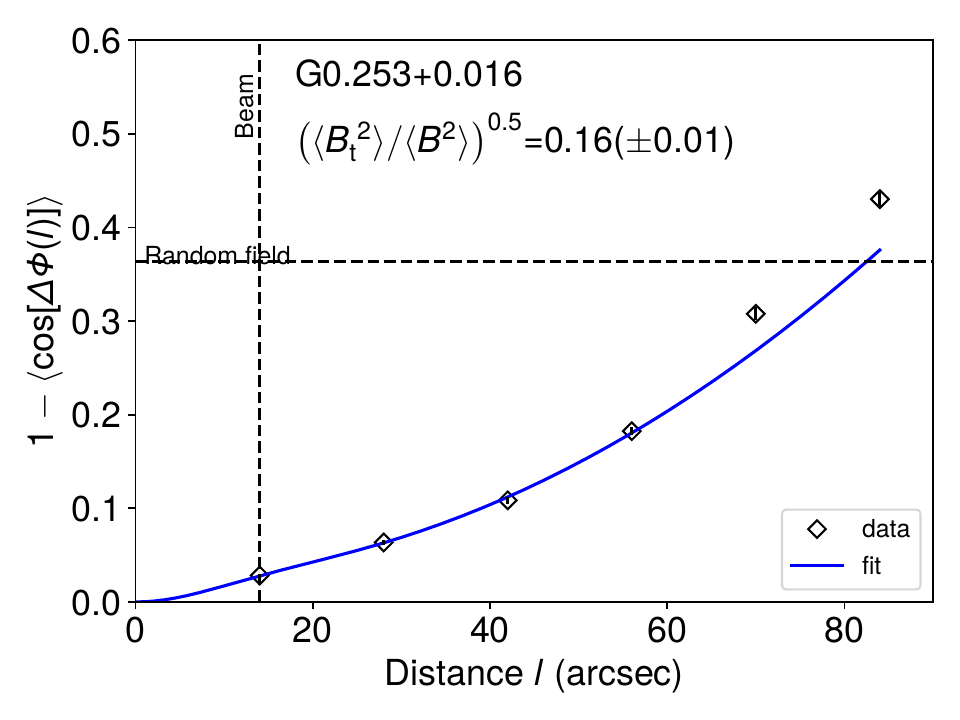}
\includegraphics[width=0.3\textwidth]{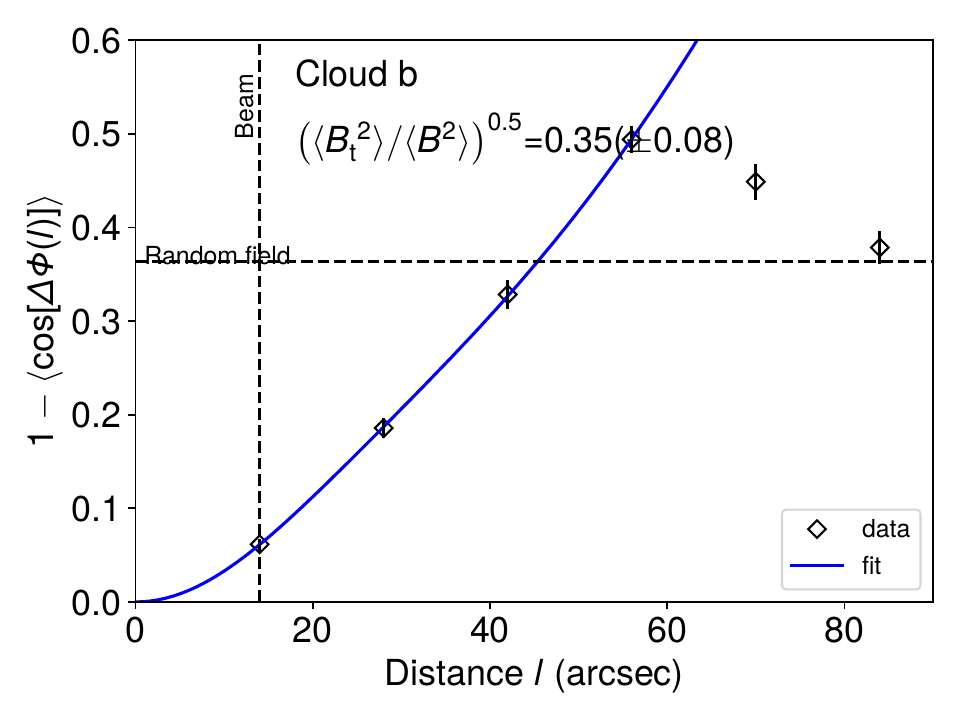}
\includegraphics[width=0.3\textwidth]{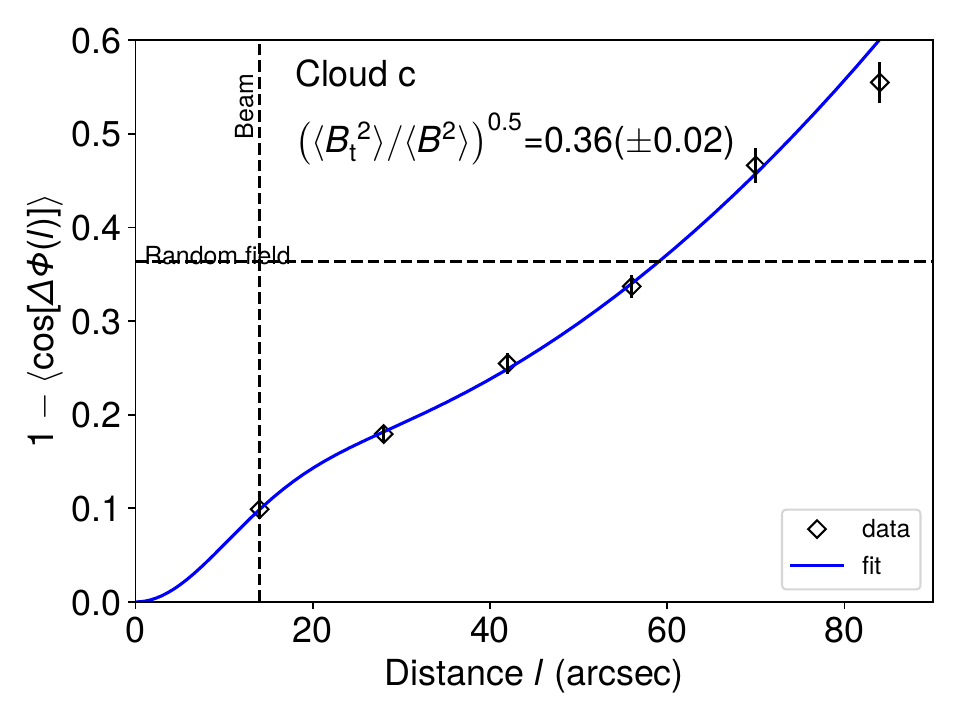} \\
\includegraphics[width=0.3\textwidth]{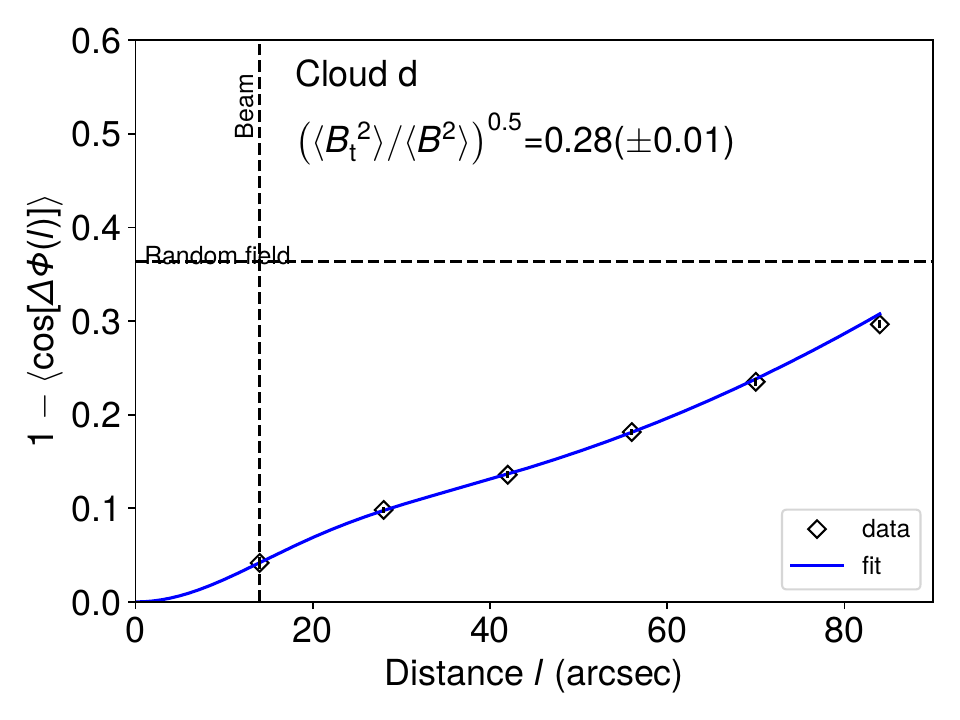}
\includegraphics[width=0.3\textwidth]{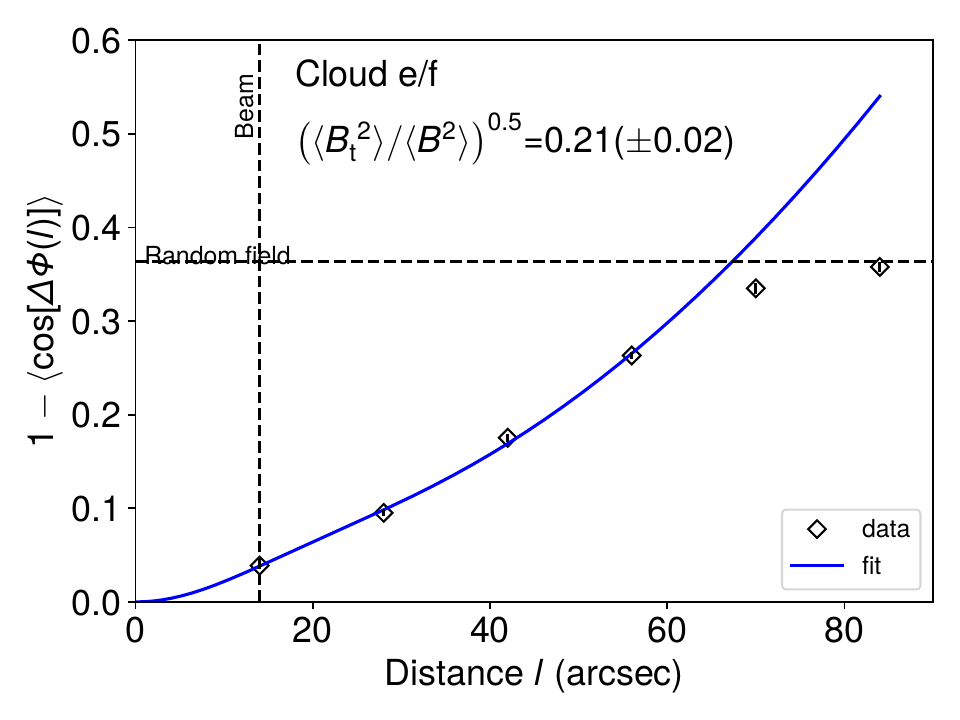}
\caption{ADFs of the 11 clouds. Uncertainties in the ADFs inherited from the scatter of the position angles are shown as vertical errorbars. The fitting results following \autoref{equ:ADF} are plotted as blue curves. The beam size (14\farcs{1}) is marked by a vertical dashed line, while the ADF value in the case of randomly distributed magnetic fields (0.36) is marked by a horizontal dashed line. The best-fit turbulent-to-total magnetic field strength ratio, as well as its uncertainty, is labeled in each panel.}
\label{app_fig:adfs}
\end{figure*}

Angular dispersion functions (ADFs) of the 11 clouds and fitting results following \autoref{equ:ADF} are presented in \autoref{app_fig:adfs}. For most data, the vertical errorbars, representing the uncertainties in the ADFs, appear smaller than the symbols. These uncertainties have been taken into account in the fitting of the ADFs, contributing to the quoted uncertainty of the turbulent-to-total magnetic field strength ratio in \autoref{app_fig:adfs}.

\section{Fitting results of the \ammthree{} spectra}\label{appd_sec:ammonia}

\begin{figure*}[!thpb]
\centering
\includegraphics[width=0.3\textwidth]{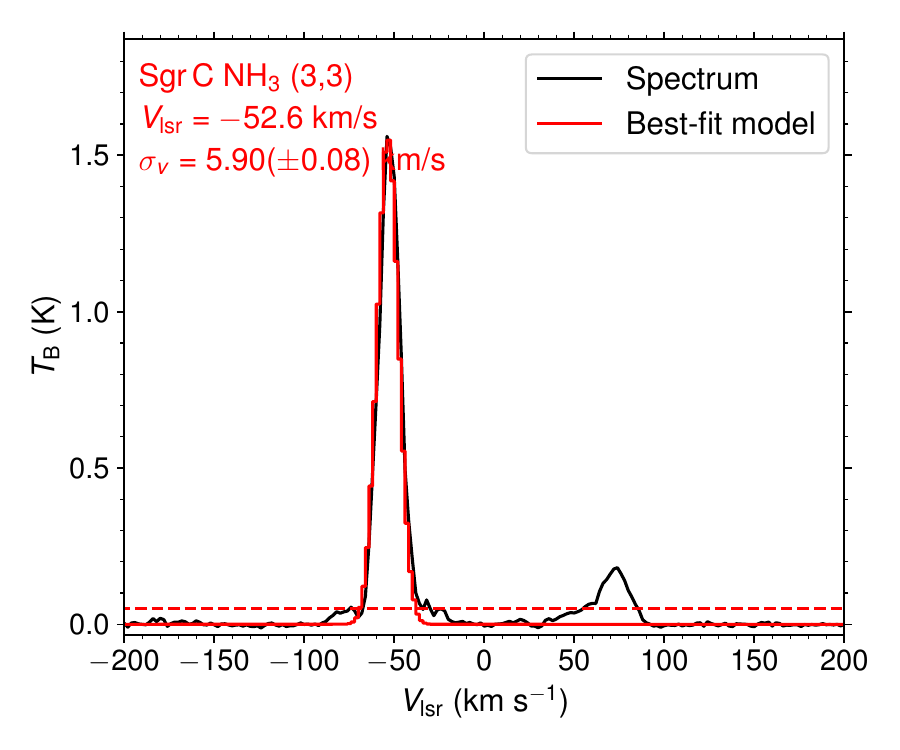}
\includegraphics[width=0.3\textwidth]{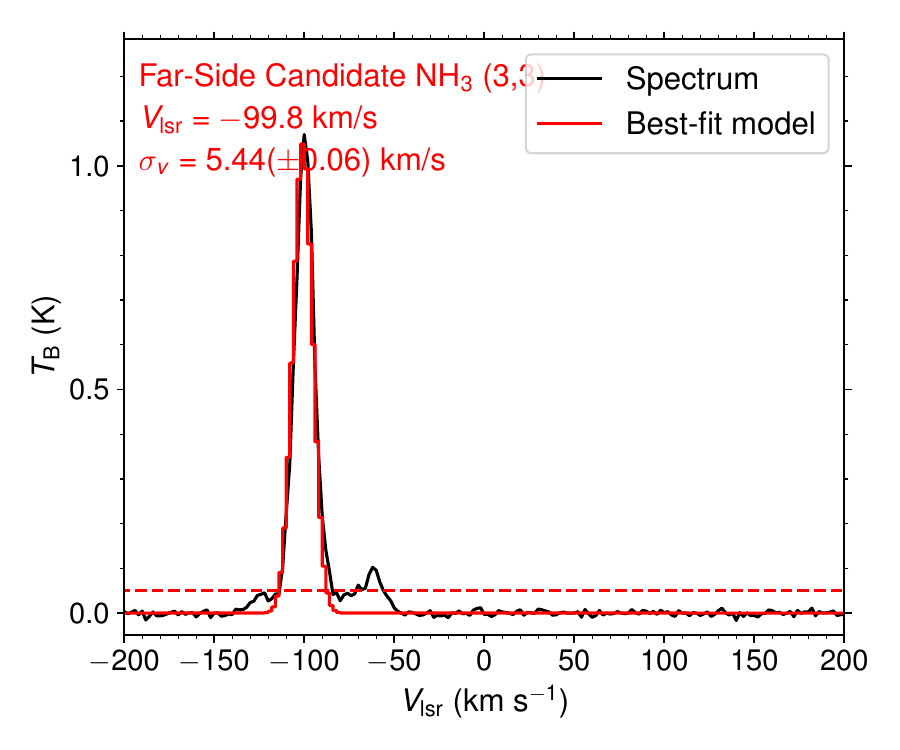}
\includegraphics[width=0.3\textwidth]{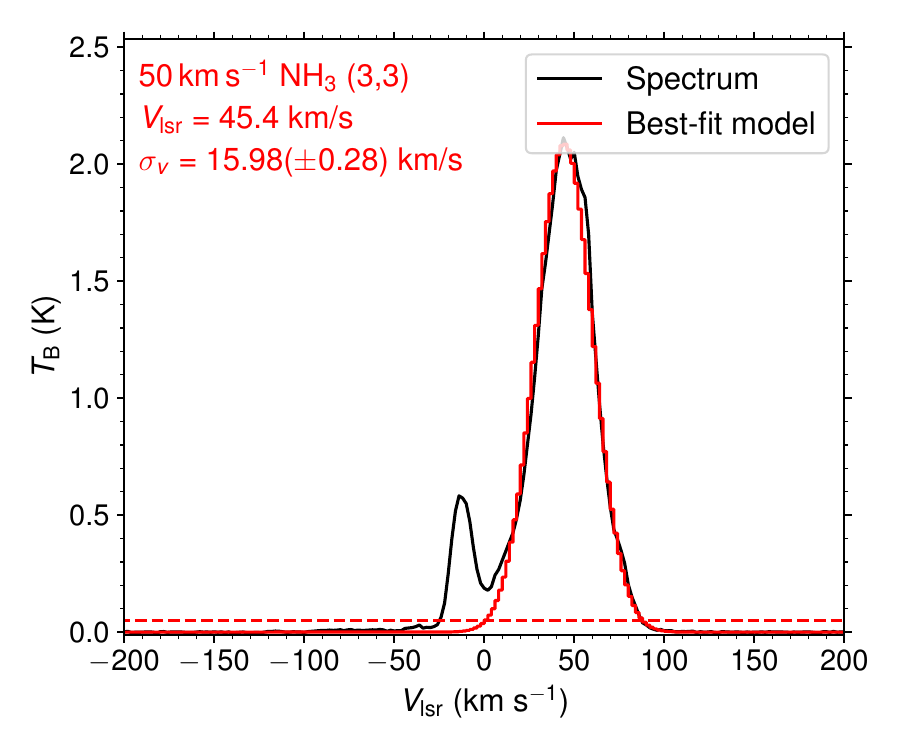} \\
\includegraphics[width=0.3\textwidth]{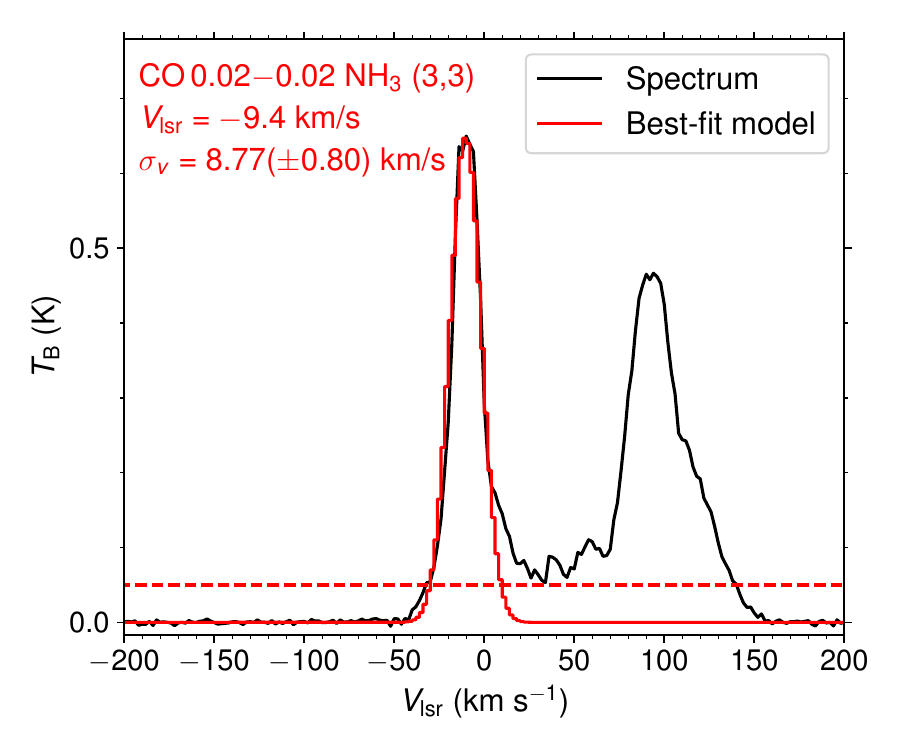}
\includegraphics[width=0.3\textwidth]{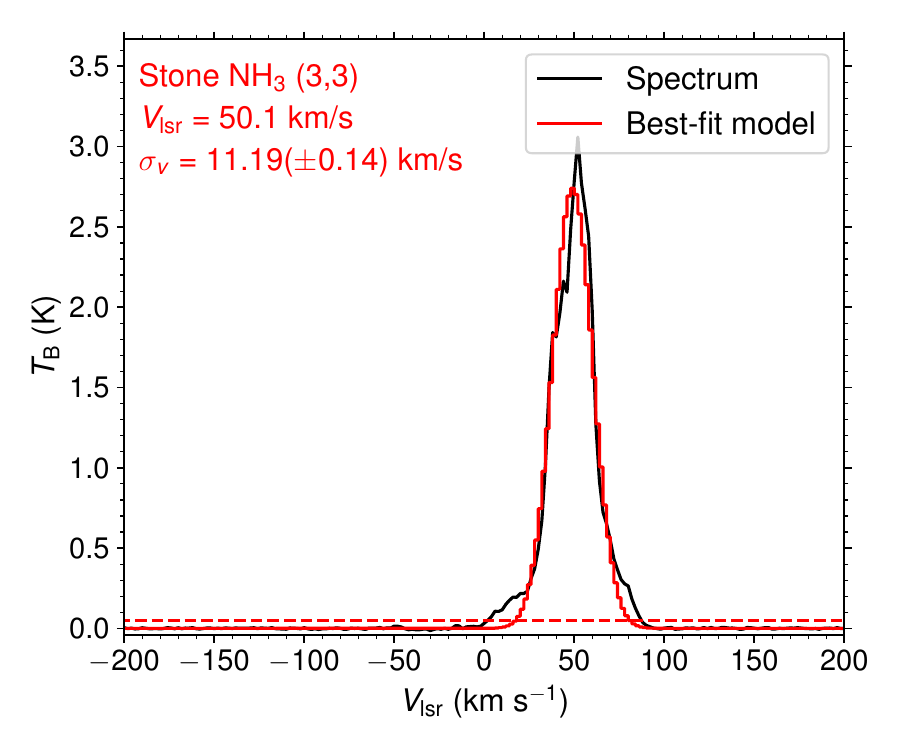}
\includegraphics[width=0.3\textwidth]{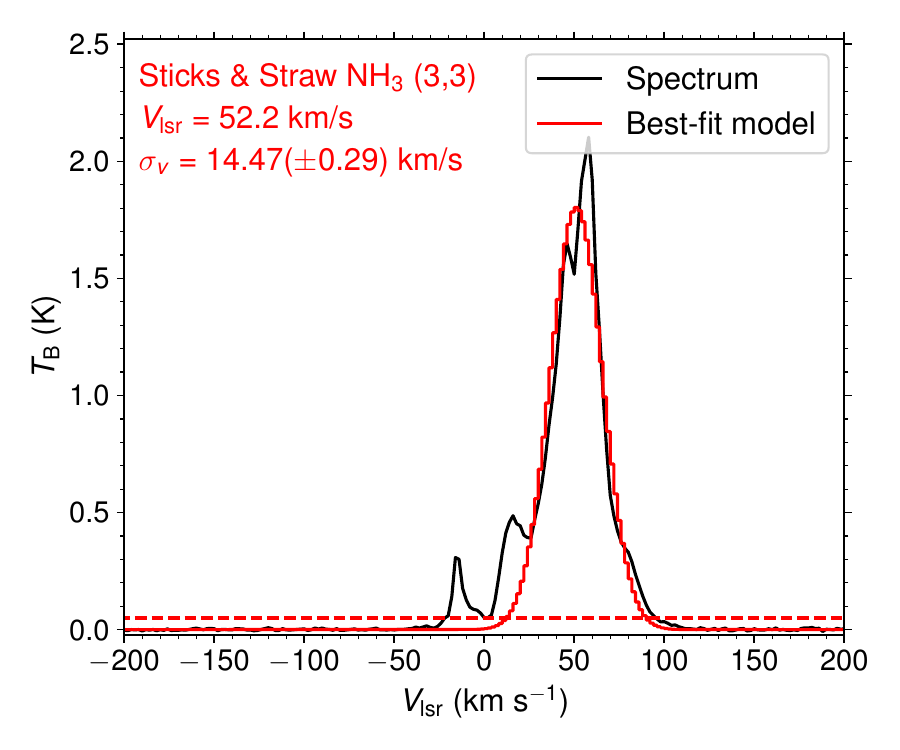} \\
\includegraphics[width=0.3\textwidth]{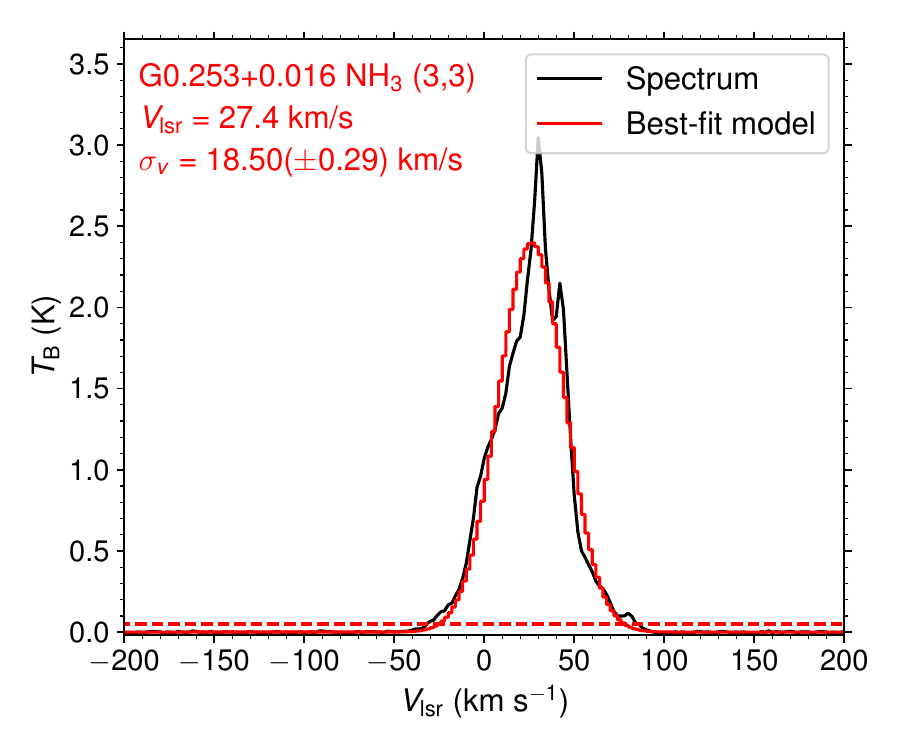}
\includegraphics[width=0.3\textwidth]{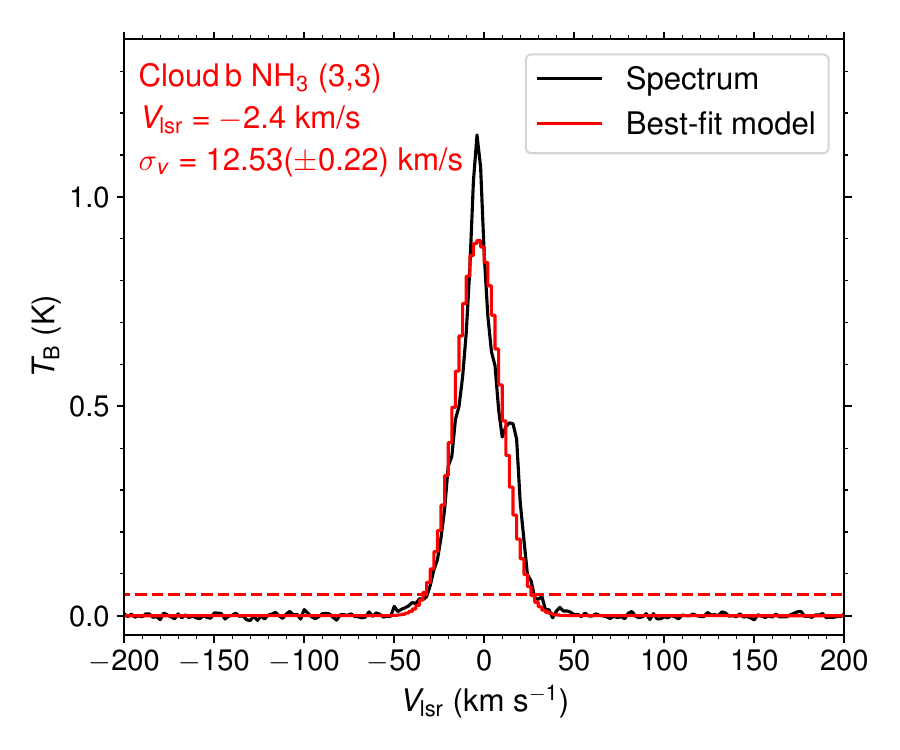}
\includegraphics[width=0.3\textwidth]{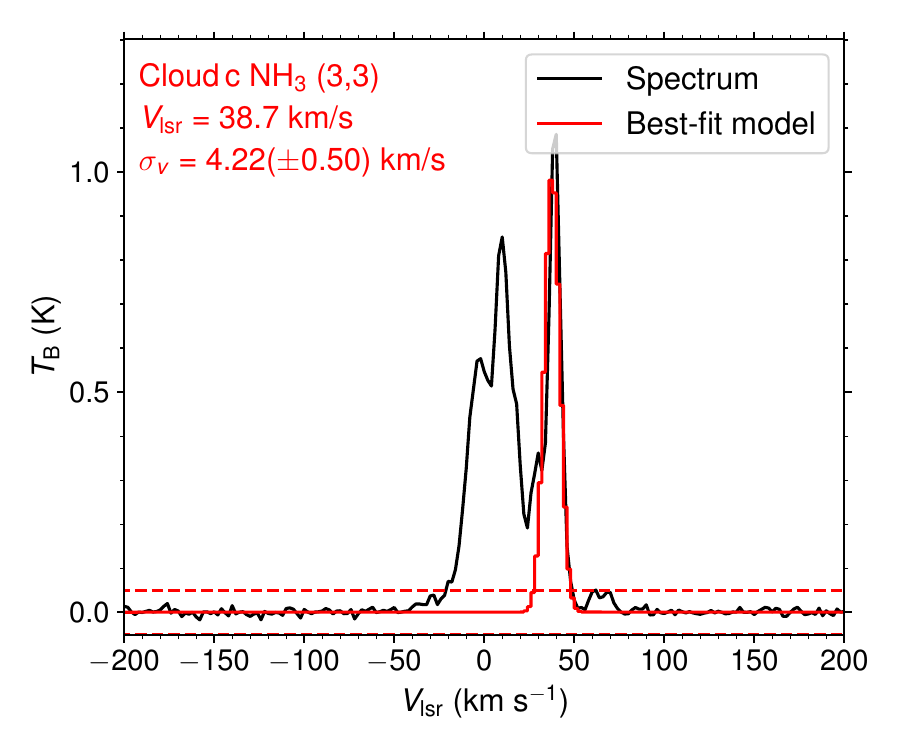} \\
\includegraphics[width=0.3\textwidth]{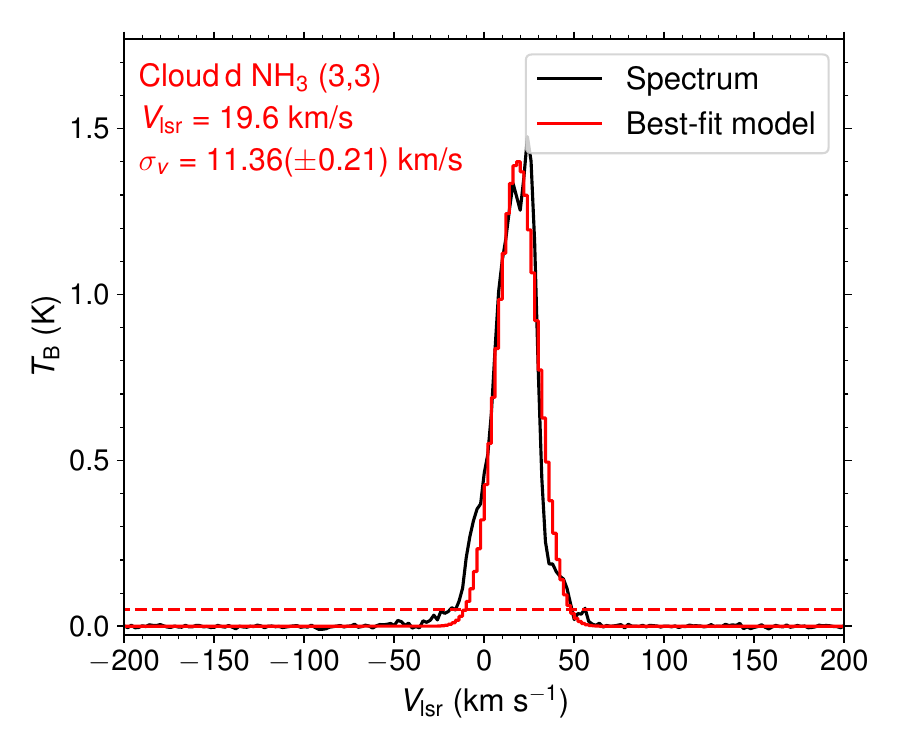}
\includegraphics[width=0.3\textwidth]{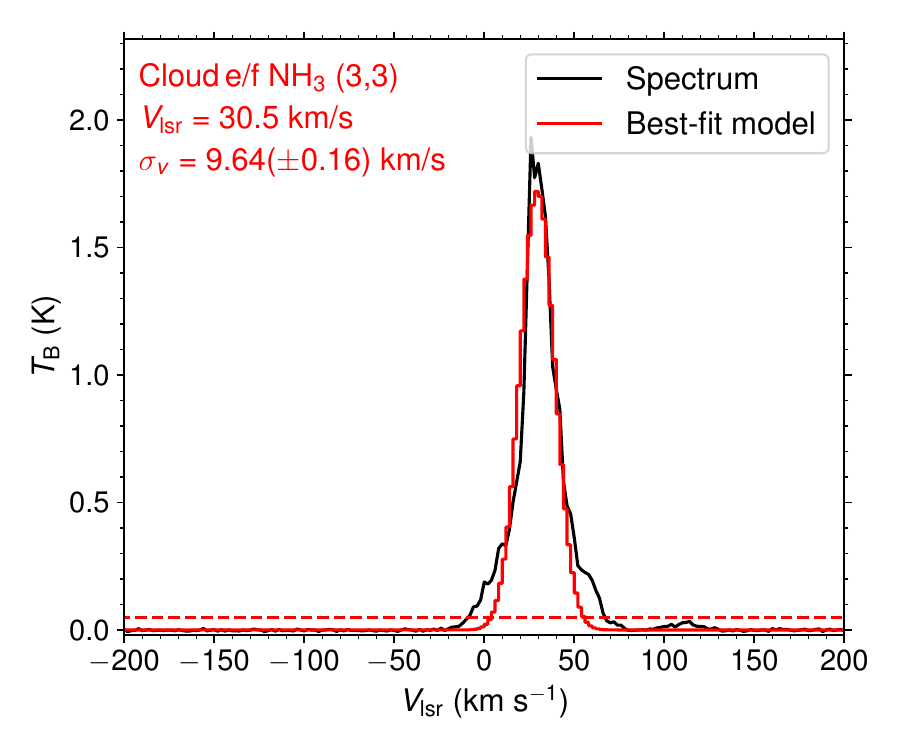}
\caption{Mean SWAG \ammthree{} spectra of the 11 clouds and single-component Gaussian fittings. The best-fit centroid velocity (\vlsr{}) and velocity dispersion ($\sigma_v$) are given in each panel.}
\label{app_fig:nh333}
\end{figure*}

\autoref{app_fig:nh333} displays the mean \ammthree{} spectra toward the clouds taken from the SWAG survey and Gaussian fittings to the spectra. The spectra are averaged over all pixels above the threshold of 0.2~\jypbm{} in the total intensity map from our data. For clouds showing multiple velocity components, the component with a higher peak is chosen to represent the principle part of the gas. We additionally cross check with observations of dense cores at higher angular resolutions to confirm that the velocity of the chosen component is consistent with those of the dense cores \citep{lu2019a,callanan2023}.

\begin{figure*}[!thpb]
\centering
\includegraphics[width=0.3\textwidth]{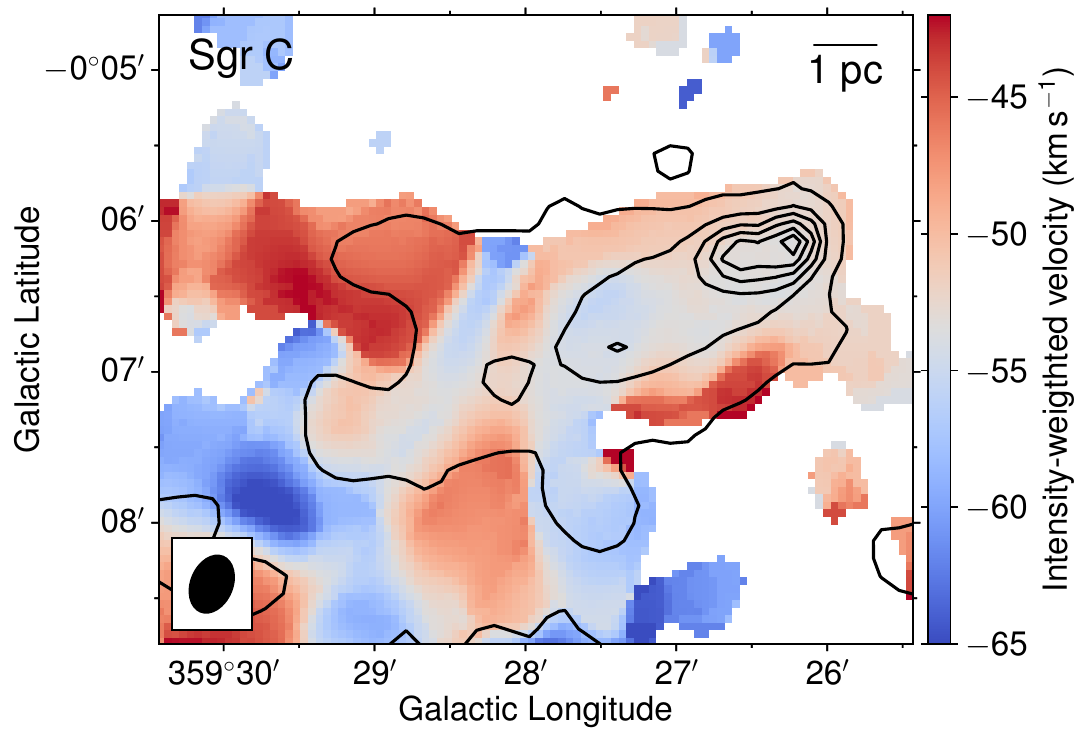}
\includegraphics[width=0.3\textwidth]{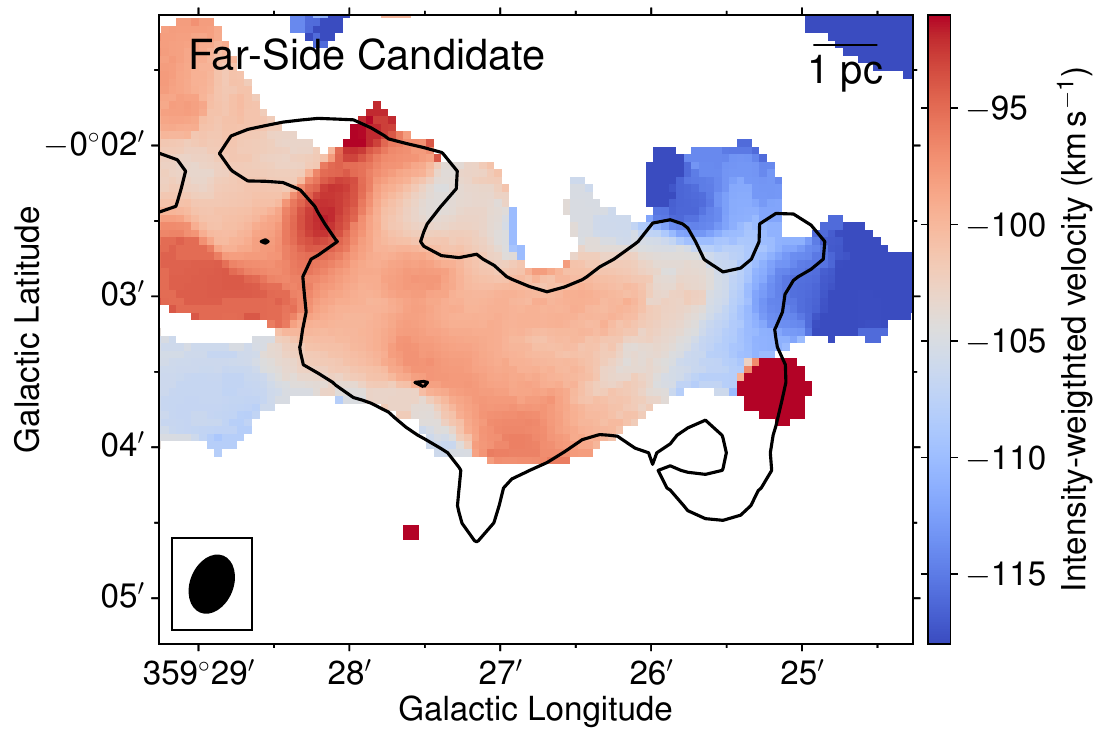}
\includegraphics[width=0.272\textwidth]{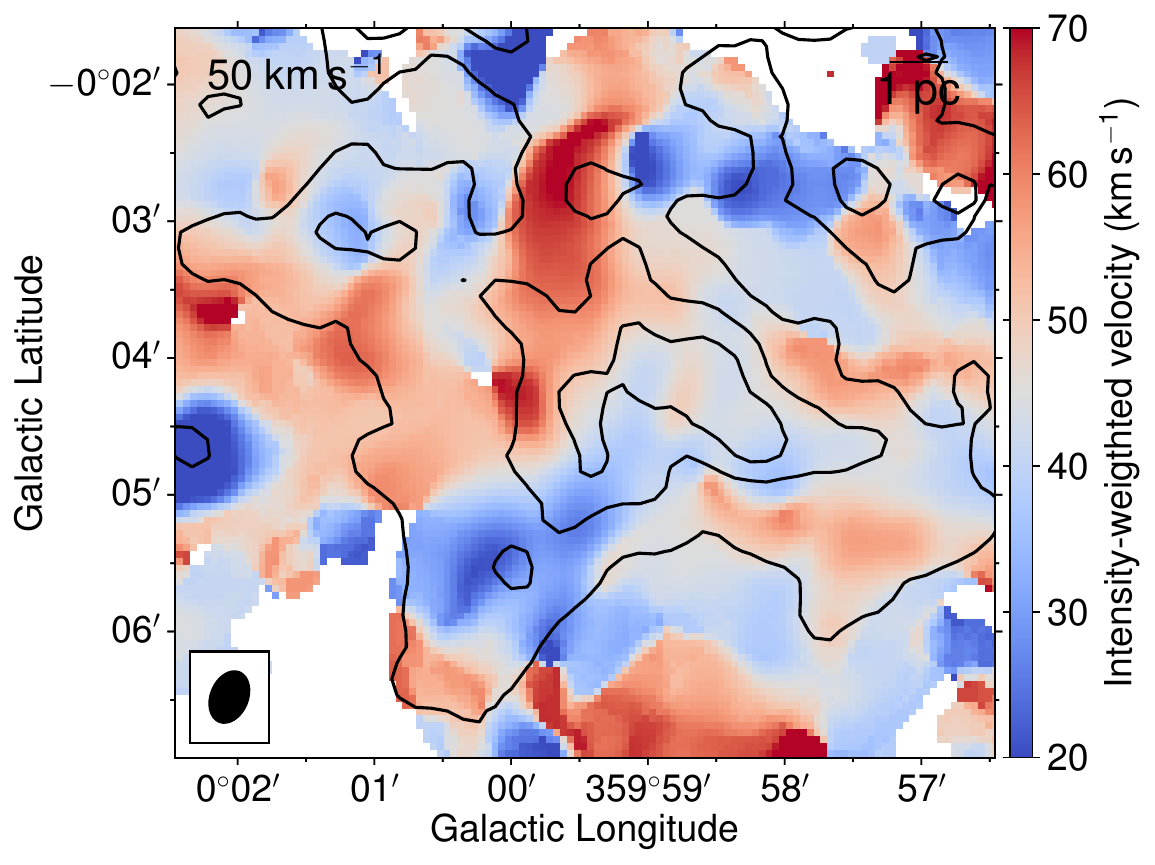} \\
\includegraphics[width=0.3\textwidth]{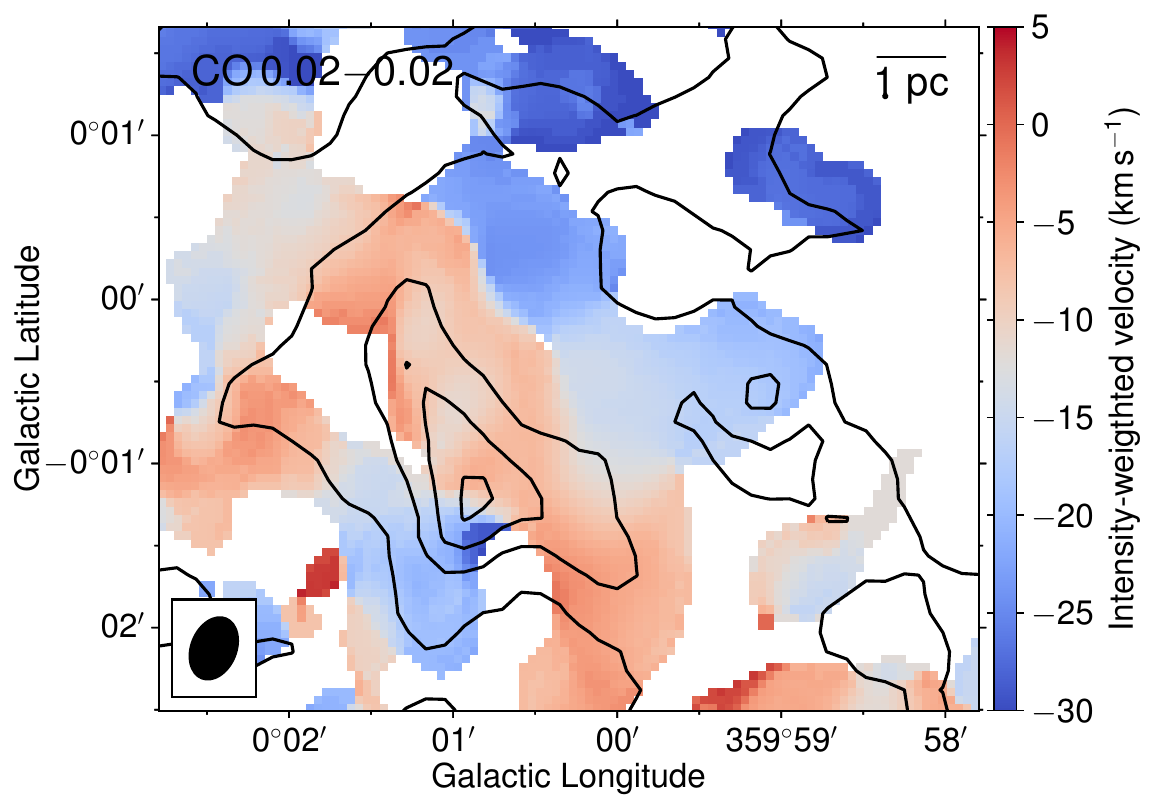}
\includegraphics[width=0.3\textwidth]{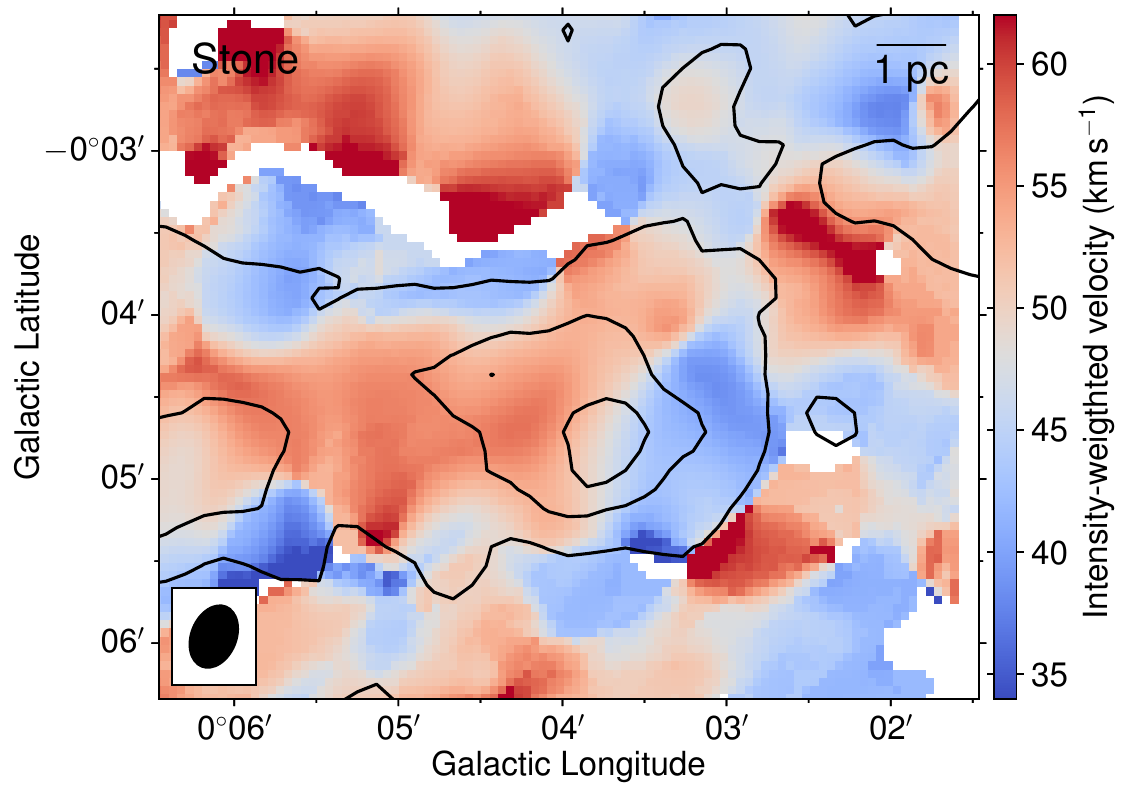}
\includegraphics[width=0.3\textwidth]{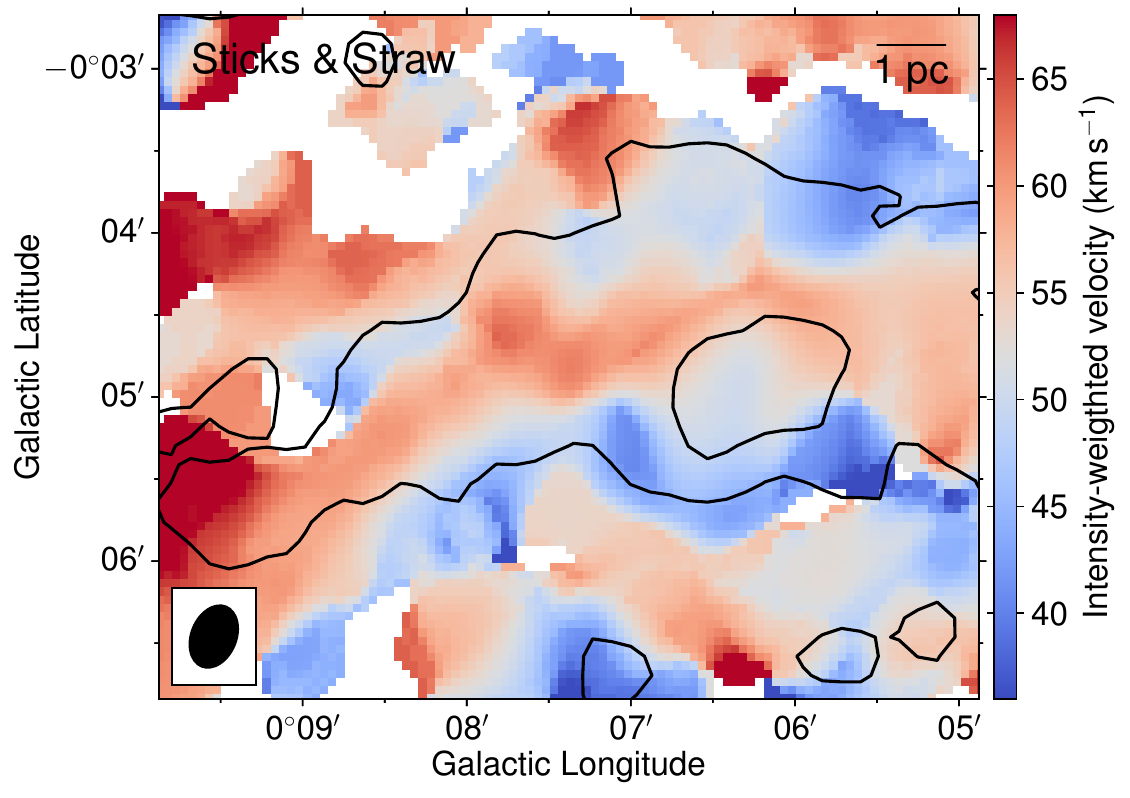} \\
\includegraphics[width=0.3\textwidth]{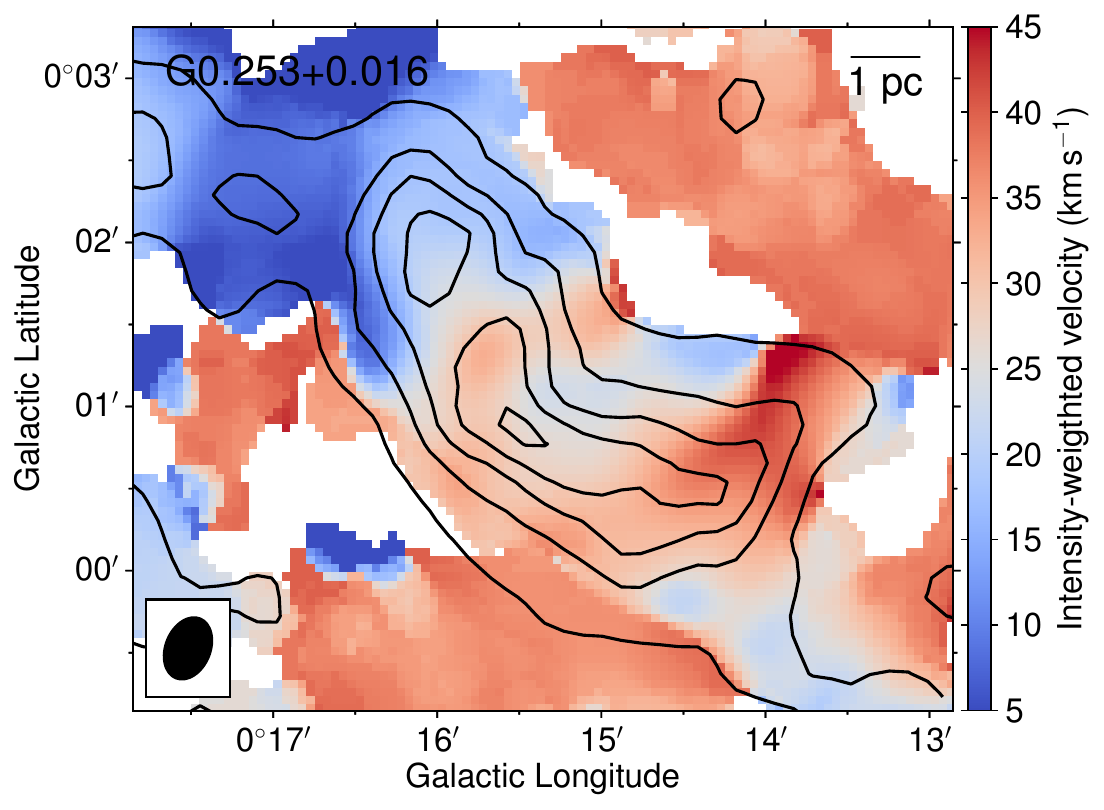}
\includegraphics[width=0.3\textwidth]{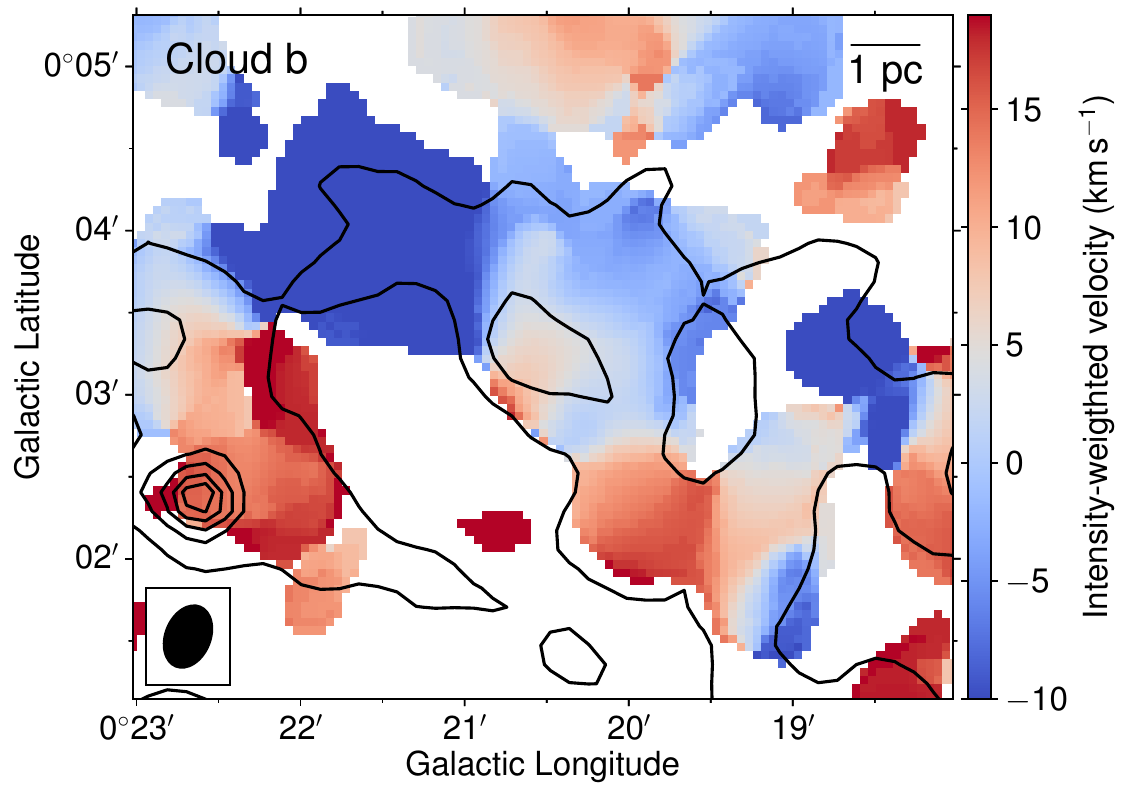}
\includegraphics[width=0.266\textwidth]{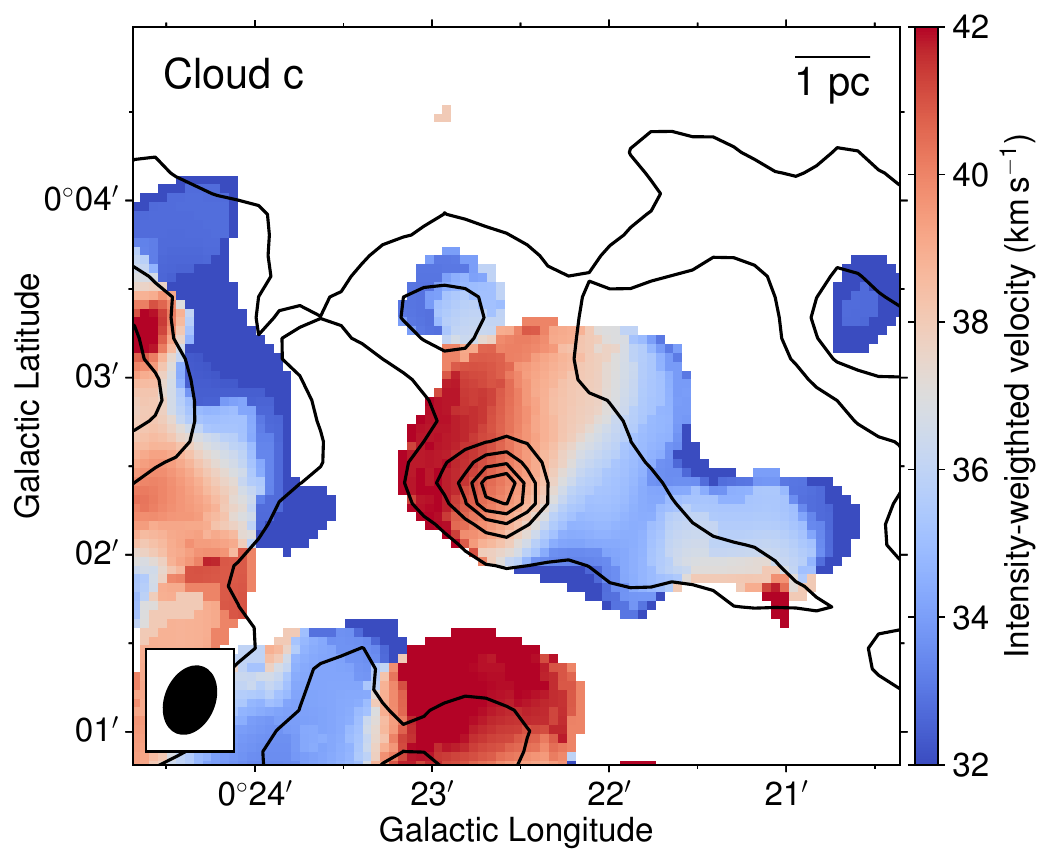} \\
\includegraphics[width=0.3\textwidth]{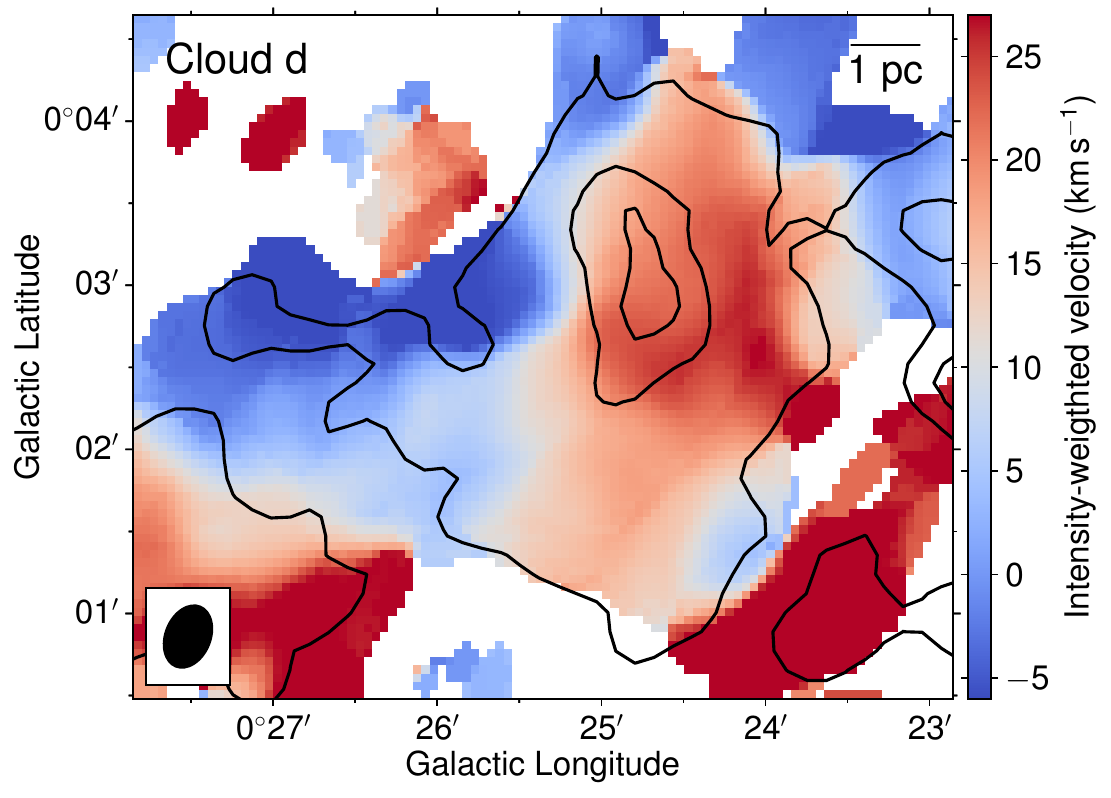}
\includegraphics[width=0.3\textwidth]{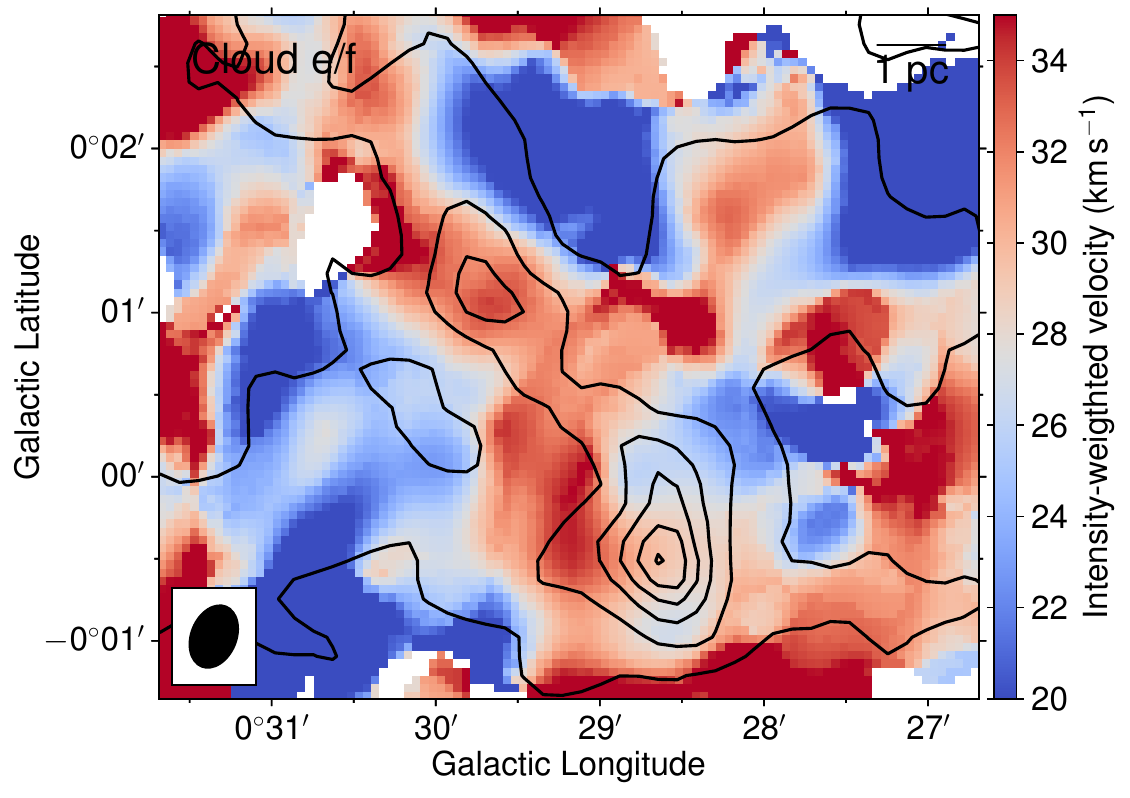}
\caption{Intensity-weighted velocity maps of the 11 clouds made from the SWAG \ammthree{} data. The synthesized beam of the ATCA observations from the SWAG program is displayed at the bottom left corner of each panel. Contours are the same as in \autoref{fig:allcmz}, representing the JCMT 850~\micron{} total intensity emission.}
\label{app_fig:nh333mom1}
\end{figure*}

We additionally show the intensity-weighted velocity (1st moment) maps of the clouds in \autoref{app_fig:nh333mom1}. The velocity ranges for making the 1st moment maps are adjusted to match those in the Gaussian fittings in \autoref{app_fig:nh333}. Some clouds, e.g., the Stone cloud, \gzp{}, and cloud~b, seem to present large velocity gradients of $>$10~\kms{} across the area where our analysis is carried out. For example, the velocity gradient in \gzp{} has been interpreted as different velocity components along the line of sight \citep{henshaw2019} or shear motions \citep{federrath2016b}. Therefore, the linewidths measured toward these clouds are likely overestimated and should be treated as upper limits.

\section{Uncertainties in the estimated physical parameters}\label{appd_sec:uncertainty}

The uncertainties in the magnetic field strengths estimated through \autoref{equ:DCF} come from two sources. Firstly, the uncertainties of the parameters on the right-hand side of the equation, including the ratio $\left(\langle B_\text{t}^{\phantom{\text{t}}2}\rangle/\langle B^2\rangle\right)^{0.5}$, the velocity dispersion $\sigma_v$, the density $\rho$, and the correction factor 0.21 can be propagated into that of the magnetic field strength. Second, the DCF method itself has inherent uncertainties due to unsatisfaction of assumptions, e.g., an overestimation of the magnetic field strength in super Alfv\'enic conditions \citep{liu2021}.

We first quantify the first source of uncertainties. The uncertainties in $\left(\langle B_\text{t}^{\phantom{\text{t}}2}\rangle/\langle B^2\rangle\right)^{0.5}$, derived from the covariance of the ADF fitting (\autoref{appd_sec:adf}), are annotated in \autoref{app_fig:adfs}. The errors in $\sigma_v$ from the Gaussian fittings can be found in \autoref{app_fig:nh333}. As discussed in \autoref{subsec:results_mass}, the uncertainties in $\rho$ is typically 50\%. The correction factor has an uncertainty of 45\% \citep{liu2021}. Propagating these errors, we obtain a characteristic uncertainty of 50\% for the magnetic field strength $B$, which is dominated by that of the correction factor.

There is another systematic uncertainty that is not included the error analysis above. The velocity dispersion $\sigma_v$ could be significantly overestimated given the presence of multiple velocity components in the clouds (see \autoref{appd_sec:ammonia}). \citet{henshaw2016} identified a mean number of Gaussian components per position in the CMZ of 1.6. Therefore, we expect that on average the measured $\sigma_v$ overestimate the true value by a factor of 1.6. This clearly dominates over the statistical uncertainties in $B$.

The inherent uncertainty from the DCF method itself, e.g., because of its unrealistic assumptions, is not straightforward to quantify, yet it often dominates the uncertainty in the estimated magnetic field. \citet{liu2021} estimated an uncertainty of at least a factor of 2, which is derived from comparison between input models and results from applying the DCF method to numerical simulations.

To summarize, the uncertainty in the magnetic field strength is dominated by the systematic uncertainty in the velocity dispersion (overestimation of a factor of $\sim$1.6) and the inherent uncertainty of the DCF method ($\gtrsim$2). As such, we adopt an uncertainty of a factor of 2 for the magnetic field strength, and note that it is a lower limit.

The physical parameters derived in \autoref{subsec:results_energies}, including the Alfv\'{e}nic Mach numbers, the mass-to-flux ratios, and the virial parameters, have dependency on the magnetic field strengths, which dominate the uncertainty. Therefore, we also adopt a lower limit for the uncertainty in these parameters of a factor of 2.

\section{Distributions of magnetic field position angles as probed by JCMT/POL2 and ACT}\label{appd_sec:bpa}

\begin{figure*}[!thpb]
\centering
\includegraphics[width=0.3\textwidth]{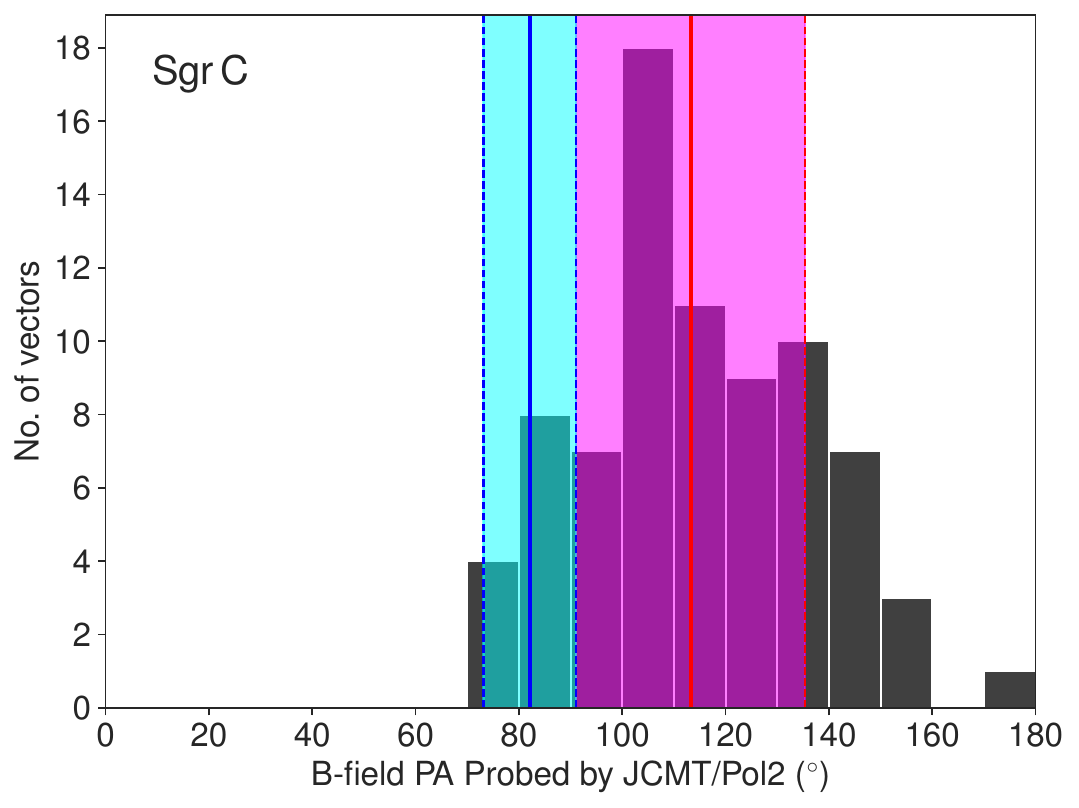}
\includegraphics[width=0.3\textwidth]{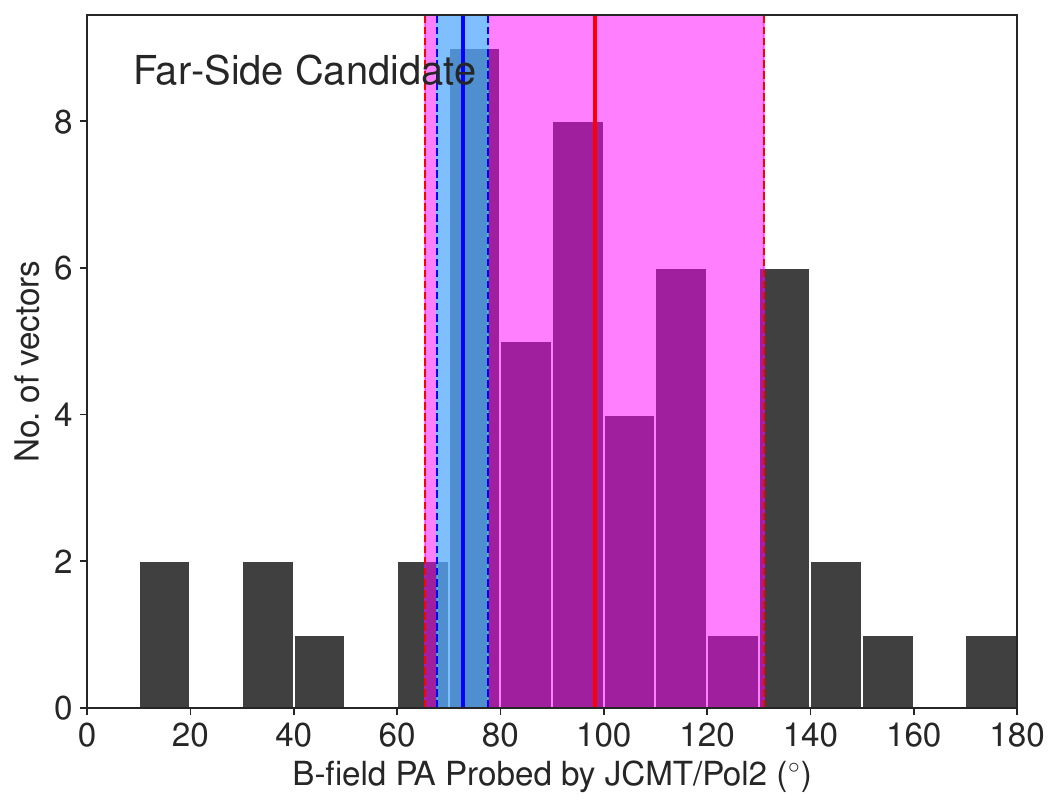} 
\includegraphics[width=0.3\textwidth]{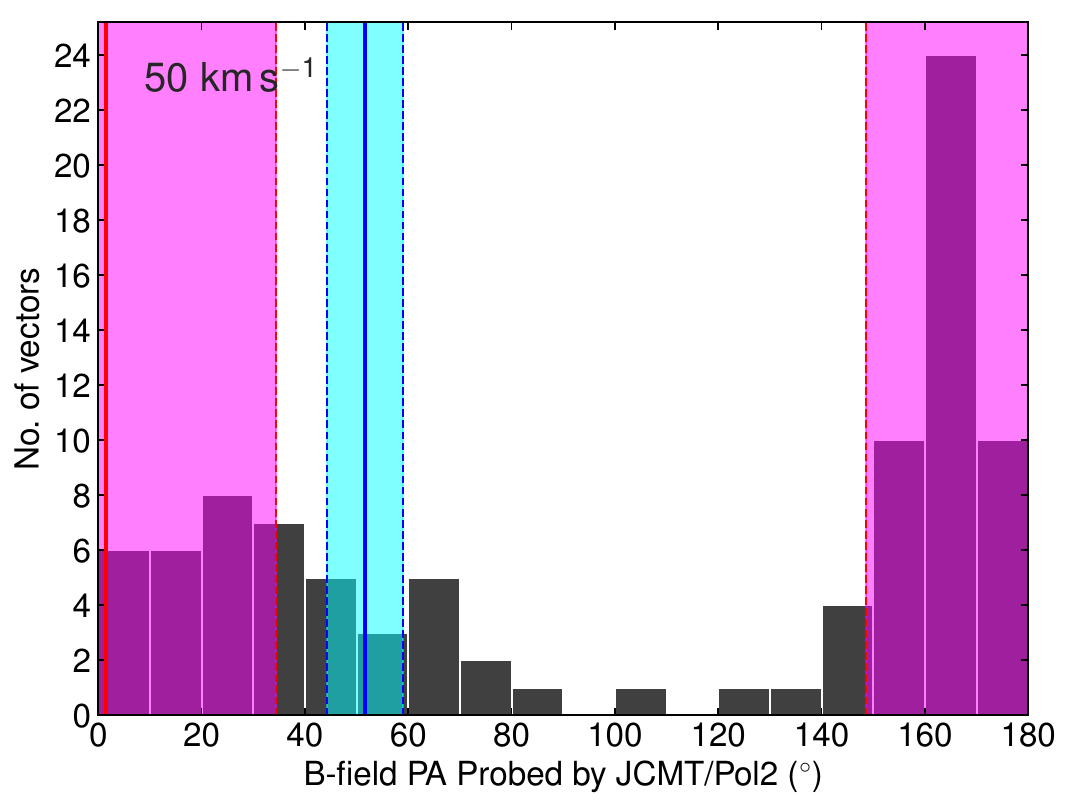} \\
\includegraphics[width=0.3\textwidth]{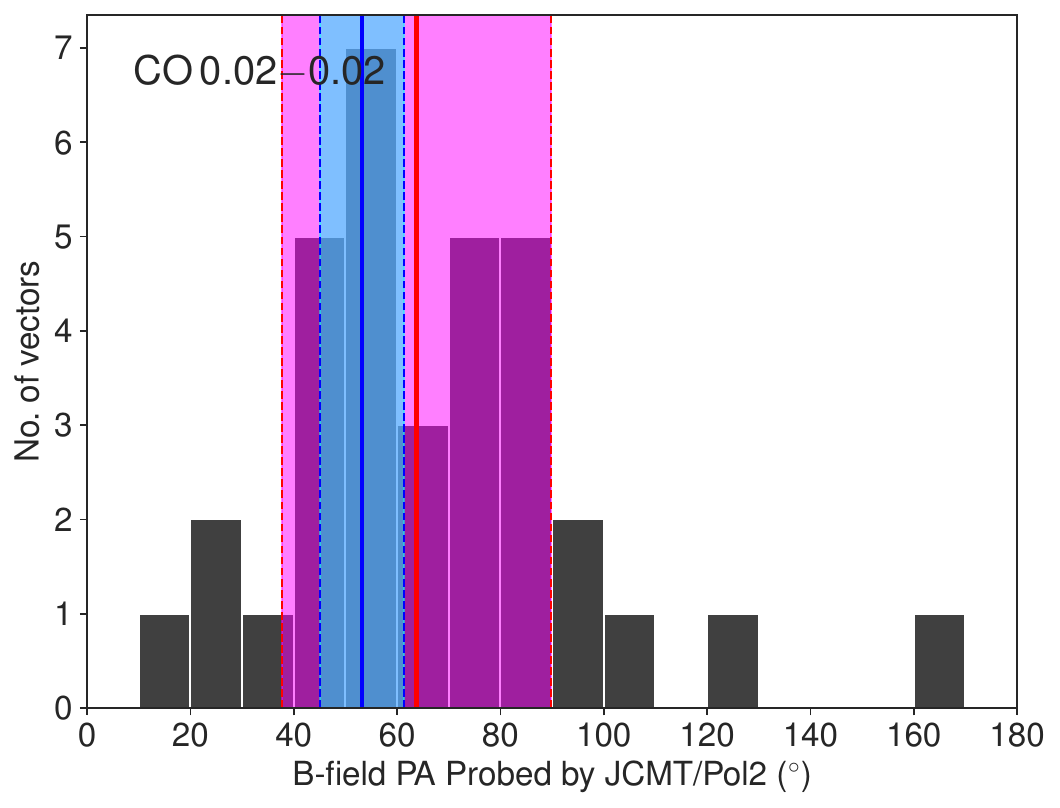}
\includegraphics[width=0.3\textwidth]{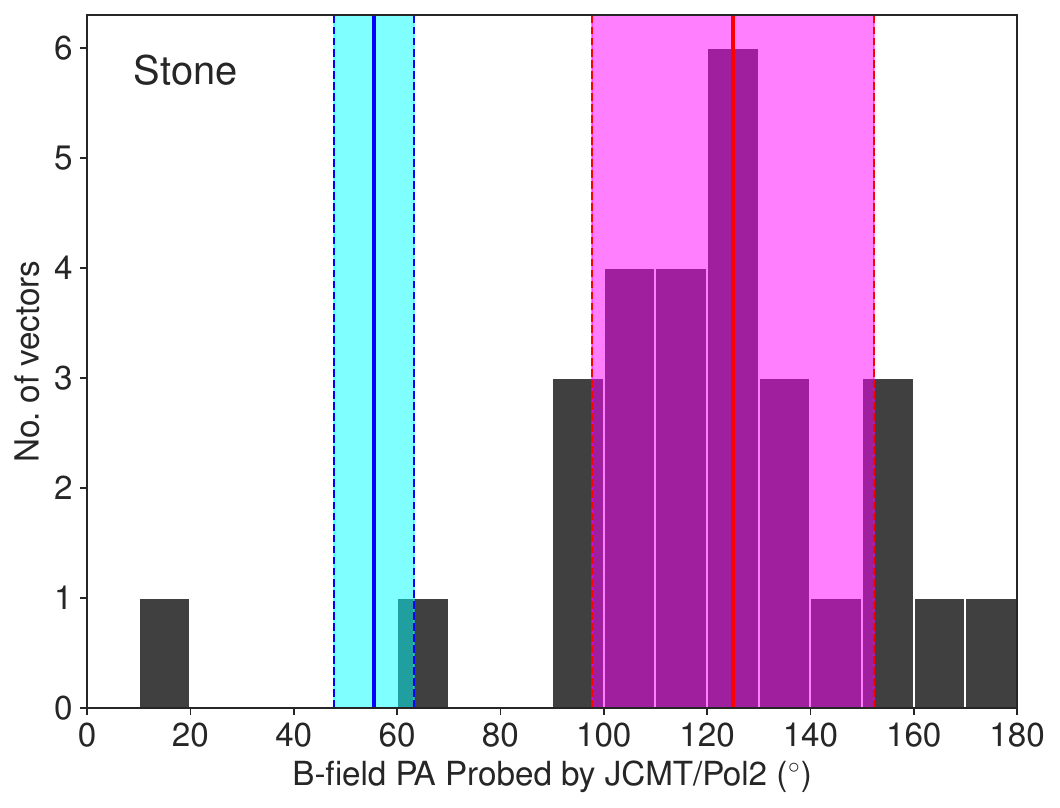} 
\includegraphics[width=0.3\textwidth]{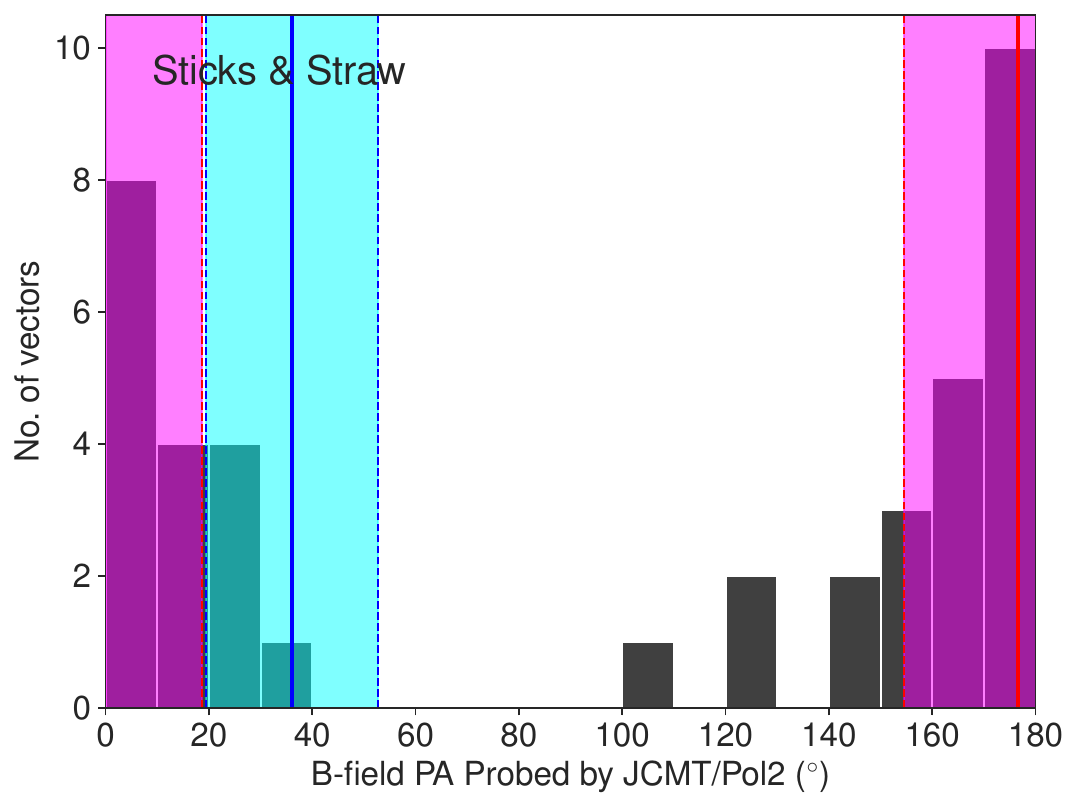} \\
\includegraphics[width=0.3\textwidth]{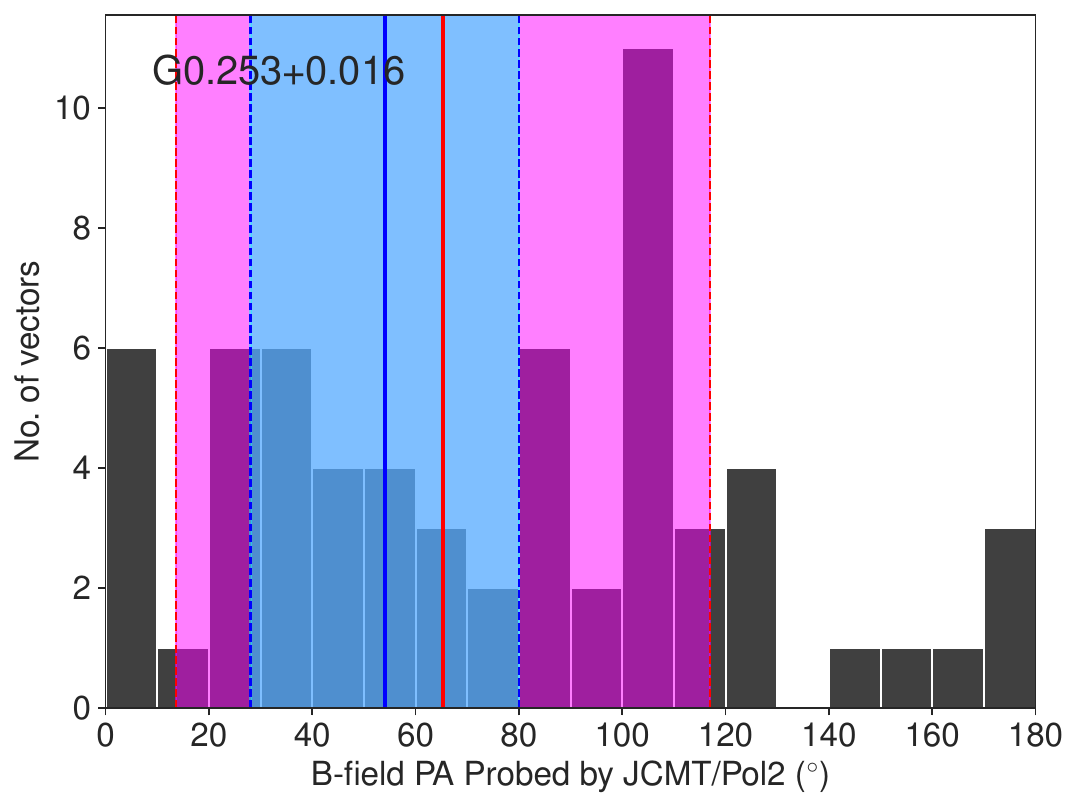}
\includegraphics[width=0.3\textwidth]{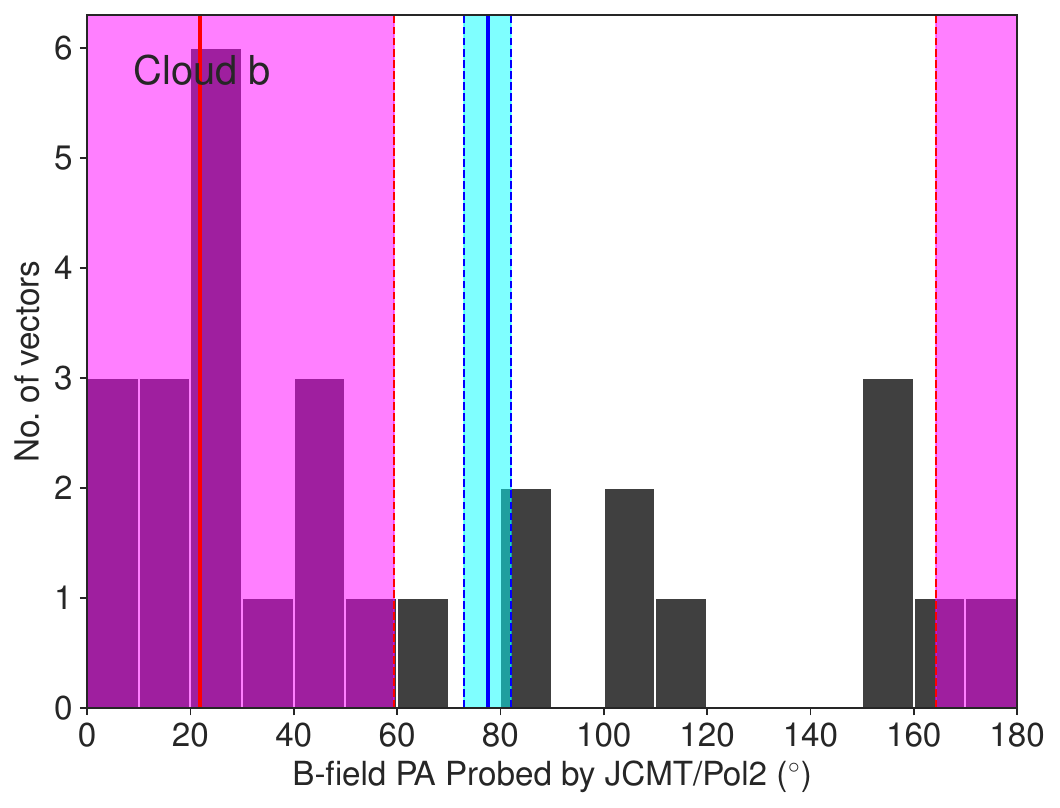} 
\includegraphics[width=0.3\textwidth]{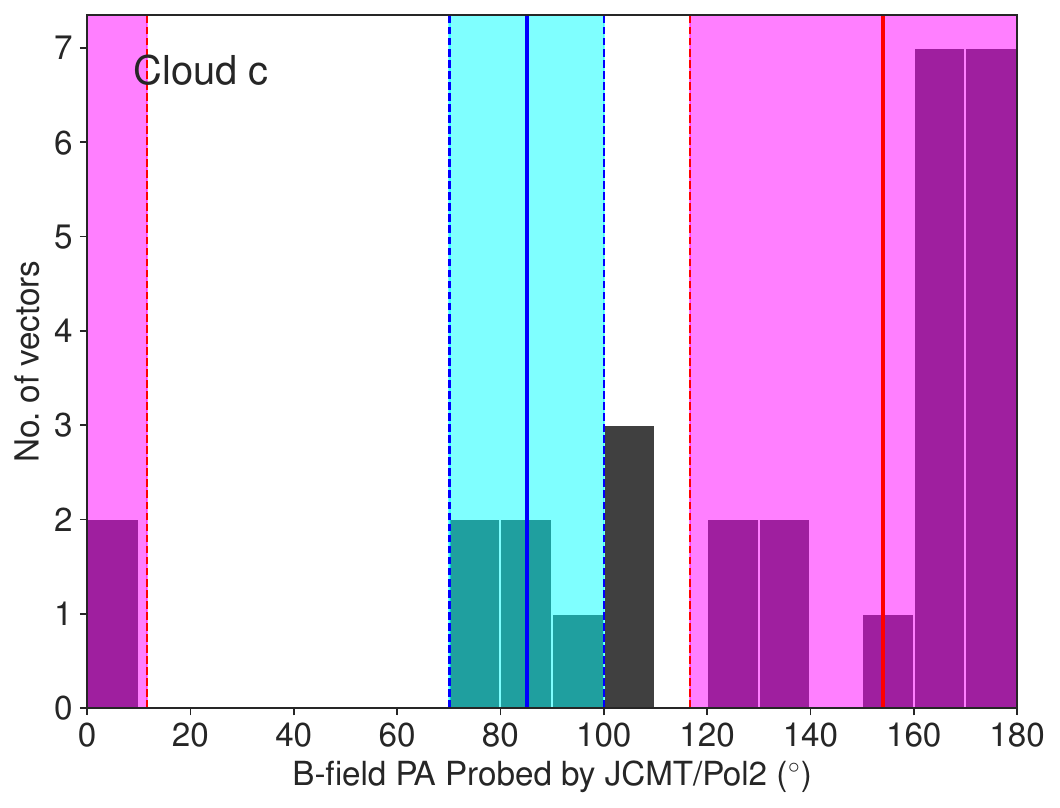} \\
\includegraphics[width=0.3\textwidth]{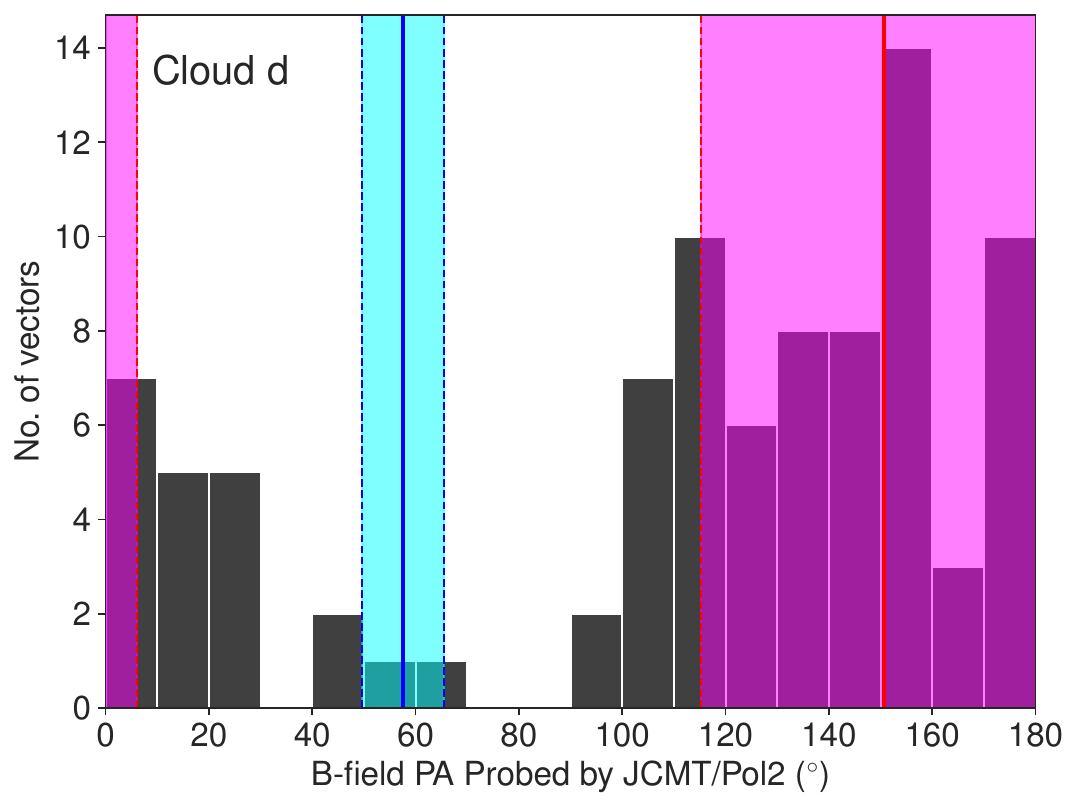}
\includegraphics[width=0.3\textwidth]{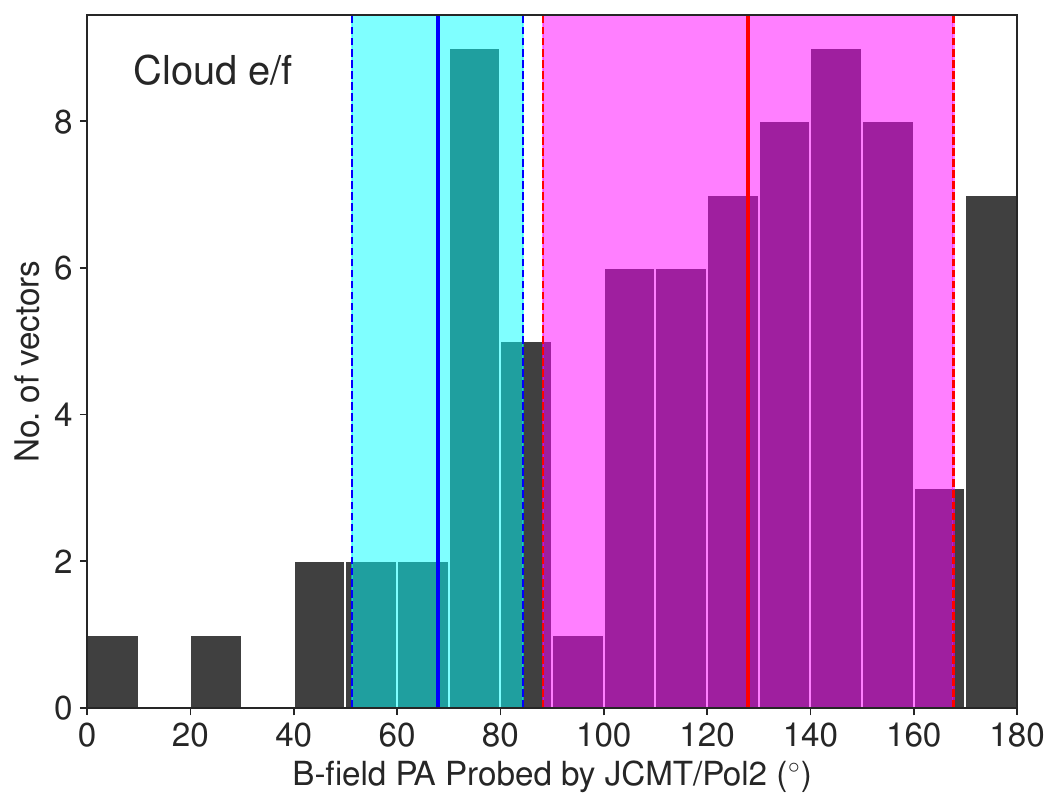} 
\caption{The histograms show magnetic field position angles as probed by JCMT/POL2. The solid red vertical lines denote the mean position angles, and the dashed red vertical lines denote the standard deviations of the position angles in the clouds. Similarly, the blue shaded areas represent the magnetic field position angles as probed by ACT \citep{guan2021}, with the central solid blue vertical lines denoting the mean position angles and the dashed blue vertical lines denoting the standard deviations of the position angles found inside the cloud boundaries. All the position angles follow the IAU-recommended  north-to-east definition.}
\label{app_fig:comparison_BPA}
\end{figure*}

\autoref{app_fig:comparison_BPA} displays histograms of magnetic field position angles as probed by JCMT/POL2 in each cloud, as well as the mean magnetic field position angle probed by ACT.

\begin{figure*}[!thpb]
\centering
\includegraphics[width=0.5\textwidth]{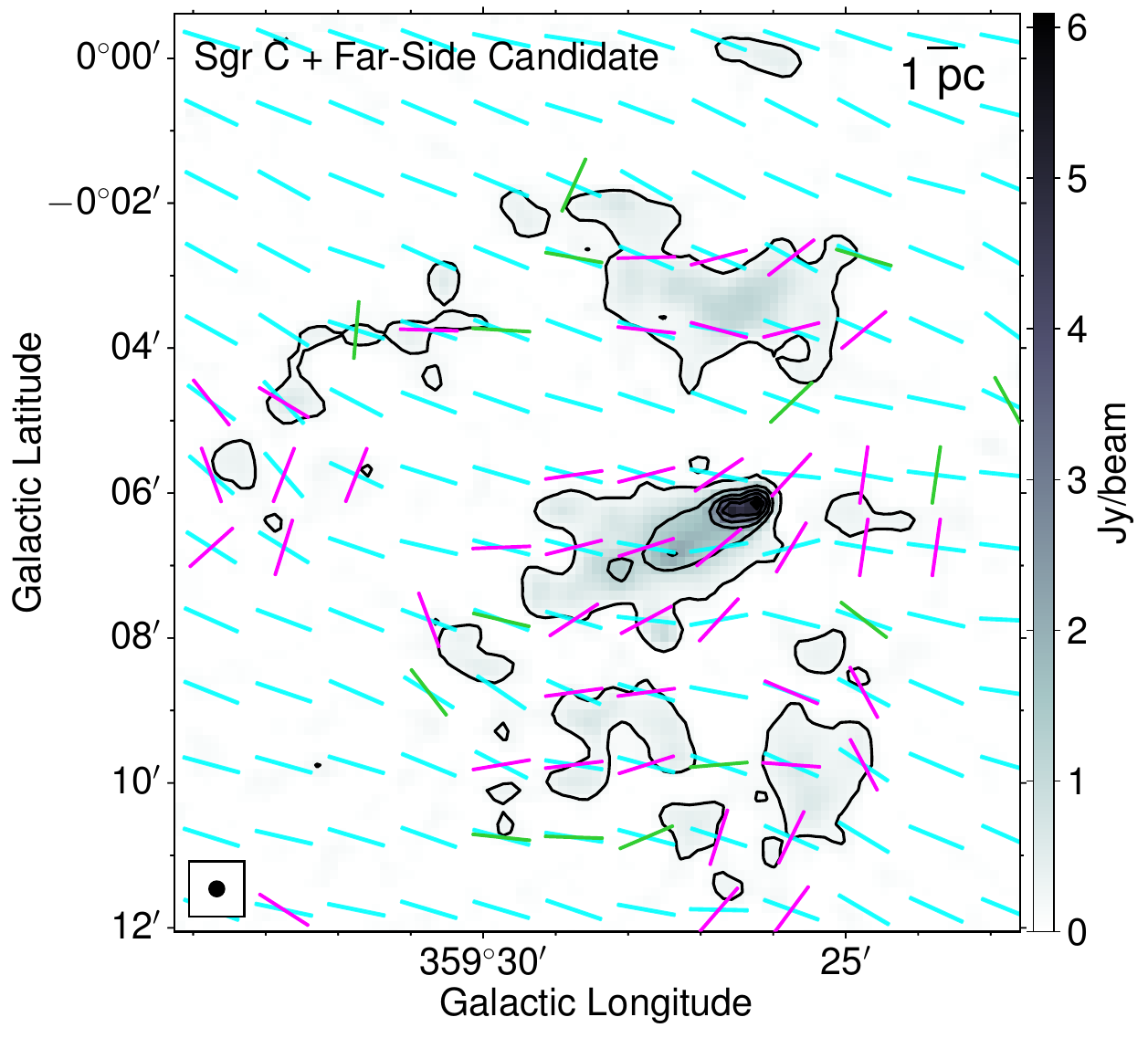} \\
\includegraphics[width=0.6\textwidth]{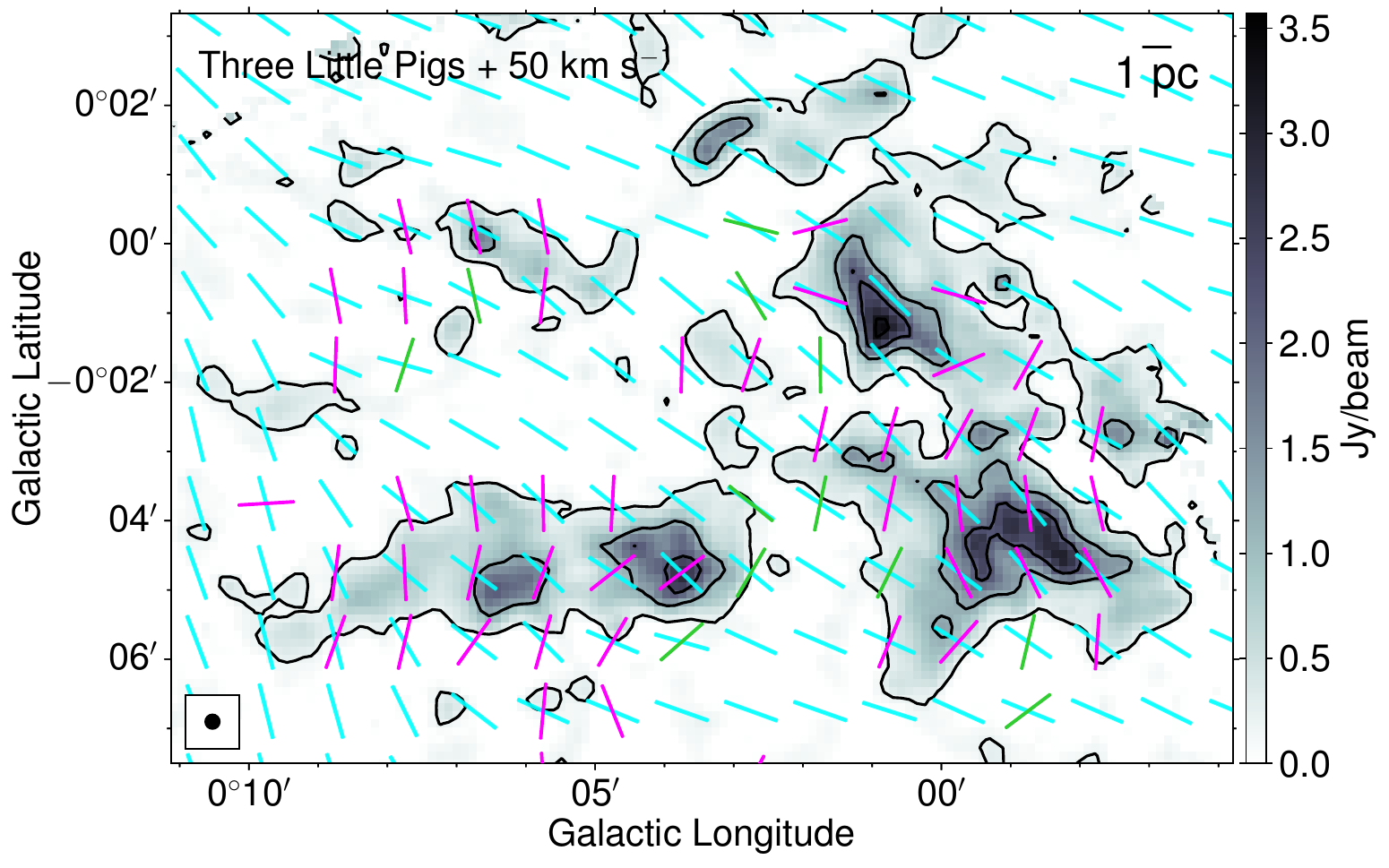} \\
\includegraphics[width=0.72\textwidth]{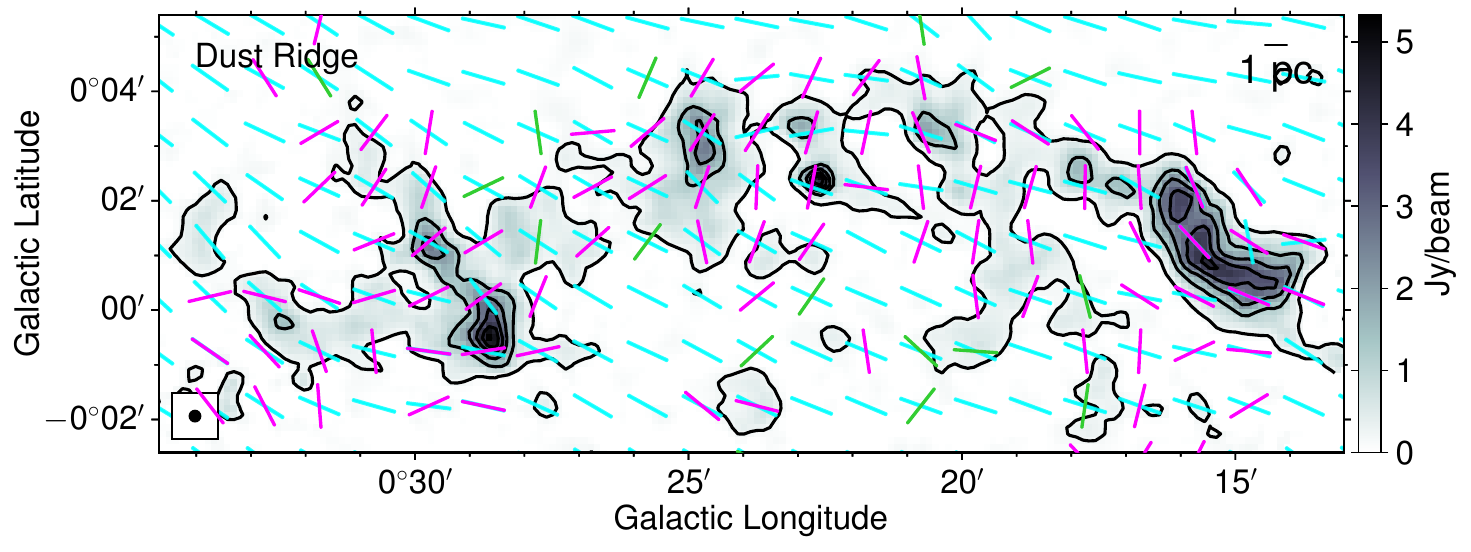}
\caption{The segments show orientations of magnetic fields probed by ACT (cyan) and by JCMT/POL2 smoothed and regridded to the same frame as ACT (magenta and green, illustrating those with $p$/d$p$$>$3 and 2$<$$p$/d$p$$<$3, respectively, following the convention in \autoref{fig:act}). Contours are the same as in \autoref{fig:allcmz}, representing the JCMT 850~\micron{} total intensity emission.}
\label{app_fig:comparison_BPAmaps}
\end{figure*}

Additionally, in \autoref{app_fig:comparison_BPAmaps} we present the magnetic field position angles derived from the JCMT/POL2 data smoothed and regridded to the same frame as the ACT observations of $1'$ resolution. Here we have smoothed the Stokes Q and U components of the POL2 data and then re-derived the magnetic field position angles. The two datasets reveal different field morphologies. Interpretations of this difference can be found in \autoref{subsec:results_morph}.

\bibliographystyle{aasjournal}
\bibliography{my_2022}

\end{CJK}
\end{document}